\documentclass[%
superscriptaddress,
twocolumn,
amsmath,amssymb,
 aps,
pra,
]{revtex4-1}

\usepackage[T2A,T1]{fontenc}
\usepackage{braket}
\usepackage{graphicx}
\usepackage{graphics}
\usepackage[colorinlistoftodos]{todonotes}
\usepackage[utf8]{inputenc}
\usepackage{amsthm}
\usepackage{amsmath}
\usepackage{physics} 
\usepackage{amsfonts}
\usepackage{slashed}
\usepackage[normalem]{ulem}
\usepackage{soul}
\usepackage{graphicx,times}
\usepackage{amstext}
\usepackage{amsmath}            
\usepackage{amssymb}            
\usepackage{latexsym}
\usepackage{bm}
\usepackage{color}
\usepackage{etoolbox}
\usepackage[FIGTOPCAP,raggedright,nooneline]{subfigure}
\usepackage{epstopdf}
\usepackage{lmodern}
\usepackage[english]{babel}
\usepackage{ae}
\usepackage{units}
\usepackage[americaninductors]{circuitikz}
\usepackage{tikz}
\usepackage{multirow}

\usetikzlibrary{arrows,snakes,calc}
\makeatletter
\let\cat@comma@active\@empty
\makeatother
\usepackage{url}
\usepackage[colorlinks]{hyperref}
\hypersetup{%
	plainpages=true,
	breaklinks=true,
	hypertexnames=false,
	pageanchor=true,
	colorlinks=true,
	linkcolor={blue},
	citecolor={red},
	urlcolor={blue},
	anchorcolor={black}
}

\newcommand{\ex}[1]{\langle #1 \rangle}

\newcommand{\arccosh}{\text{arccosh}}

\newcommand{\beq}{\begin{eqnarray}}
\newcommand{\eeq}{\end{eqnarray}}

\def\<{\langle}
\def\>{\rangle}

\def \info#1{}

\newcommand{\be}{\begin{equation}}
\newcommand{\ee}{\end{equation}}
\newcommand{\bea}{\begin{eqnarray}}
\newcommand{\eea}{\end{eqnarray}}

\begin{document}

\title{Bound states in ultrastrong waveguide QED}

\author{Juan Román-Roche}
\affiliation {Instituto de Ciencia de Materiales de
  Aragón and Departamento de Física de la Materia Condensada ,
  CSIC-Universidad de Zaragoza, Pedro Cerbuna 12, 50009 Zaragoza,
  Spain}
  
\author{Eduardo Sánchez-Burillo}

\affiliation{Max-Planck-Institut für Quantenoptik, D-85748 Garching, Germany}
  
\author{David Zueco}
\affiliation {Instituto de Ciencia de Materiales de
  Aragón and Departamento de Física de la Materia Condensada ,
  CSIC-Universidad de Zaragoza, Pedro Cerbuna 12, 50009 Zaragoza,
  Spain}
\affiliation{Fundación ARAID, Campus Río Ebro, 50018 Zaragoza, Spain}

\date{\today}

\begin{abstract}
We discuss the   properties of bound states in finite-bandwidth waveguide QED beyond the Rotating Wave Approximation or excitation number conserving light-matter coupling models. 
Therefore, we extend  the \emph{standard} calculations to a broader range of light-matter strengths, in particular,  in the so-called ultrastrong coupling regime.
We do this using the Polaron technique.
Our main results are as follows.
We compute the spontaneous emission rate, which is renormalized as compared to the Fermi Golden Rule formula.
We generalise the  existence criteria for bound states, their properties and their role in the qubits thermalization.
We discuss  effective spin-spin interactions through both vacuum fluctuations and bound states.
Finally, we
sketch a perfect state-transfer protocol among distant emitters.
\end{abstract}

\maketitle

\section{Introduction}

Photons are weakly coupled to matter, so they rarely interact, making them perfect information carriers.
Weak coupling constitutes, however, a double-edged sword, as it also hinders the readout process when the time comes to access the information being carried.
\emph{The} trade off is optimized in waveguide QED, where the photons are confined in one dimensional waveguides to enhance the light-matter coupling \cite{roy2017, gu2017}.
So far, different experimental platforms have been used to implement this coupling between quantum emitters (typically two level systems or qubits) and a one dimensional quantized electromagnetic field.
Examples are superconducting circuits \cite{astafiev2010,van2013, liu2017}, optical waveguides \cite{faez2014} among others \cite{lodahl2015,chang2018}.
Waveguide QED  can serve to control light-matter emission, to induce photon-photon interactions or to route the photons in quantum networks. 
Besides, by engineering the guides, more exotic interfaces can be implemented for quantum simulation \cite{arguello2019}, topological photonics \cite{bello2019}, chirality \cite{lodahl2017, sanchez2019b} or quantum computing \cite{zheng2013}.
Consequently, waveguide QED may be a quantum technological solution.

Trying to optimize the light-matter coupling, several experiments have reached the so-called ultrastrong coupling  regime (USC)  between light and a single quantum emitter, both in cavity  \cite{Niemczyk2010,forn2010} and waveguide QED \cite{forn2017, martinez2019, leger2019}.
The USC is the regime where higher order processes,  than the creation (annihilation) of \emph{one} photon by annihilating (creating) \emph{one} matter excitation play a role.
Two main phenomena are paradigmatic of USC.  
The Rotating Wave Approximation (RWA) for the interaction breaks down and the atomic bare parameters get renormalized, either  the Bloch-Siegert shift  \cite{shirley1965} in cavity QED or the  renormalization  due to the coupling to the continuum electromagnetic (EM) field in waveguide QED \cite{Leggett1987}. Besides, the ground state becomes nontrivial \cite{Ashhab2010}.
This has interesting consequences.  Some of them are the possibility of transforming virtual onto real photons by perturbing the ground state \cite{Ciuti2005, Stassi2013, QiKai2018, DeLiberato2017}, doing nonlinear optics with zero photons \cite{Stassi2017}.
Further phenomenology in cavity QED can be found in recent reviews \cite{kockum2019, forn2019}.
In this  work we are interested in the USC regime in waveguide QED. 
Apart from the qubit frequency renormalization, there exist the localization-delocalization transition \cite{peropadre2013, Shi2018},  particle production \cite{gheeraert2018},  non-linear optics at the single photon limit \cite{sanchez2014, sanchez2015} or vacuum light emission \cite{sanchez2019}.

In \emph{conventional} waveguide QED, {\emph i.e.} when the RWA can be performed, the main objective is to control atom-atom interactions mediated by the waveguide's EM-fluctuations \cite{dzsotjan2011,gonzalez2011, zueco2012,zheng2013b, manzoni2017}.
Propagating photons induce long range but dissipative interactions.
Dissipative because the information is lost in the travelling wavepackets.
On the other hand, dressed atom-field eigenstates localized around the quantum emitter, called bound states \cite{John1984,John1987,john1990,John1991,john1994}, generate non dissipative but exponentially-bounded interactions \cite{GonzalezTudela2015,Douglas2015,shi2016,calajo2016,GonzalezTudela2017,GonzalezTudela2017b,GonzalezTudela2018,GonzalezTudela2018b,GonzalezTudela2018d,shi2018c,bello2019,sanchez2019b}.
These exact non-propagating eigenstates lie within the band gap (hence \textit{non-propagating}).  
Besides, bound states modify the spontaneous emission \cite{Khalfin1958,Bykov1975,Fonda1978,Onley1992,gaveau1995,Garmon2009,Garmon2013,Garmon2013b,Lombardo2014,sanchez2017b} which makes them an interesting resource for engineering quantum photonics.

\emph{In this work}, we discuss the physics of bound states in the USC regime of waveguide QED.
We focus on the lowest energy ones, discussing their existence and  role in the spontaneous emission and thermalization. 
We also discuss the effective spin-spin models emerging when several emitters are ultrastrongly coupled to the EM field and envision protocols for perfect state transfer between distant atoms.
To do this, we face a technical difficulty.
The light-matter coupling is modelled via spin(s)-boson type Hamiltonians, a paradigmatic example of a non exactly solvable  model \cite{Weiss}.
Different techniques are available in the literature to deal with it.
Matrix-product states (MPS)  \cite{peropadre2013, sanchez2014, sanchez2015} , density matrix renormalization group (DMRG) \cite{prior2010} or path integral approaches \cite{grifoni1998, lehur2010}, comprise the toolbox of numerical techniques.
Analytical treatments are also used.
They are based on different varational anstatzs: Polaron-like \cite{silbey1984, Bera2014, Camacho2016, Shi2018, zueco2019, sanchez2019} or Gaussian ones \cite{shi2018b}. 
In this manuscript we will use a Polaron-type approach that has been shown to be accurate in a wide  range of parameters, including couplings well inside the USC. 

The rest of the manuscript is organized as follows.
In the next section, Sect. \ref{sec:model},  we will introduce the system, its model and the Polaron picture.
In Section \ref{sec:single}, we treat the single emitter case.  We discuss the ground state properties and the lowest bound state, discussing its existence conditions, energy, and localization length. 
Section \ref{sec:multi} develops the multiqubit case with emphasis in the effective tight-binding model and in protocols for perfect state transfer.
We finish with some conclusions. Several technical issues are sent to the appendices.
Finally, the link to the python codes used in the numerical calculations is given in App. \ref{ap:code}. 
\section{Light-matter interaction and the Polaron picture}
\label{sec:model}

\subsection{Model}

\begin{figure*}[!htb]
    \centering
    \includegraphics[width=0.70\textwidth]{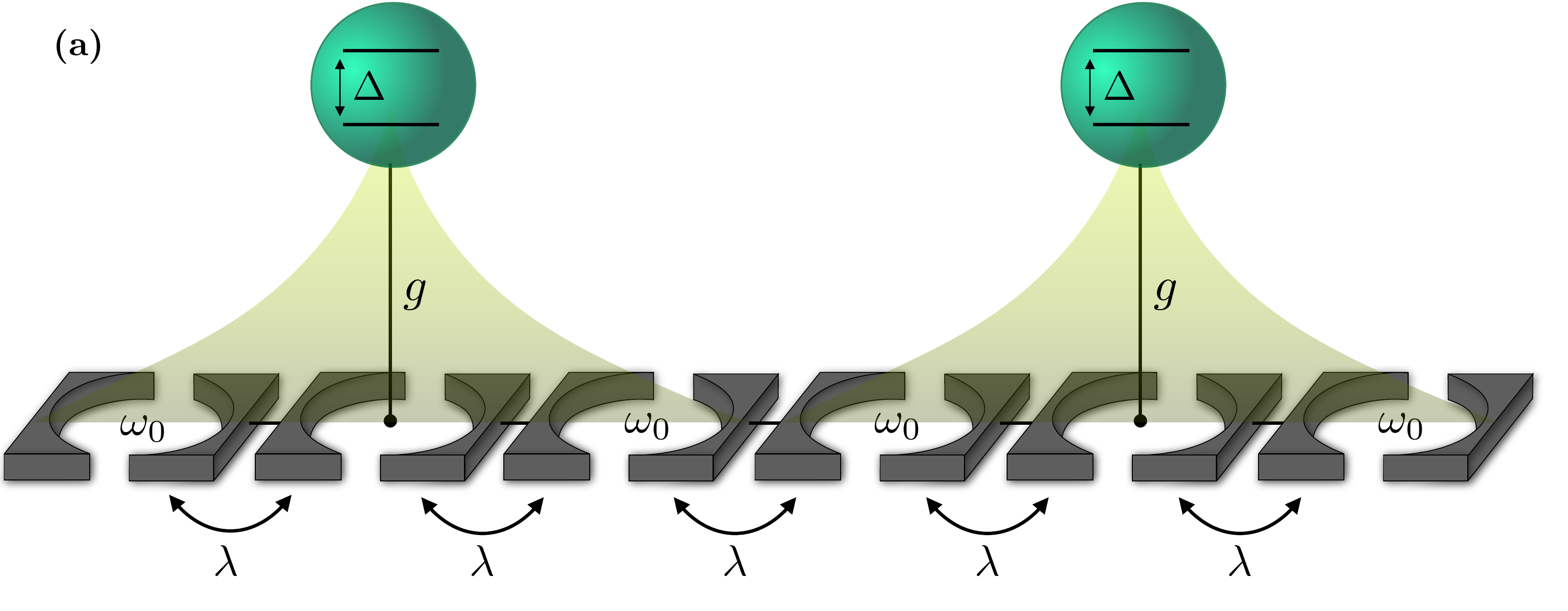}
    \includegraphics[width=0.25\textwidth]{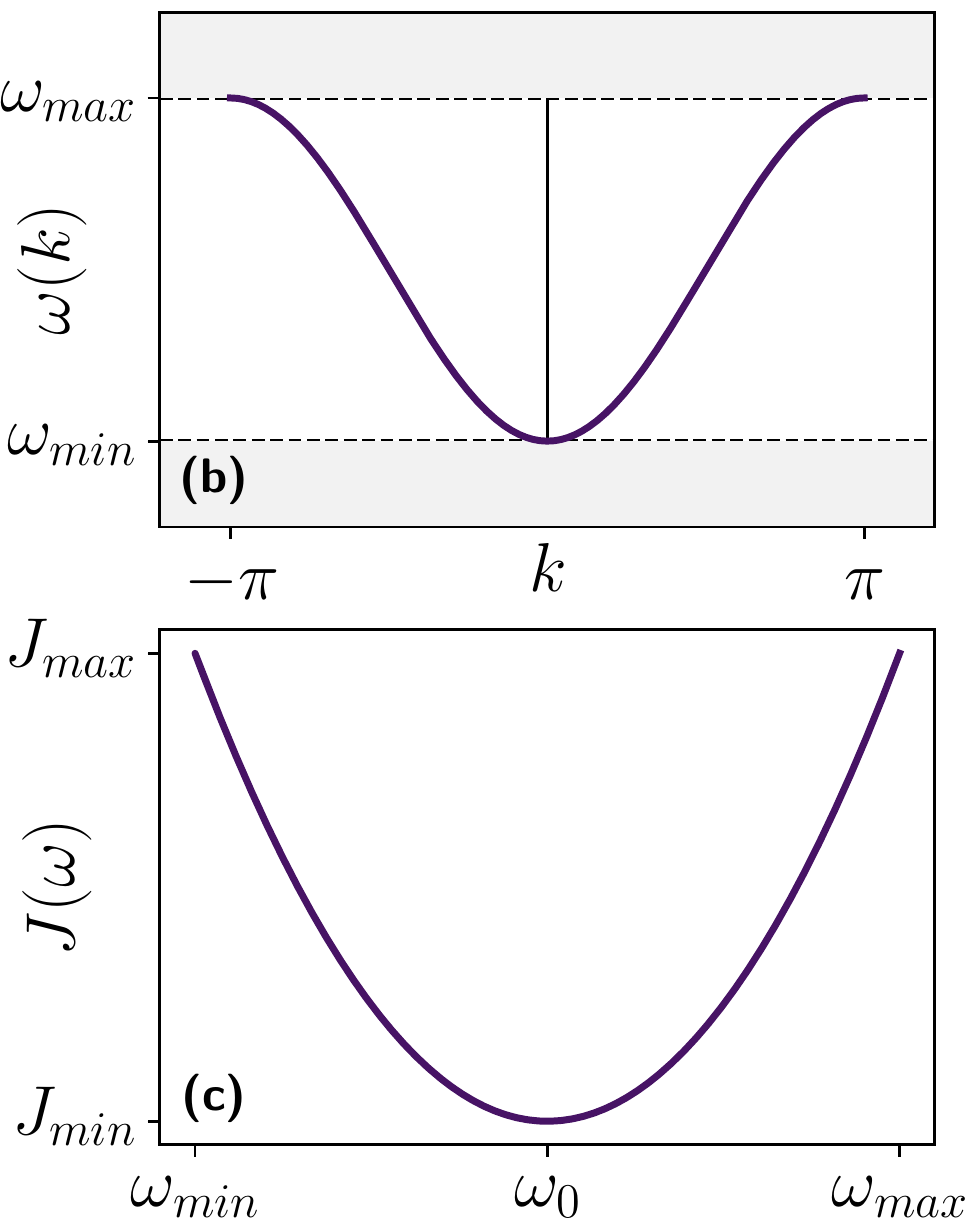}
    \caption{(a) Schematic depiction of two qubits coupled to specific sites of a linear cavity array. Where $g$ is the coupling constant, $\Delta$ is the energy difference between the two states of the qubits, $\omega_0$ is the resonance frequency for photons in the cavity (omitted in the coupled cavities for aesthetic purposes) and $\lambda$ is the hopping constant for photons travelling between cavities. The yellow shades represent localized-photon clouds. (b) Finite-band dispersion relation of the model. (c) Spectral density function for the model.}
    \label{fig:model}
\end{figure*}

In this manuscript we study the system sketched in Fig. \ref{fig:model}(a).
Several qubits are coupled to a cavity array forming the photonic medium.
In the dipole gauge \cite{di2019}  and assuming that each qubit is coupled to a single cavity, the model is ($\hbar=1$ is set through the paper)
\begin{align}
\label{H0}
\nonumber
    H & = \sum_{j=1}^{N_{q}} \frac{\Delta}{2} \sigma_j^z + \omega_0 \sum_n^N b_n^\dagger b_n  -\lambda \sum_n^N \left(b^\dagger_n b_{n-1} + {\rm H.c.}\right) 
    \\
    & + g \sum_{j=1}^{N_{q}} \sigma_j^x \left(b_{x_j}^\dagger + b_{x_j}\right) \; .
\end{align}
Here, $N_q$ is the total number of qubits with level splitting $\Delta$ (let us assume that all the atoms are identical). $N$ is the number of sites; we will consider the thermodynamic limit $N\to\infty$ in our analytical treatment.  $x_j$ is the site to which the $j$th-qubit  is coupled. Operators $b_n^\dagger$ and $b_n$ correspond to the bosonic creation and annihilation operators at site $n$, and $\sigma^z$ and $\sigma^x$ are the $z$ and $x$ Pauli matrices.  
To avoid extra parameters, we will consider that the qubit-resonator coupling is the same for all the qubits, $g$.
The photonic medium (second and third term in Eq. \eqref{H0}), which is a cavity array, is  diagonalized introducing the bosonic operators in momentum space, which are the Fourier transform of their spatial counterparts: $b_k =\frac{1}{\sqrt{N}} \sum_n e^{i k n} b_n$  obtaining:
\begin{align}
    \nonumber
    H &= \frac{\Delta}{2} \sum_{j=1}^{N_q} \sigma_j^z + \sum_k \omega_k b_k^\dagger b_k
    \\ 
    & + \sum_{j=1}^{N_q} \sigma_j^x \sum_k c_k \left( b_k^\dagger e^{ikx_j} + b_k e^{-ikx_j} \right).
    \label{Hk}
\end{align}
The dispersion relation, sketched in Fig. \ref{fig:model}(b), is
\begin{align}
    \omega_k = \omega_0 - 2 \lambda \cos k, \label{eq:dispersion} 
\end{align}
and the coupling per spin in momentum space is given by
\begin{align}
    c_k = \frac{g}{\sqrt{N}} \label{eq:coupling}.
\end{align}
The dispersion relation is a finite band of width $4 \lambda$ centred around $\omega_0$. Besides, the coupling constant is independent of the photonic mode and proportional to $g$, which justifies why, through out this work, we refer to both $g$ and $c_k$ indistinctly as the \textit{coupling constant}.
Finally, it is convenient to define the spectral density, plotted in Fig. \ref{fig:model}(c), 
\begin{equation}
\label{Jw}
    J(\omega) = 2 \pi \sum_k |c_k|^2 \delta (\omega - \omega_k)
    \; ,
\end{equation}
conveniently rewritten in terms of the density of states
\footnote{
Take a function $\mathfrak{f}(\omega)$. Then, $\int d \omega J(\omega)\mathfrak{f}(\omega)= 2 \pi g /\sqrt{N} \sum \mathfrak{f}(\omega_k) 
\to 2 \pi g \int d \omega (d\omega_k/ dk)^{-1} \mathfrak{f}(\omega)$. This yields Eq. \eqref{Jwca} in the main text.}
\begin{equation}
\label{Jwca}
J(\omega) =  2 \pi g^2 \left ( \frac{d \omega_k}{dk} \right )^{-1}
\; .
\end{equation}

\subsection{A brief comment on the RWA}
\label{sec:RWA}

If the coupling constant is small enough, the Rotating Wave Approximation can be used, by which the interaction term [last term in Eq. \eqref{H0}] becomes
\begin{equation}
   \sum_{j=1}^{N_q} \sum_k c_k \left(\sigma_j^- b_k^\dagger + \sigma_j^+ b_k\right).
\end{equation}
It is clear now that the state $ \ket{0_1, 0_2, \dots , 0_{N_q}; {\bf 0}} $ with $\sigma^z_j |0_j\rangle = - |0_j\rangle$  and $b_k |{\bf 0}\rangle = 0$ is the (trivial) ground state (GS) of the system and that the Hamiltonian preserves the number of excitations $N$, $\left[H, N \right] = 0$ with $N = \sum_k b_k^\dagger b_k + \sum_{j=1}^{N_q}\sigma_j^+ \sigma_j^-$. 
Consequently, within the RWA, the dynamics are split in subspaces with a fixed number of excitations which makes the low-energy dynamics amenable, at least numerically.
%

\subsection{Polaron picture}

It has been shown that in the low-energy sector of a spin(s)-boson model [\eqref{Hk}] is well approximated by an effective, excitation-number-conserving Hamiltonian derived from a Polaron transformation \cite{Bera2014, Camacho2016}. The basic idea is to construct a unitary transformation that disentangles the TLS from the bath.
This unitary transformation depends on some parameters that are found with the variational principle. 
In this case, the \emph{ansatz} is,
\begin{equation}
    | \Psi_{GS} [f_k, c_{\bf s}] \rangle 
    = U_P [f_k] \;  | {\bf 0} \rangle \otimes \sum_{s_j = 0,1} c_{\bf s} |s_1, ..., s_{N_p} \rangle \; .
\end{equation}
Here $|{\bf 0} \rangle$ is the photon vacuum state ($ b_k |{\bf 0} \rangle= 0$ for all $k$) and the spin state is arbitrary.
The varational parameters are the $c_{\bf s}$-coefficients and $\{f_k\}$, the $N$ parameters in the unitary $U_P$.
Building up on previous work from McCutcheon et al. \cite{McCutcheon2010} and Zheng et al. \cite{Zheng2015}  we used a natural extension of the single-qubit Polaron transform  valid for arbitrarily distant qubits
\begin{equation}
\label{UP}
    U_P = \exp[-\sum_{j=1}^{N_q} \sigma_j^x \sum_k (f_k b_k^\dagger e^{i k x_j} - f_k^* b_k e^{-i k x_j})].
\end{equation}
Provided there is no privileged direction of travel, we can assume that for each boson with wavenumber $k$ there will be another with $-k$, so that $|f_k| = |f_{-k}|$. From that, and the fact that the sine is odd, Eq. \eqref{UP}  factors as
\begin{equation}
\label{Upfactor}
    U_P = \bigotimes_{j=1}^{N_q} \; U_j \; ,
\end{equation} 
with
\begin{equation*}
    U_j = \exp[-\sigma_j^x \sum_k (f_k b_k^\dagger e^{i k x_j} - f_k^* b_k e^{-i k x_j})].
\end{equation*} 

It turns out that minimizing the energy of the spins-boson [\eqref{Hk}],
\begin{equation}
\label{minegs}
    \epsilon_{GS} = {\rm min}_{f_k, c_{\bf s}} \{
    \langle  \Psi_{GS} [f_k, c_{\bf s}]  | H |  \Psi_{GS} [f_k, c_{\bf s}]  \rangle \}
\end{equation}
is done by finding the ground state of the \emph{effective} spin model
\begin{equation}
\label{Hs}
    {\mathcal H}_S = \frac{\Delta_r}{2} \sum_{j=1}^{N_q} \sigma^z_j + \sum_{i<j} \mathcal{J}_{ij} \sigma_i^x \sigma_j^x + N_q \sum_k f_k (w_k f_k - 2 c_k)
\end{equation}
with
\begin{equation}
 \mathcal{J}_{ij} = 2 \sum_k f_k (2 c_k - \omega_k f_k) \cos[k(x_i-x_j)]  \;,
 \label{eq:Jij}
\end{equation}
and the renormalized qubits frequency
\begin{equation}
\label{Dr}
    \Delta_r= \Delta  \exp[-2 \sum_k |f_k|^2] \; .
\end{equation}
In the next section we will work explicit expressions in the case of one and two qubits.
A generalized Polaron transformation is discussed in App. \ref{ap:PTgeneral}, where
it is shown that the much less cumbersome Eq. \eqref{UP} is sufficiently good.  

\section{Single qubit case}
\label{sec:single}

In the single qubit case, $N_q=1$, Hamiltonian [\eqref{Hk}] is nothing but the spin-boson model \cite{Leggett1987} \cite[Chap. 3]{Weiss}.
In this section, we tackle the ground-state properties, the single-qubit bound states, and the spontaneous emission within the USC regime for the cavity array model [Eq. \eqref{Jwca}].

\subsection{Ground state}

\begin{figure*}
\centering
  \includegraphics[width = \textwidth]{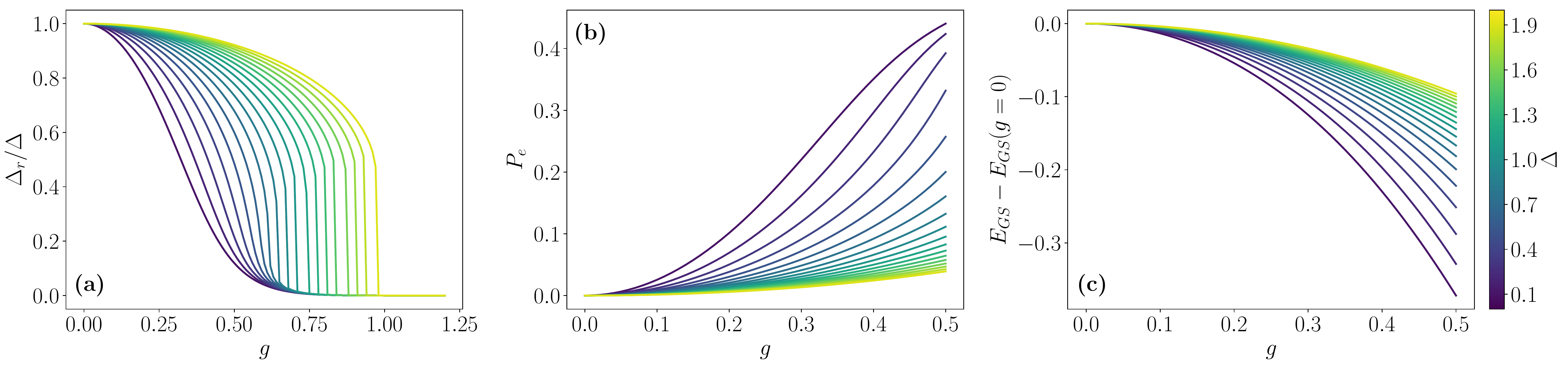}
    \caption{(a) Renormalized frequency in units of the bare qubit frequency as a function of $g$ for several values of $\Delta$. (b) $P_e$ as a function of $g$ for several values of $\Delta$. (c) Dependence of the ground state energy with $g$ for several values of $\Delta$, plotted with respect to the GS energy of an uncoupled qubit and the bath.}
    \label{fig:1q_delta_renorm}
\end{figure*}

Setting $N_q=1$, Eqs. \eqref{minegs} and \eqref{Hs} yield that the minimum of the energy is reached when $\sigma_z  |s_1 \rangle = - |s_1 \rangle$ [Cf. \eqref{Hs}] with
\begin{equation}
    \bar E_{GS} = - \frac{\Delta_r}{2} + \sum_k w_k |f_k|^2 - \sum_k c_k \left(f_k + f_k^*\right),
\end{equation}
which is minimum when  \cite{silbey1984},
\begin{equation}
f_k = \frac{c_k}{\Delta_r + \omega_k}.
\label{eq:1q_fk}
\end{equation}
Putting together Eqs. \eqref{Dr} and \eqref{eq:1q_fk},  we realize that the qubit frequency renormalizes to zero as the coupling strength increases.  This is a well known result \cite{Leggett1987}.
Besides, this renormalization is the responsible for the   localization-delocalization phase transition  that corresponds to  the ferromagnetic-antiferromagnetic phase transition in the Kondo model \cite{guinea1998}.
The delocalized phase corresponds to $\Delta_r \to 0$, then the qubit state can be in either the symmetric or antisymmetric superpositions of the eigenstates of $\sigma^z$. 
On the other hand, if $c_k=0$, the spin is at an eigenstate of $\sigma_z$, which corresponds to the localized sector \footnote{ \label{foot:loc} For those who have condensed-matter background the  spin-boson is paradigmatic in impurity models. In those formulations that naturally lead to a double-well interpretation of the TLS, the roles of $\sigma^x$ and $\sigma^z$ are switched in the Hamiltonian. In that case, $c_k = 0$ is viewed as the delocalized regime whereas  $\Delta = 0$ is viewed as the localized regime.}.
Using Eqs. \eqref{Jw} and \eqref{eq:1q_fk} we can rewrite Eq. \eqref{Dr} as,
\begin{equation}
\Delta_r = \Delta \exp{- \frac{1}{\pi} \int d \omega \frac{J(\omega)}{(\Delta_r + \omega)^2}}   \; .  \end{equation}
Having a phase transition  depends  on $J(\omega)$ \cite{Spohn1985}. In our system it is not expected to have critical behaviour \cite{Lowen1988}.  It is not within the aspirations of this work to study (the absence of) this phase transition, partly because it is not clear that the PT is valid in these ranges, so we will restrict our study to the so-called \textit{ultra-strong coupling} region, $g \in (0, \sim 0.5)$, where we are confident that the Polaron \textit{ansatz} works \cite{zueco2019, sanchez2019}. As we can see from Fig. \ref{fig:1q_delta_renorm}(a), this region is characterized by a significant, albeit not complete, shrinkage of the tunneling frequency ($\Delta$), and as such we expect predictions from RWA to fail.  \\
We can further characterize the GS by computing spin observables as is $P_e = \expval{\sigma^+ \sigma^-}$, the probability of having the spin excited. This is an insightful observable because it relates a measurable quantity, $P_e$, to the renormalized frequency, $\Delta_r$:
\begin{equation}
\label{Pe}
     P_e = \expval{\sigma^+ \sigma^-}{GS} = \frac{1}{2} \left(1 - \frac{\Delta_r}{\Delta}\right).
\end{equation}
Here we used $\sigma^+ \sigma^- = \frac{1}{2} (\sigma^z + I)$ together with $\expval{\sigma^z}{GS} = \expval{U_P^\dagger \sigma^z U_P}{0} = -\frac{\Delta_r}{\Delta}$.
In Fig. \ref{fig:1q_delta_renorm}(b) we show the dependence of $P_e$ with $\Delta$ and $g$, alongside is the GS energy plotted in the same parameter range, in Fig. \ref{fig:1q_delta_renorm}(c).
These are  signatures of RWA failure, since within the RWA both $P_e$ and $E_{GS}$ are zero.

\begin{figure}[!htb]
  \includegraphics[width = \columnwidth]{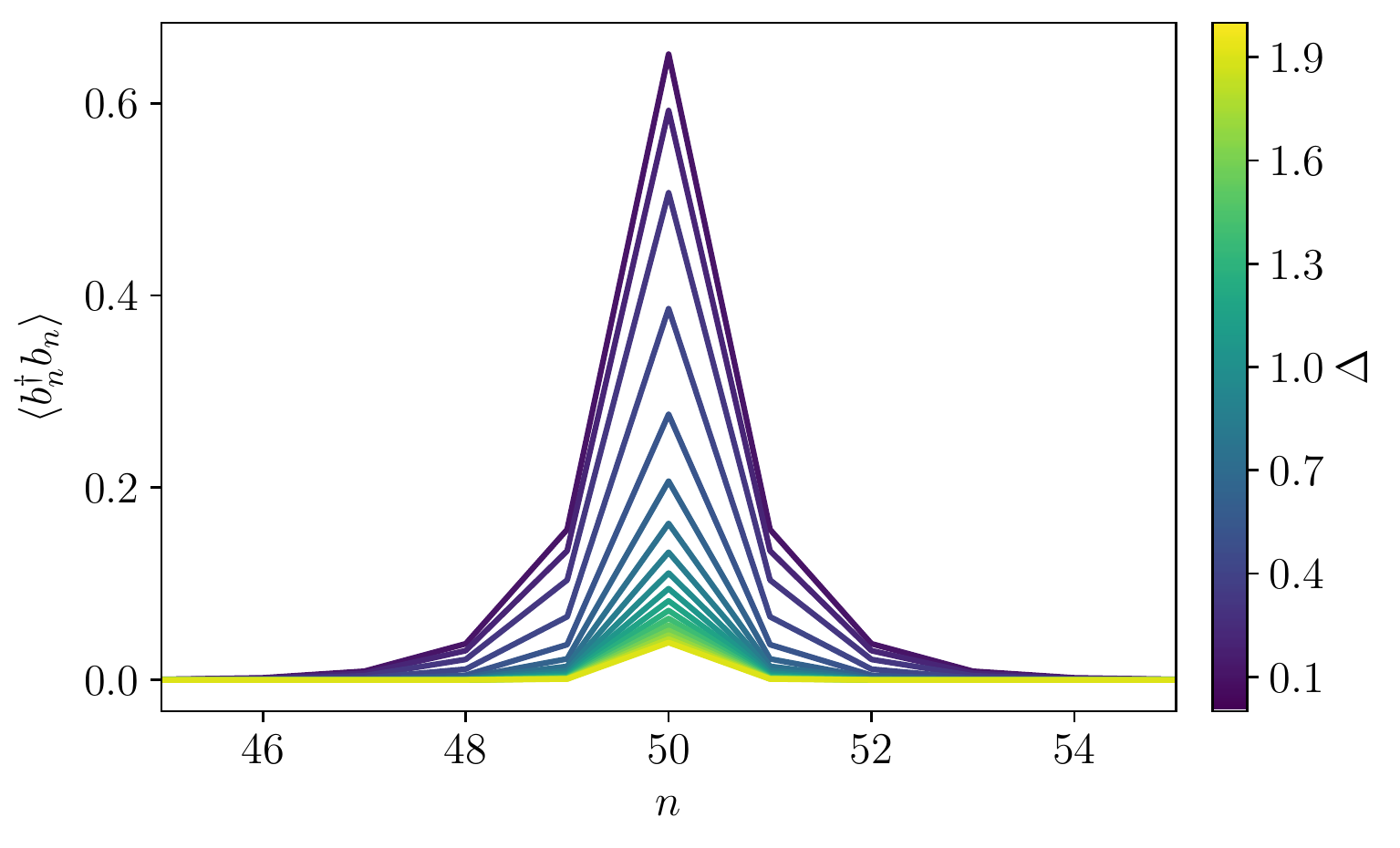}
    \caption{Comparison of the spatial distribution of GS photons for $g = 0.5$.}
    \label{fig:1q_rwa_vs_polaron}
\end{figure}

Another interesting observable is the spatial distribution of the photons, $\expval{b_n^\dagger b_n}$. Some algebra (fully done in App. \ref{ap:number_photons}) yields 
\begin{align}
\label{eq:1q/n_photons}
    \expval{b_n^\dagger b_n} = f_n^2
\end{align}
with $f_n=\frac{1}{\sqrt{N}} \sum_k e^{i k (n - N/2)} f_k$ being the Fourier transform of $f_k$.  We center the transformation at the qubit position that it is understood to be at the middle of the chain. Notice that $f_n$ has the clear interpretation of being the real-space variational amplitudes for the Polaron transformation.
Figure \ref{fig:1q_rwa_vs_polaron} shows the spatial distribution of photons as calculated in Eq. \eqref{eq:1q/n_photons}. We observe that they are well localized around the impurity, exhibiting exponential decay $f_n \sim \exp \{-\kappa_{GS} (n- N/2) \}$  with localization lenght (See App. \ref{ap:1q_exp_localisation} for a proof)
\begin{equation}
\label{kappa}
    \kappa_{GS}^{-1} = \arccosh ^{-1}
    \left( 
    \frac{\omega_0 + \Delta_r}{2 \lambda}
    \right).
\end{equation} 
The photons dressing the impurity are commonly named \textit{virtual photons} in reference to their special properties of being non-propagating and exponentially localized, that distinguish them from real photons \cite[Chap. 1.3]{nano_optics}. Again, this is an effect of being in the USC regime and is in stark contrast with the GS found with the RWA which is trivially $\ket{0_1 ;{\bf 0}}$.

\subsection{Bound states}

We discuss now the single excitation bound states (SEBS), which are the basis for creating effective interactions between the qubits. 
Before moving to the USC regime, let us summarize the existence of bound states within the RWA approximation
where the number of excitations is conserved, see Sect. \ref{sec:RWA}. %
In this case, the lowest energy bound states are localized eigenstates in the single excitation subspace.
Its energy must be outside of the single-photon band. Given a general photonic model, its existence is not guaranteed; \emph{i.e.}, the eigenvalue equation may not have solutions for energies outside of the dispersion relation \cite{gaveau1995,shi2016}.
Notice that photons in these states cannot propagate.
They can be thought of as particles trying to enter a potential barrier greater than their energy, and as such, their wavefunction must be exponentially decaying with the distance from the qubit.
It turns out that, within the RWA, Hamiltonian [\eqref{H0}] or [\eqref{Hk}], always accepts two exponentially localized eigenstates: one with energies above and other below the photonic band. \\
In the full model [\eqref{Hk}] the number of excitations is not conserved and we cannot work in the single-excitation subspace.
On the other hand, in the Polaron picture 
the effective Hamiltonian $H_P = U_P^\dagger H U_P$ is approximately number-conserving  (see App. \ref{ap:1q_Heff} for details on the derivation), 
\begin{align}
\label{Hp}
H_{P} &=  \Delta_r \sigma^+ \sigma^- 
+ \sum_k \omega_k b_k ^\dagger b_k
\\ \nonumber
& + 2 \Delta_r \big ( \sigma^+ \sum_k f_k b_k + {\rm h.c.} \big ) -
    2 \Delta_r \sigma_z \sum_{k, p} f_k f_p b_k^\dagger b_p 
\\ \nonumber    
   & + E_{ZP} +  \rm{h.o.t}
\; .    
\end{align}
Here, h.o.t. stands for higher-order terms of order ${\mathcal O} (f^3)$ with two and more excitations.
 $E_{ZP}=-\frac{\Delta_r}{2} + \sum_k f_k (w_k f_k - 2 c_k)$ is the constant term in $H_P$. 
Thus, in the Polaron picture, $H_P$ conserves the number of excitations and becomes tractable with the same techniques as RWA models; in particular we can compute the single-excitation eigenstates.
It is interesting to note that the GS obtained from the variational method is an eigenstate of $H_P$ with eigenvalue equal to the GS energy. This gives us a sense of consistency that confirms the effective RWA model is accurate:  If the GS is well caught, one expects that the first excitations are single particle (quasiparticles) excitations over it. \\
\begin{figure*}[!htb]
\centering
  \includegraphics[width = \textwidth]{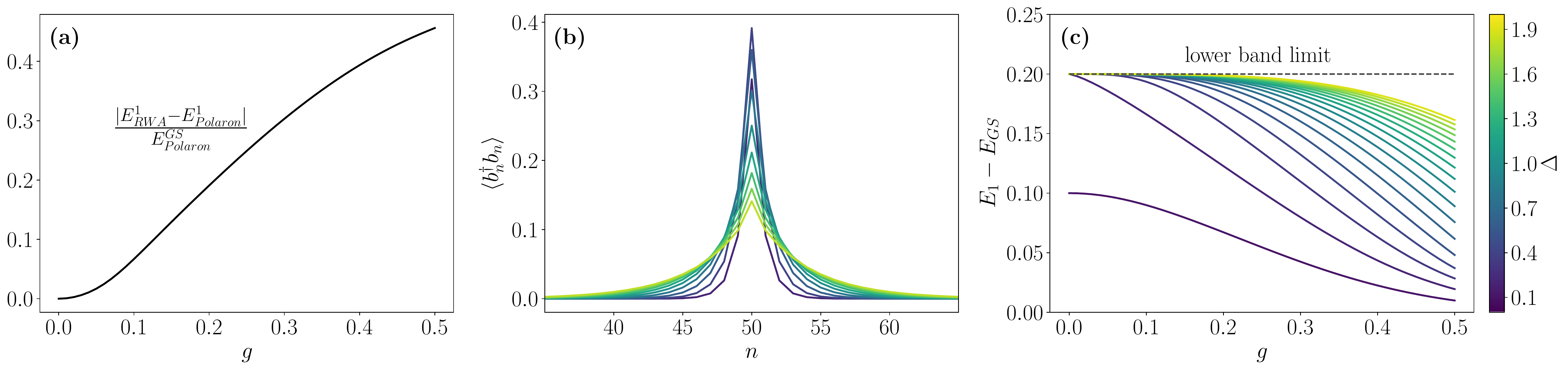}
    \caption{(a) Relative energy difference between the SEBS as predicted by the PT ($E^1_{\text{Polaron}}$) and the RWA ($E^1_{\text{RWA}}$). The ground state energy used is the one calculated with the PT for $\Delta = 0.3$. (b) Spatial distribution of the photons for $g = 0.3$ as a function of $\Delta$. (c) Energy difference between the first excited state and the ground state in comparison with the lower band limit as a function of $g$ and $\Delta$.}
\label{fig:1q_nphotons_bound}
\end{figure*}

%
In App. \ref{ap:existence} we show that Eq. \eqref{Hp} admits a bound state below the band, with energy $E_1$, and that its localization length is given by, 
\begin{equation}
\label{ksebs}
  \kappa_{\rm SEBS}^{-1}= \max   ( \kappa_{\rm GS}^{-1}, \kappa^{-1} )
\end{equation}
with 
    $\kappa \cong \arccosh \left ( 
    \frac{\omega_0-E_1}{2 \mathcal{J}}
    \right ) $. 
See App. \ref{ap:localization} for the proof.
In Fig. \ref{fig:1q_nphotons_bound}(b) we observe the exponential tails.
We observe that, the higher the qubit bare-frequency is the peaks become shorter and they also get broader.
Qualitatively, we can understand this as follows.
The Polaron transformation is local in space \cite{sanchez2019}, thus, for discussing the tails  we can argue in the Polaron picture.
Because the total number of excitations is $1$ in this subspace, the sum of the values of the number of photons in each site must add up to 1, minus the amount taken up by the qubit (which is expected to decrease as the bare-$\Delta$ increases). That is why, as the peaks become smaller, they also get broader, in order to preserve the number of excitations. Figure \ref{fig:1q_nphotons_bound}(c) shows the de-excitation energy for several values of $\Delta$, referred to the lower band limit, which means that there exists a bound state below the band for all values of $\Delta$.

\begin{figure*}
    \centering
    \includegraphics[width = \textwidth]{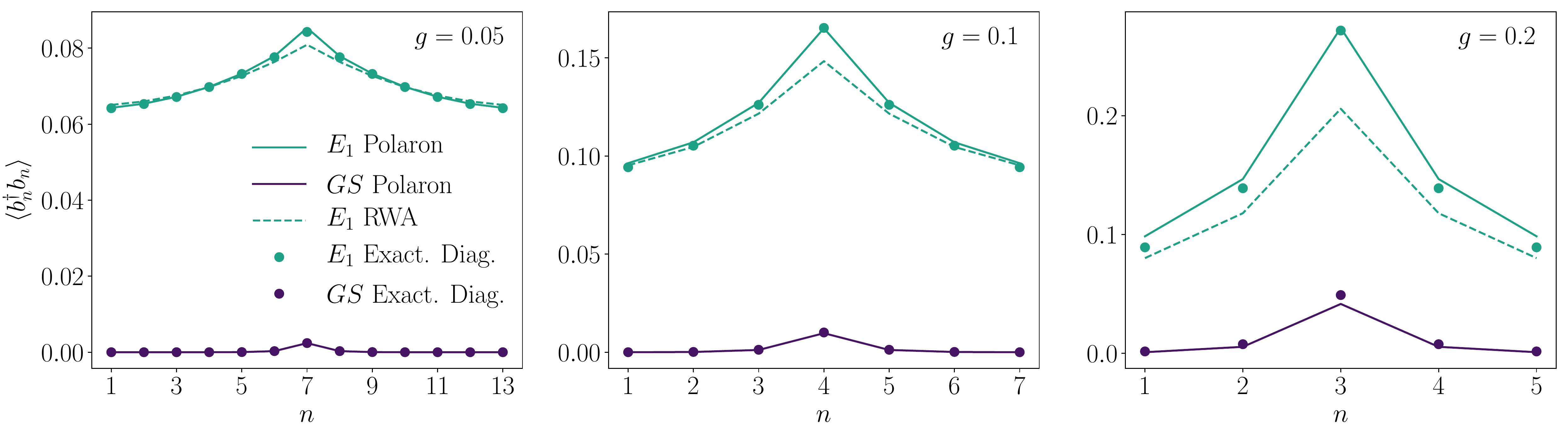}
    \caption{Comparison of spatial boson distributions from the PT, the RWA and exact diagonalisation. They correspond to $\Delta = 0.3$, and $g = 0. 05$, $g = 0.1$ and $g = 0.2$ respectively from left to right. Solid lines are used to indicate Polaron results, dashed lines for RWA results and dots for exact diagonalisation results.}
    \label{fig:1q_mps}
\end{figure*}

We can now compare the difference between the results provided by the Polaron transform to those obtained using the RWA. Figure \ref{fig:1q_nphotons_bound}(a) shows the relative energy difference between the SEBS calculated with each method. The difference increases with $g$, becoming significant for $g \sim 0.1$. We have shown only one value of $\Delta$ for clarity, as all values behaved similarly, being the difference greater the smaller the value of $\Delta$. The fact that the PT predicts different bound state energies is not sufficient to declare it superior to the RWA, it could be the case that these results were worse than those provided by the RWA. The definitive confirmation comes from Fig. \ref{fig:1q_mps}, where we compare the bosonic spatial distributions from the PT and the RWA with those generated by  exact diagonalisation of the Hamiltonian for different coupling strengths (see App. \ref{ap:n_photons_bound} for details on the calculation). 
The results from the PT are in agreement with the numerical results, both in the ground  and excited states. In addition, we see how the RWA  and PT coincide for low values of $g$, but the RWA immediately begins to underestimate the number of photons when the value of $g$ increases beyond $g = 0.1$. Exact diagonalisation is very limited because the state-space grows exponentially with the number of elements. In addition, \textit{exact} might be an overstatement considering that one must limit the number of excitations per site in order to have a finite size Hamiltonian. That is why only $12$ sites were used in the benchmark for $g = 0.05$, a number that had to be reduced for greater values of $g$ in order to accommodate more excitations per site while maintaining the state-space size allowed by our numerical capabilities. \\
Finally, let us show that the bound state above the single-photon band that exists in the RWA ($\ket{E_1^u}$) \cite{Longo2010,Longo2011,shi2016, calajo2016},  \emph{does not} exist in general in the full model [\eqref{Hk}]. 
First of all, there are numerical evidences that the model [\eqref{Hk}] has, at least, an even bound state $\ket{E_2}$ \cite{sanchez2014}. 
One can define a band of one-photon states over $\ket{E_2}$: $\ket{k,E_2}$ \cite{SanchezBurillo2018}. The parity of these states is odd. The hypothetical bound state $\ket{E_1^u}$ would also be odd, since it has one excitation in the RWA limit. This implies that, in order for this state to exist, it cannot be embedded in the band formed by $\ket{k,E_2}$, since otherwise they would hybridize.
A necessary condition is:
\begin{equation}
\label{noex}
   \omega_0 \geq 4\lambda \, . 
\end{equation}
Otherwise,  $\ket{E_1^u}$ does not exist.
To demonstrate the latter, we note that
 the bound state energy is such that $E_1^u - E_{GS} > \omega_0  +  2\lambda $.  On the other hand, $E_2- E_{GS}  <  2 (\omega_0 - 2 \lambda)$ (\emph{i.e.} the two photon band).  The overlap occurs (and thus the non-existence) if $E_2 + \omega_0 - 2 \lambda < E_1^u $. Putting it all together we arrive to the condition for existence given by Eq. \eqref{noex}.
 It seems a paradox, since this state does exist in the RWA for all $\omega_0$ and $\lambda$. The puzzle is solved by noting that in the full model this state becomes a resonance with a lifetime that diverges in the RWA limit.

\subsection{Spontaneous emission}

\begin{figure}[!b]
	\centering
	\includegraphics[width=\columnwidth]{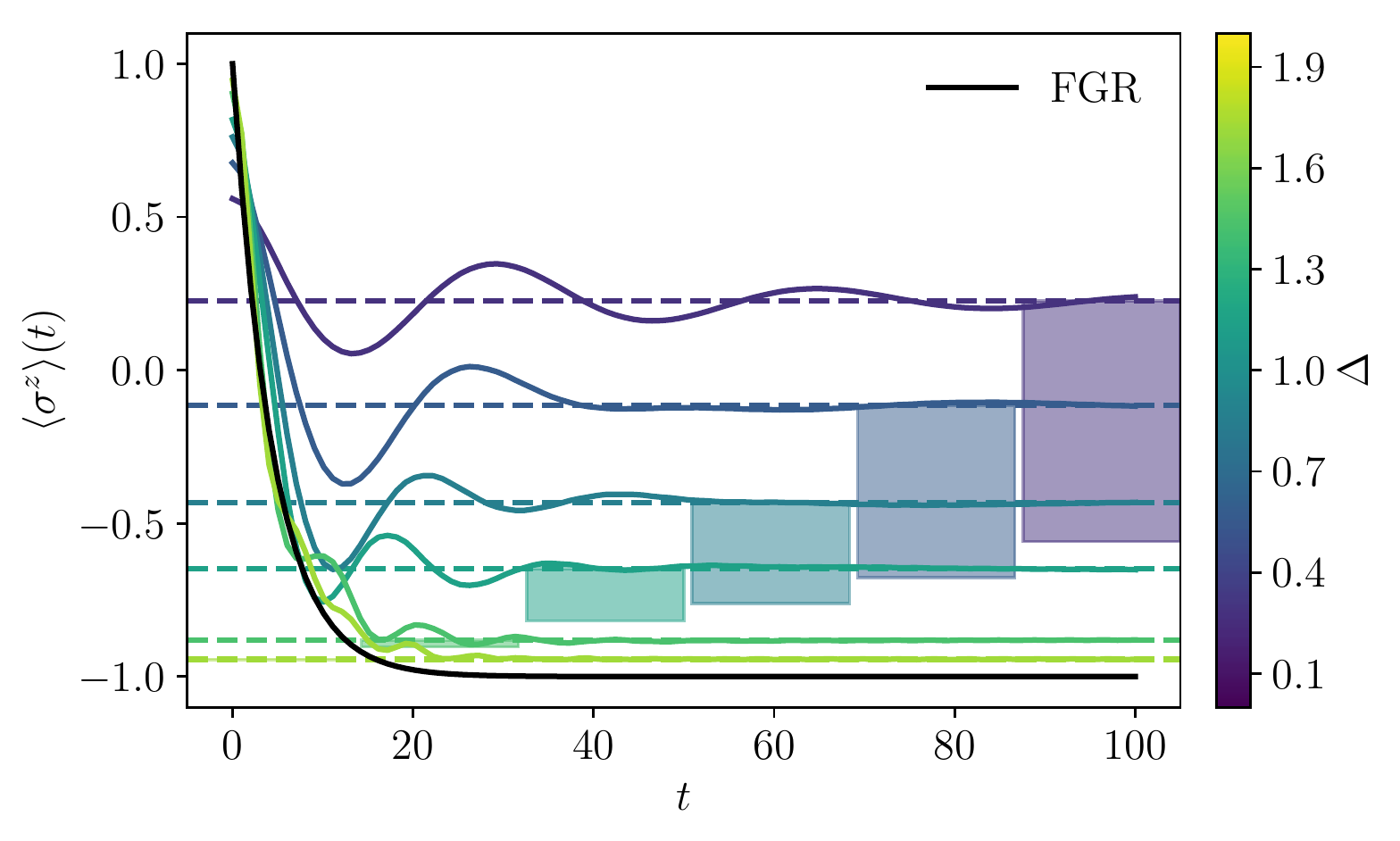}
	\caption{Evolution of $\ex{\sigma^z}$ (magnetization) for an initially excited qubit as a function of $\Delta$ for a fixed $g = 0.3$. Solid coloured lines represent the simulated evolution while dashed coloured lines mark the stationary value predicted analytically, Eq. \eqref{eq:1q_stationary}. For contrast, the solid black line corresponds to a Markovian evolution calculated by applying the FGR to the excited and ground states. Coloured shaded boxes have been used to showcase the difference between the stationary magnetisation for each $\Delta$ and the corresponding ground state magnetization, $\expval{\sigma^z}_{GS}$.}
	\label{fig:sp_emission}
\end{figure}

To end our analysis of the single-qubit model we discuss the behaviour of the system during spontaneous emission.
We assume the atom-waveguide at the GS, then the qubit is driven within a $\pi$-pulse. 
After the $\pi$-pulse, the wavefunction is given by $\ket{\Psi (0)} = \sigma^+ |GS \rangle$.  
Since $[\sigma_x, U_P]=0$, we may work in the single excitation manifold in the Polaron picture.  
Employing the single excitation ansatz $|\psi \rangle_P = ( \beta \sigma^+ + \sum_k \beta_k a_k^\dagger ) |0; {\bf 0} \rangle $, the solution is obtained as the inverse Laplace transform $\beta (t) = {\mathcal L}^{-1} [\beta(s)]$ with,
\begin{equation}
\label{betas}
    (s+i\Delta_r) \beta (s) =  1 - \sum_k \frac{| \langle 0; {\bf 0} | a_k H_P \sigma^+ | 0; {\bf 0} \rangle|^2}{s+ i \langle 0; {\bf 0} | a_k H_P a_k^\dagger| 0; {\bf 0} \rangle}  \;  \beta(s) \; .
\end{equation}
The properties of the (inverse) Laplace transform determine the spontanteous emission.
In particular, since $\langle 0; {\bf 0} | a_k H_P \sigma^+ | 0; {\bf 0} \rangle|^2= 2 \Delta_r^2 |f_k|^2$, in the continuum limit the sum in Eq.  \eqref{betas} can be converted to an integral over the spectral density $J(\omega)$.
Let us discuss the two main contributions to this integral.
Far from the band limits, $J(\omega)$ is sufficiently smooth and the main contribution comes from the poles in the sum, yielding the exponential decay $\exp[- J(\Delta_r) t]$.
Notice, that this is analogous to the RWA result (where the spontaneous emission is $J(\Delta)$) but now it is renormalized \cite{zueco2019}.
The other important feature is the long time dynamics of $\beta(t)$ which accounts for the qubit thermalization process.  
The final value theorem, $\lim_{s \to 0} s \beta (s) = \lim_{t\to \infty} \beta (t)$, tells that $\beta \neq 0$ if some divergence occurs in that integral.  
This occurs if bound states exist.
Physically, this means that the initially excited state overlaps with the bound state  \cite{john1990, john1994}.  
This is conveniently calculated by chosing as a basis in the single excitation manifold,
\begin{equation}
    \left\{ \ket{E_1}, \ket{E_1}^\perp_p \right\}.
\end{equation}
Where $\ket{E_1}$ is the bound state and $\ket{E_1}^\perp_p$ are all other eigenstates orthogonal to it. We recall that the bound state can be written in terms of the original states spanning the one-excitation subspace
\begin{equation}
    \ket{E_1} = \lambda_0 \ket{1} \ket{0} + \sum_k \lambda_k \ket{0} \ket{1_k}.
\end{equation}
The first term corresponds to the initial state of the system $\ket{\psi_0} = \ket{1} \ket{0}$, which indicates that the initial state has some projection on to the bound state,
\begin{equation}
    \ket{\psi_0} = \lambda_0 \ket{E_1} + \sum_p \lambda_p \ket{E_1}^\perp_p.
\end{equation}
The projection onto the orthogonal basis states will contribute to the continuum and as such, it will not contribute the long time dynamics.  The projection onto the bound state is responsible for the divergence and thus for the nonzero value for $\beta (t \to \infty)$.  Doing the algebra and computing the observable (notice our return to the lab frame) we obtain that:
\begin{equation}
    \ex{\sigma^z (t\to \infty)} = \lambda_0^2 \expval{U_P^\dagger \sigma^z U_P}{E_1} - (1 - \lambda_0^2) \frac{\Delta_r}{\Delta}.
    \label{eq:1q_stationary}
\end{equation}
In Fig. \ref{fig:sp_emission} we confirm this expression. The evolution converges to the stationary value predicted by Eq. \eqref{eq:1q_stationary}. Also shown in Fig. \ref{fig:sp_emission} is the difference with $\ex{\sigma^z}_{GS}$, which becomes significant as the ratio $g / \Delta$ increases, that is, as the system progresses into the USC regime.  
Let us emphasize that these results show that our theory is able to deal with the dynamics in USC confirming the peculiarities of the thermalisation process when both the light-matter coupling is non-perturbative and there exist excited bound states.
\section{Two-qubit case}
\label{sec:multi}
We tackle the case of two qubits coupled to the cavity array.
Much like in the single qubit case, we first report the results for the ground state  continuing with the bound states properties.
We put emphasis in the qubit-qubit interactions mediated by the cavity array.
As an application, we devise a simple state transfer between two distant qubits that uses those interactions. %
\subsection{Ground state}
Setting $N_q = 2$ and $x = x_1-x_2$ in Eqs. \eqref{Hs} and \eqref{eq:Jij} yields (see Apps. \ref{ap:2q_commutation_relations}- \ref{ap:H_B}) a spin model
\begin{equation}
    {\mathcal H}_S = \frac{\Delta_r}{2} \left(\sigma^z_1 + \sigma_2^z \right) - \mathcal J \sigma_1^x \sigma_2^x + 2 \sum_k f_k (w_k f_k - 2 c_k)
\end{equation}
with $\mathcal J = 2 \sum_k f_k (2 c_k - \omega_k f_k) \cos(kx)$
which can be diagonalized to yield a ferromagnetic GS of the form
\begin{equation}
    \ket{GS}_S = \cos \theta \ket{00} + \sin \theta \ket{11},
    \label{eq:2q_spin_gs}
\end{equation}
where $\ket{00} \equiv \ket{s_1 = 0, s_2 = 0}$ and the coefficients are
\begin{align}
    \cos \theta = \frac{\Delta_r + \sqrt{\Delta_r^2 + \mathcal J^2}}{\sqrt{\left( \Delta_r + \sqrt{\Delta_r^2 + \mathcal J^2} \right)^2 + \mathcal J^2}}\\
    \sin \theta = \frac{\mathcal J}{\sqrt{\left( \Delta_r + \sqrt{\Delta_r^2 + \mathcal J^2} \right)^2 + \mathcal J^2}}.
\end{align}

By Eq. \eqref{minegs}, the GS mean energy is 
\begin{equation}
    \bar E_{GS} = - \sqrt{\Delta_r^2 + \mathcal J^2} + 2 \sum_k f_k (w_k f_k - 2 c_k).
\end{equation}
which is minimum for, see also Refs. \cite{McCutcheon2010, Zheng2015} and App. \ref{ap:f_k} for a detailed derivation,
\begin{equation}
    f_k = c_k\frac{\mathcal E + \mathcal J \cos(kx)}{\omega_k \mathcal E  + \omega_k \mathcal J \cos(kx) + \Delta_r^2}.
    \label{eq:2q_f_k}
\end{equation}
We have introduced the constant $\mathcal E = \sqrt{\Delta_r^2 + \mathcal J^2}$ to ease notation. It is immediate to check that, should the interaction constant ($\mathcal J$) vanish, we would recuperate the expression of $f_k$ that we found in the single-qubit case.
This indeed happens when we set the qubits infinitely apart, as will be shown shortly.
It is also evident that $f_k$ is even with respect to $k$, which matches the restriction we imposed so that the PT could be factored, see Eq. \eqref{Upfactor}. 
\begin{figure}
  \includegraphics[width=\columnwidth]{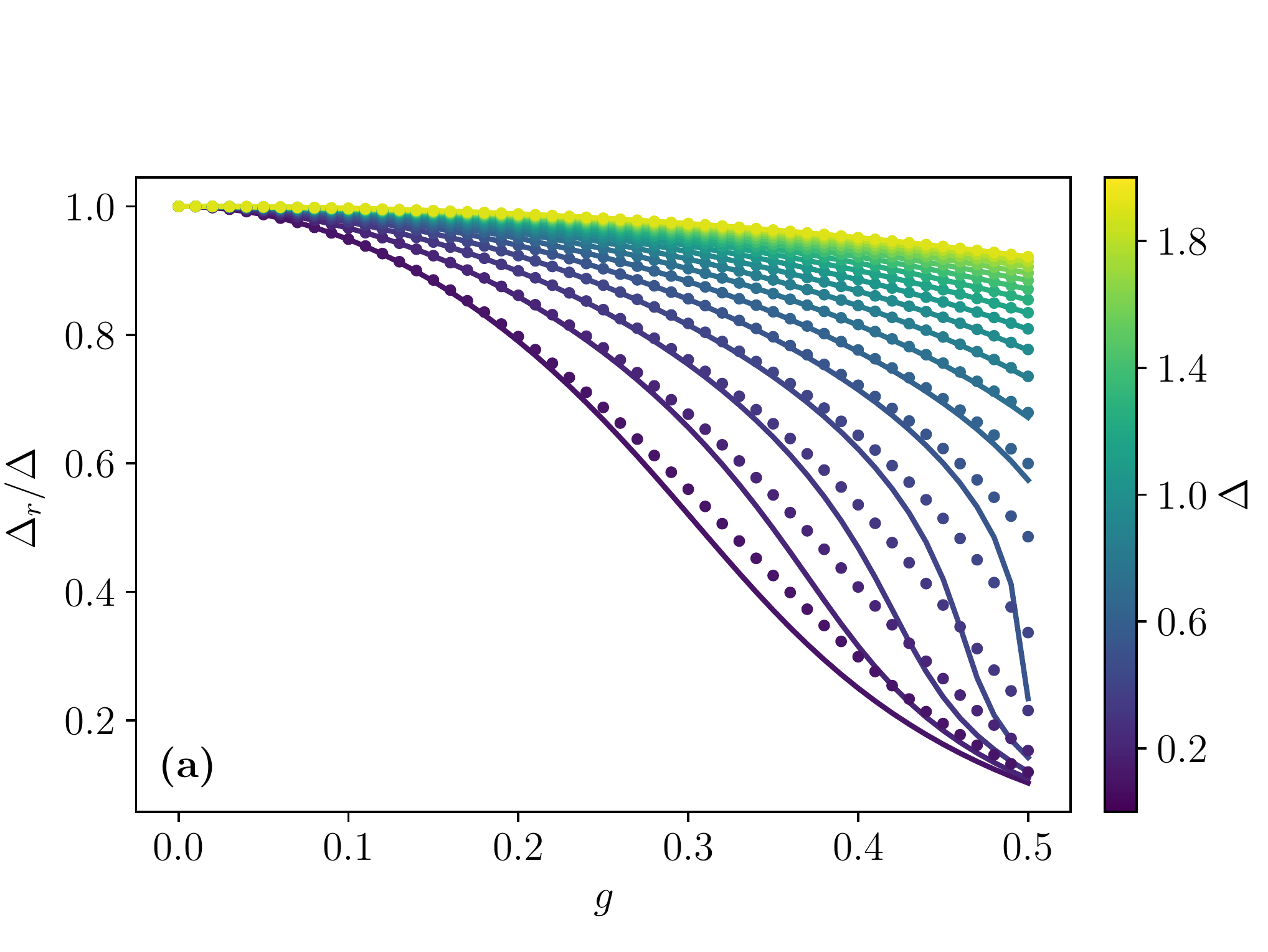}
  \caption{Renormalization of the bare frequency for $x = 2$ as a function of $g$ and $\Delta$. Solid lines represent two-qubit results. Dots represent single-qubit results.}
\label{fig:2q_J_with_x}
\end{figure}

\begin{figure}[!htb]
  \includegraphics[width = 0.95\columnwidth]{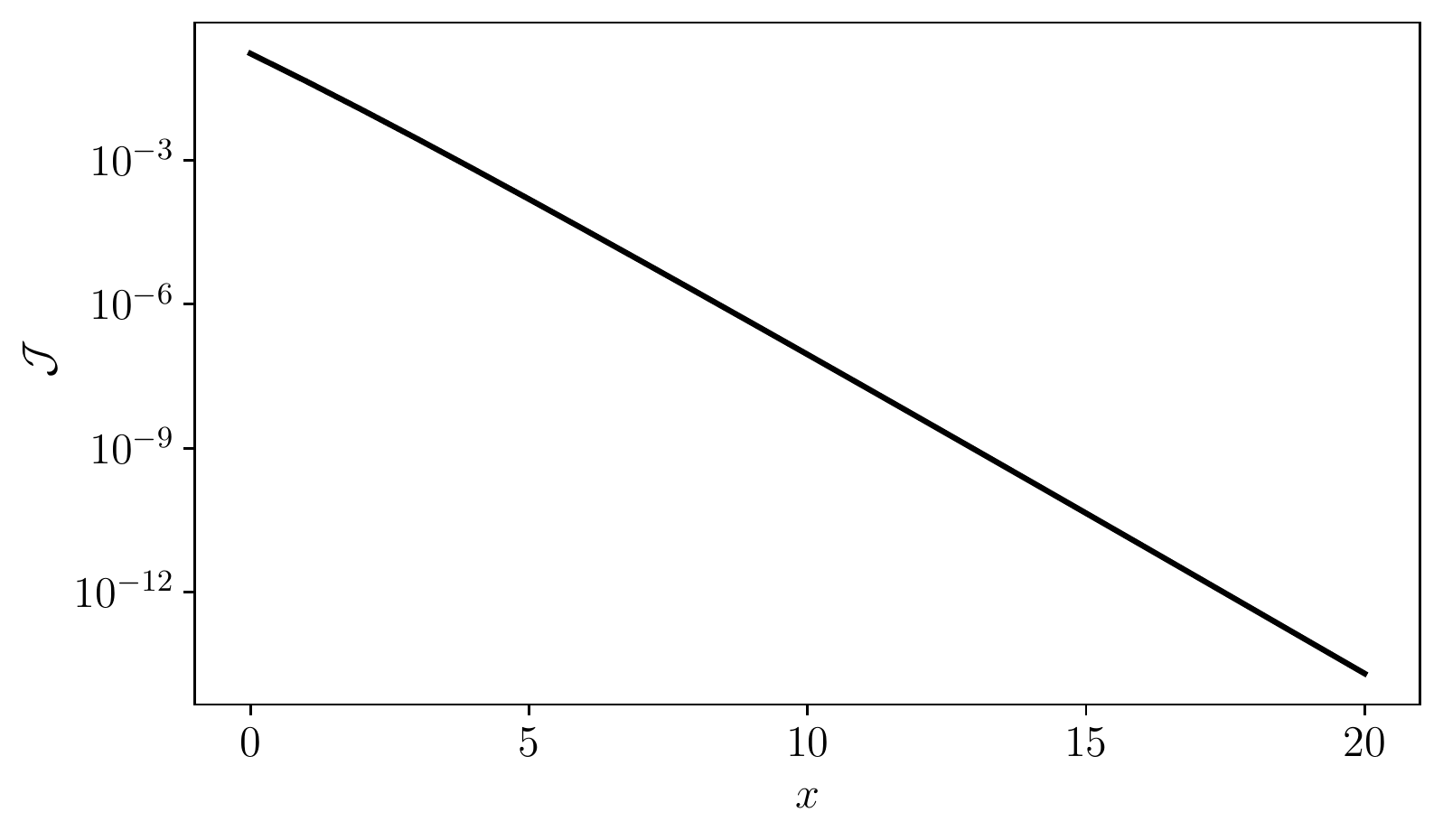}
    \caption{Dependence of the Ising constant, $\mathcal J$, with the distance between the qubits, $x$. The behaviour is analogous for all values of $g$.}
    \label{fig:J_with_x}
\end{figure}

Figure \ref{fig:J_with_x} shows that the dependence of $\mathcal J$ with $x$ is exponential. This implies that the GS is a ferromagnetic state in a short-range Ising model. As such, in a multi-qubit scenario, only the interaction with first-nearest neighbours would have to be taken into account. Following our analysis of the single-qubit case, it is useful to study the renormalization of the bare frequency $\Delta$ with $g$. We have used a distance of $n=2$ sites to illustrate the deviation from the results obtained in the one-qubit scenario.
 Figure \ref{fig:2q_J_with_x}(b) shows that the influence of the neighbouring qubit sharpens the renormalization process, making the system go into full renormalization at lower values of $g$. Granted, this effect vanishes if one places the qubits further apart. Due to the exponential decay of $\mathcal J$, we have found that at distances of around $20$ sites the results obtained for one and two qubits are indistinguishable.\\
\begin{figure}
\centering
\begin{subfigure}
  \centering
  \includegraphics[width=\columnwidth]{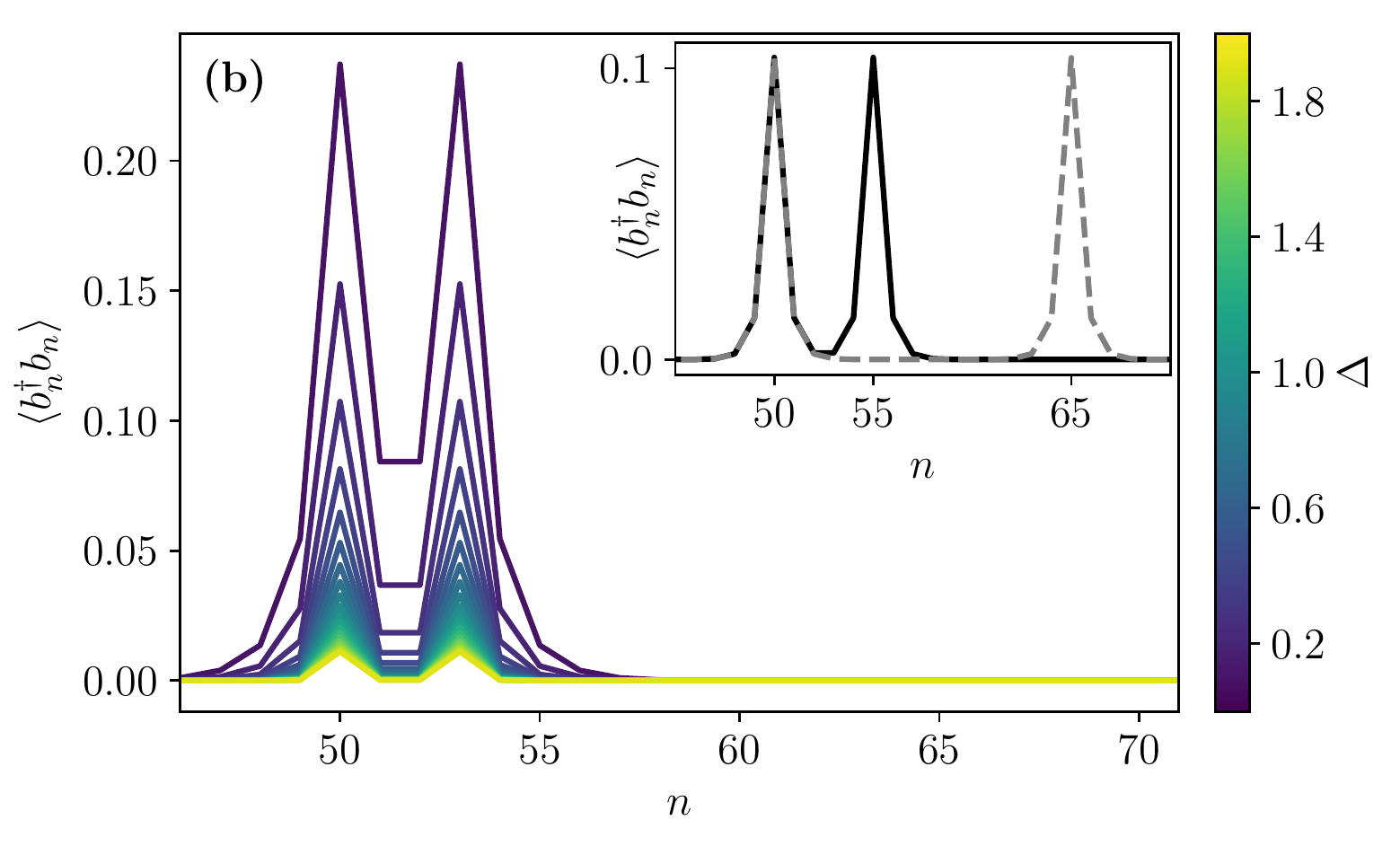}
\end{subfigure}
\caption{Spatial distribution of GS the photons for $g = 0.3$ and $x = 3$ as a function of $\Delta$, with inset showing the difference in photon cloud localisation when the qubits are placed at distances $x = 5$ and $x = 15$ for $g = 0.3$ and $\Delta = 0.3$.}
\label{fig:2q_photon_concentration}
\end{figure}
Back when we studied the single-qubit system we showed that the probability of having an \textit{excited} spin state was an observable directly related to the renormalization of the bare frequency. The extension to two qubits is straightforward
\begin{equation}
    P_e = \frac{2 + \expval{\sigma_1^z + \sigma_2^z}{GS}_S}{2} = 1 - \frac{\Delta_r}{\Delta} \left(\cos^2{\theta} - \sin^2{\theta} \right)
    \label{eq:2q_Pe}
\end{equation}
Where $\ket{GS}_S = \cos \theta \ket{00} + \sin \theta \ket{11}$ now.
Notice that at large distances, as $\mathcal J\to 0$ then $\cos \theta \rightarrow 1$ and $\sin \theta \rightarrow 0$, so Eq. \eqref{eq:2q_Pe} reduces to twice the probability found for a single qubit [Eq. \eqref{Pe}]. The effect of the Ising interaction is revealed at short distances where $P_e$ deviates from the single-qubit result, no longer equating to the sum of two non-interacting spins. We can again probe the spatial localisation of the bosonic cloud. Following the scheme presented in the single-qubit case, we obtain
\begin{equation}
    \expval{b_n^\dagger b_n} = |f_{n, 1}|^2 + |f_{n, 2}|^2 + 2 \cos \theta \sin \theta \Re{f^*_{n, 1} f_{n, 2}},
\end{equation}
where
\begin{align}
    f_{n, j} = \frac{1}{\sqrt{N}} \sum_k e^{i k x_j} f_k e^{-i k (n - N/2)}.
\end{align}
See App. \ref{ap:GS_nphotons} for details on this calculation. It is interesting to see the overlap between the two bosonic clouds surrounding each qubit. Figure \ref{fig:2q_photon_concentration} shows this phenomenon for a value of $n = 3$ where the overlap is significant. In the same figure, the inset monitors the effect of the coalescence of the clouds as the two qubits approach each other. 
\subsection{Bound states}
Analogously to the single-qubit case, we seek an effective Hamiltonian for the two-qubit model that is a good approximation of the full Hamiltonian but conserves the number of excitations, allowing us to restrict our search for bound states to the one-excitation subspace. We have discussed how $f_k$  for two-qubits converges to the expression of $f_k$ for a single qubit, so if we assume $\mathcal J$ to be small, we can write
\begin{equation}
    f_k = f_k^0 + \delta(f_k).
\end{equation}
This allows us to reach the effective Hamiltonian
\begin{equation}
\label{Heff2q}
    \begin{split}
        H_P =& \frac{\Delta_r}{2} \left(\sigma_1^z + \sigma_2^z \right) - 2 \Delta_r \sum_j \sigma_j^z \sum_{k,p} f_k f_p e^{i (k - p) x_j} b_k^\dagger b_p \\
         + & \sum_{j, k} \left (2 \Delta_r f_k^0 + \delta(f_k)(\Delta_r - \omega_k) \right ) \left( \sigma_j^- b_k^\dagger e^{i k x_j} + {\rm h.c.} \right)\\
        + & \sum_k \omega_k b_k^\dagger b_k - \mathcal J \sigma_1^x \sigma_2^x + E_{ZP}.
    \end{split}
\end{equation}
By construction, the GS obtained by applying the variational method is an eigenstate of $H_P$ and the eigenvalue also coincides with the variational energy. 

\begin{figure}
    \centering
    \includegraphics[width = \columnwidth]{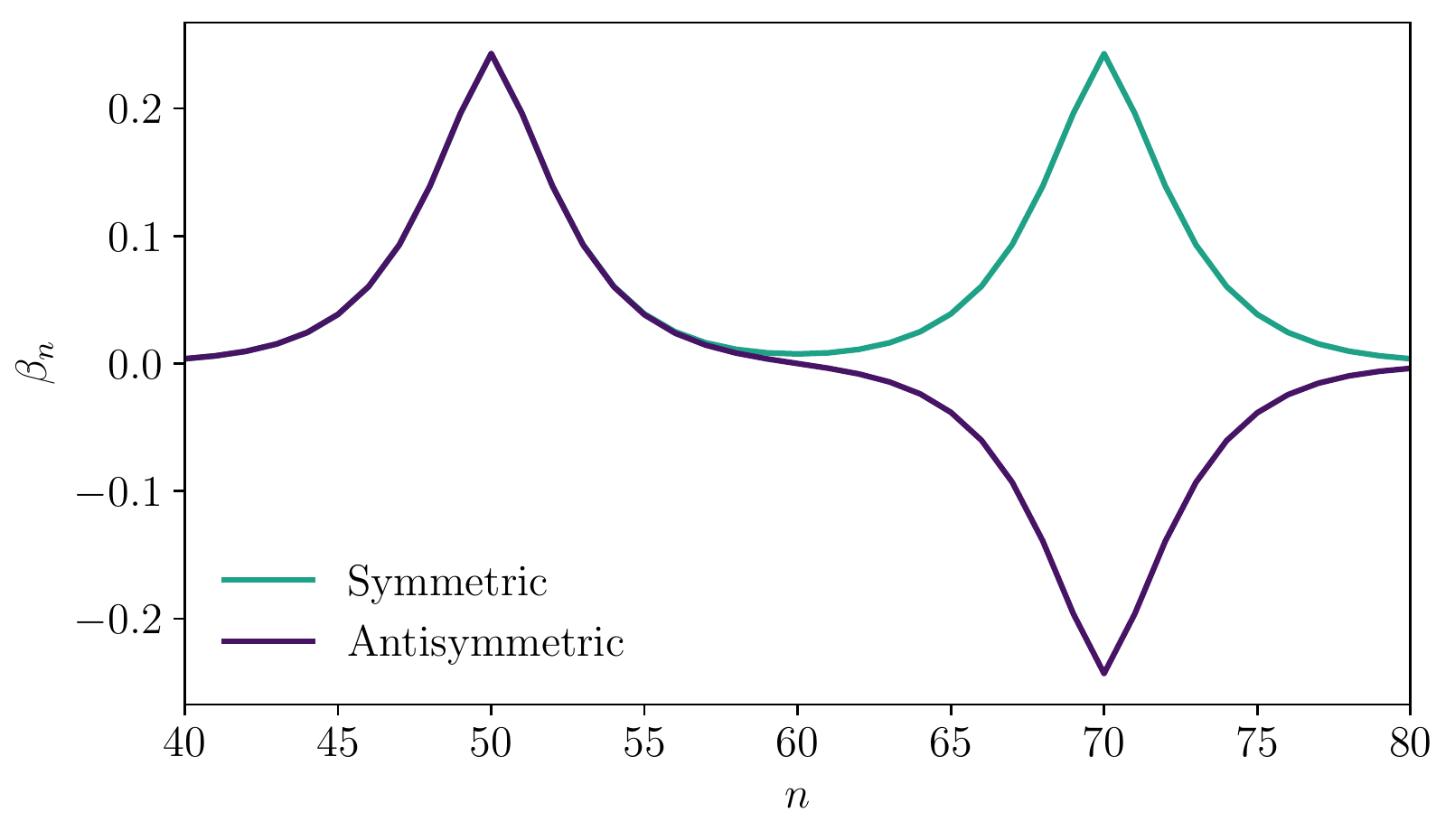}
    \caption{Symmetric and antisymmetric bound state wave functions for $g = 0.3$, $\Delta = 0.3$ and $x = 20$. Solid lines represent Polaron results and dashed lines represent RWA results.}
    \label{fig:2q_symmetric_antisymmetric}
\end{figure}

\begin{figure}
    \centering
    \includegraphics[width = \columnwidth]{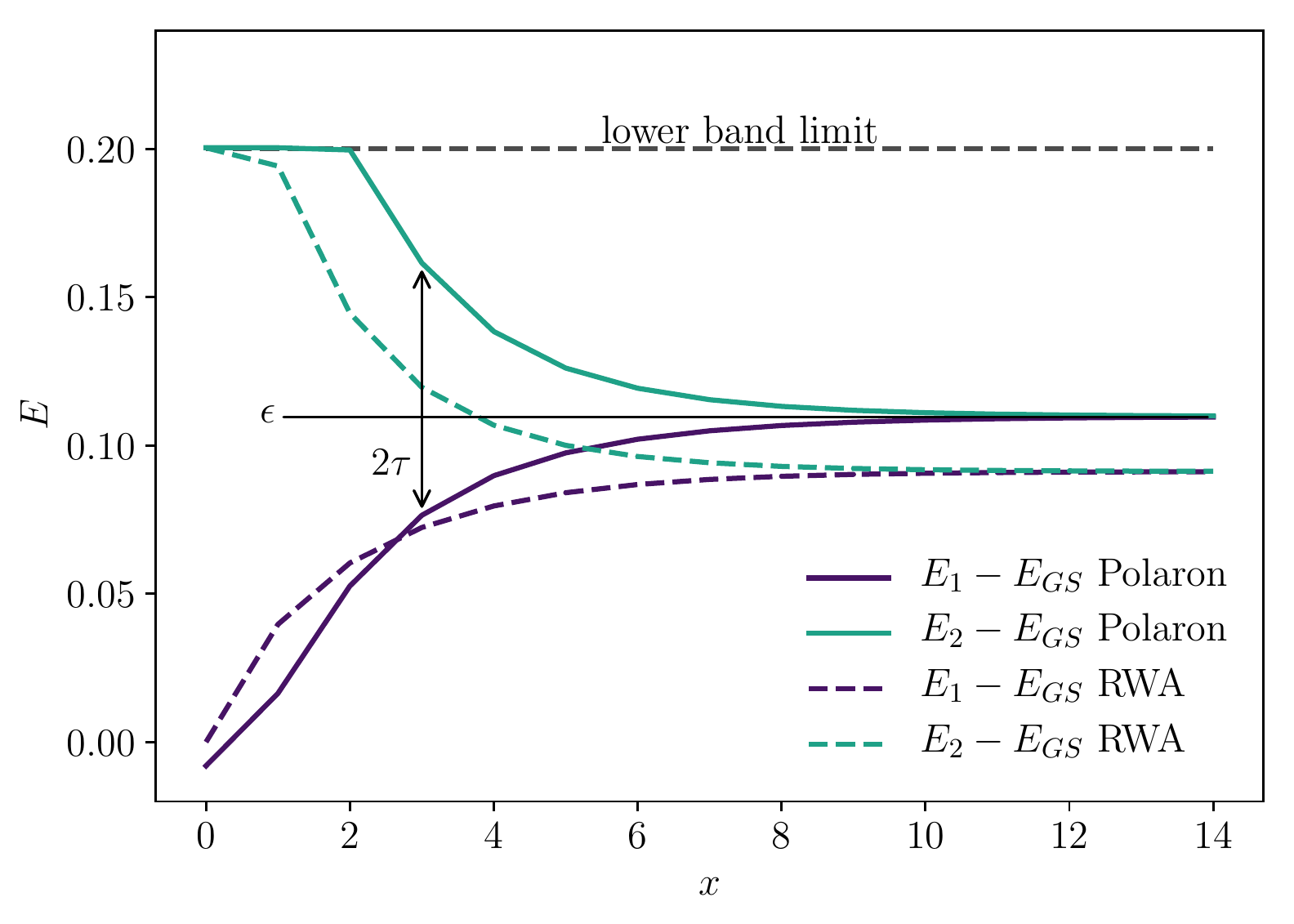}
    \caption{Energy diference between symmetric and antisymmetric bound states as a function of $x$, the distance between qubits for $g = 0.3$ and $\Delta = 0.3$. Dashed lines represent RWA results and solid lines represent Polaron results.}
    \label{fig:2q_polaron_rwa}
\end{figure}

Once again, we can diagonalise the restriction of $H_P$ in search of states whose energy lies below the band limit and are, thus, bound. In this case, we expect to find two bound states, corresponding to the symmetric and antisymmetric combination of the wavefunctions corresponding to each single-qubit bound state (See Fig. \ref{fig:2q_symmetric_antisymmetric}).
A general eigenstate of $H_P$ has the form
\begin{equation}
|\psi \rangle_P = ( \beta_1 \sigma_1^+ + \beta_2 \sigma_2^+ + \sum_k \beta_k a_k^\dagger ) \ket{GS}.
    \label{eq:2q_general_state}
\end{equation}
 The reader might recall that in the single-qubit case the one-excitation subspace was spanned by $\ket{1}\ket{0}$ and $\ket{0} \ket{1_k}$ and wonder why we cannot substitute $\ket{S}\ket{1_k}$ by $\ket{00}\ket{1_k}$ in the two-qubit basis. In that sense, it must be clarified that we seek to work in a subspace that is one excitation above the GS, regardless of however many excitations the GS contains.
 The proposed state, Eq. \eqref{eq:2q_general_state}, on the other hand, comes with a problem since the subspace spanned by Eq. \eqref{eq:2q_general_state}
 is not closed under the action of the Hamiltonian, \emph{e.g.} $\sigma_1^- a_k^\dagger |10 \rangle | 0_k\rangle  = \cos (\theta) |GS\rangle_S |1_k \rangle + \sin (\theta) (-\sin(\theta) |00 \rangle + \cos (\theta) |11 \rangle |1_k\rangle)  $.  Fortunately, this second contribution  is  of second order in $f_k$. Besides, the terms containing $\sigma_1^- a_k^\dagger $ in Hamiltonian [\eqref{Heff2q}] are of the order of $f_k$.  Thus they are h.o.t that, consistently with Eq. \eqref{Hp}, are discarded.  \\
Figure \ref{fig:2q_symmetric_antisymmetric} shows $\beta_n$ for the two lowest energy eigenstates. Where $\lambda_n$ is obtained by Fourier transforming $\lambda_k$ in Eq. \eqref{eq:2q_general_state}. In the single-qubit case, we showed that there exists a bound state in the form of a cloud of virtual photons localised around the qubit. We have also shown that two sufficiently distant qubits do no interact and, as such, their wavefunctions do not overlap, contributing two bound states of equal energy to the spectrum. As the two qubits approach, we expect the increasing overlap to break the degeneracy, and split the two bound states into different energies. If that is the case, the interaction can cause the energy of the antisymmetric state to rise above the lower band limit, forcing it to no longer be bound (nor antisymmetric), as the corresponding photons have an allowed frequency in the waveguide, and as such they no longer exhibit exponential decay. These oscillating eigenstates are referred to as scattering states \cite{sanchez2017b}. Figure \ref{fig:2q_polaron_rwa} shows the aforementioned effect. 
In the figure we also compare our results wich the ones obtained within the RWA.
We conclude that the latter understimates the interaction between the two bound states (see below).
In Fig. \ref{fig:2q_cloud_coalescence}
a comparison between the spatial distribution of photons for two different distances is drawn. As the two qubits approach, the difference in profiles becomes significant, and, should they reach $n = 2$, the antisymmetric bound state would cease to exist as it enters the allowed frequency band. See App. \ref{ap:bounds_nphotons} for a calculation of $\expval{b^\dagger_n b_n}$. \\
\begin{figure}
    \centering
    \includegraphics[width = \columnwidth]{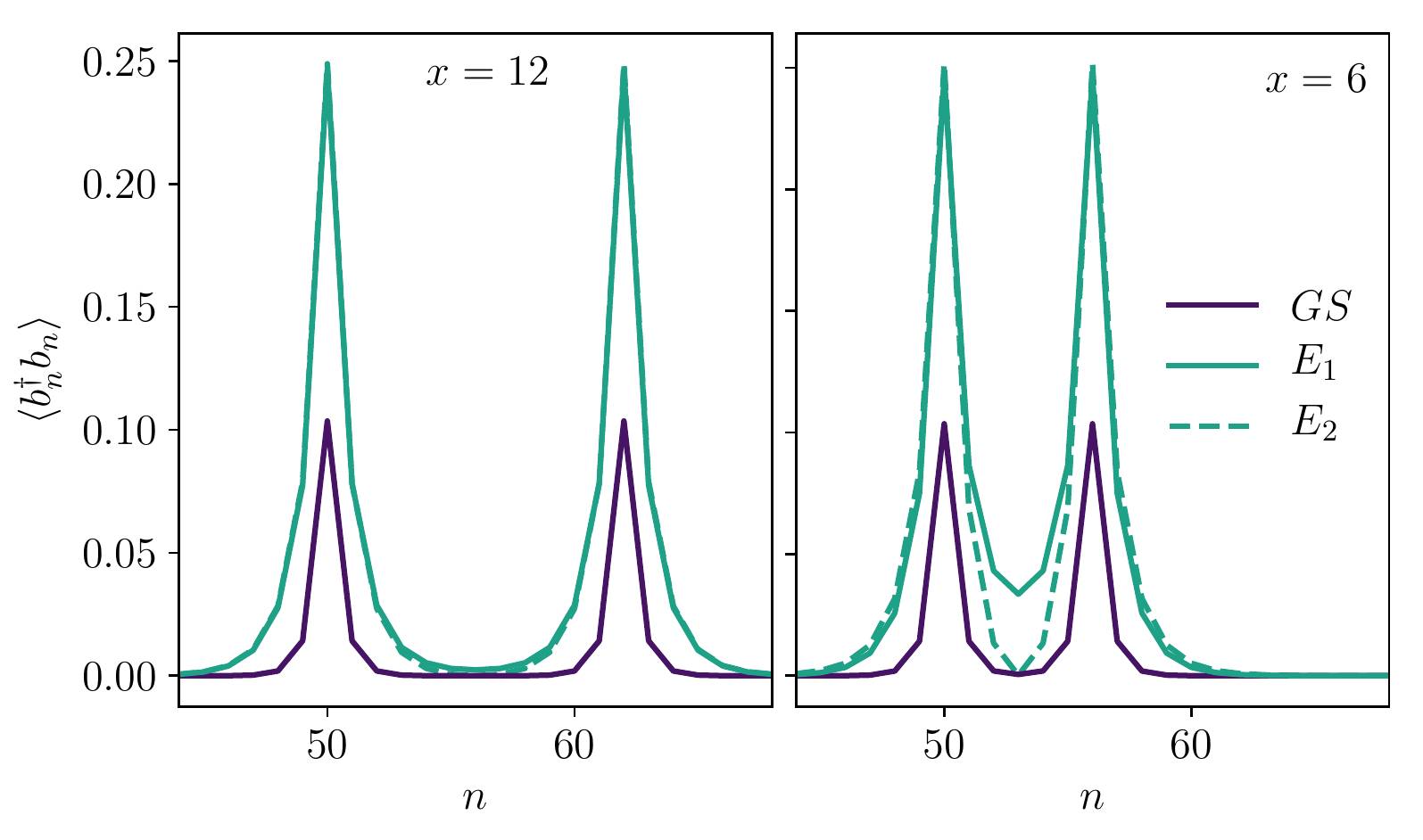}
    \caption{Comparison of spacial photon distributions of the symmetric ($E_1$) and antisymmetric ($E_2$) bound states for $g = 0.3$ and $\Delta = 0.3$ at two different distances, $x = 12$ and $x = 6$.}
    \label{fig:2q_cloud_coalescence}
\end{figure}

\subsection{State transfer}
\begin{figure*}
    \centering
    \includegraphics[width=0.9\textwidth]{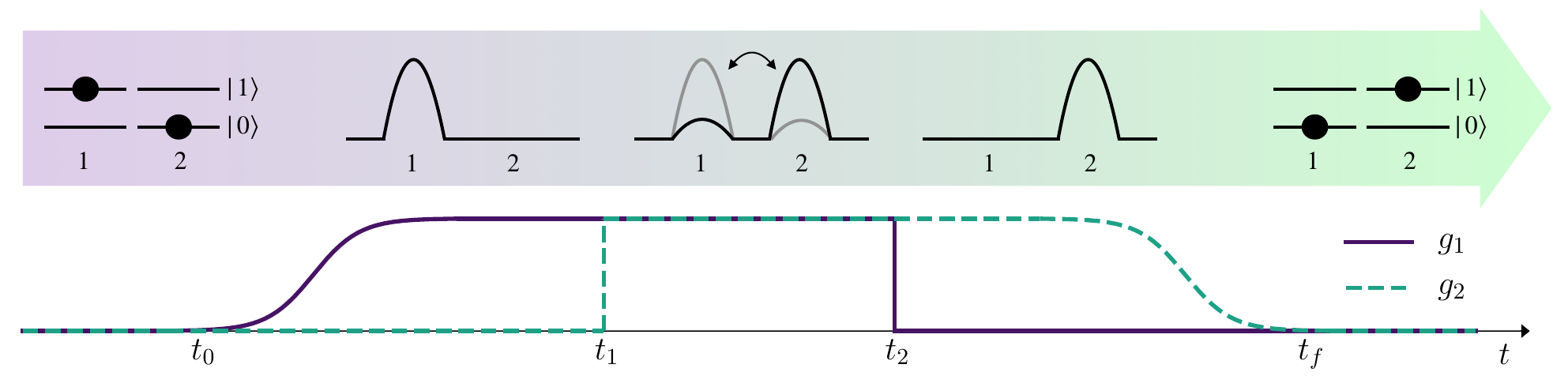}
    \caption{State transfer protocol between two qubits coupled to the same linear cavity array with distinct tunable coupling constants $g_1$ and $g_2$. Bound states are depicted as parabolic instead of exponential for aesthetic purposes.}
    \label{fig:protocol}
\end{figure*}
Inspired by Fig. \ref{fig:2q_polaron_rwa} and restricting ourselves to the two bound states we can define a tight-binding Hamiltonian
\begin{equation}
\label{tb}
    H_{TB} = \sum_{i = {L, R}} \epsilon \ket{i} \bra{i} + \sum_{\substack{i, j = L, R\\
    i \neq j}} \tau \ket{i} \bra{j},
\end{equation}
where $\ket{L}$ represents the bound state of the left-most qubit and $\ket{R}$ represents the bound state of the right-most one. The eigenstates of  $H_{TB}$ are the symmetric and antisymmetric combinations of $\ket{L}$ and $\ket{R}$, provided $\tau \neq 0$, with respective
energies $\epsilon - \tau$ and $\epsilon + \tau$. This simplified model is the basis for the study of effective interactions between bound states, which, as we introduced, provides a means to engineer lossless state transfer protocols through virtual photons, one of the main objectives of our work. Real, propagating photons can be used to transport information between distant qubits, but, even in one-dimensional arrays where an emitted photon travels non-dissipatively, there exist undesired losses intrinsic to the emission process. The reason is simple, in the absence of anisotropies, it is equally likely that a radiated photon will travel in the direction of the neighbouring qubit as it is for it to travel in the opposite direction, and thus be lost. The transmission of information via virtual photons bypasses this limitation. By virtue of them being non-radiative, there is no loss, information is shared between close qubits through the overlap of their photonic clouds.\\
To exemplify perfect lossless state-transfer using bound states, we propose the protocol shown in Fig. \ref{fig:protocol}. Its purpose is to transfer the excited state from one qubit to the other deterministically.
First, the left-most qubit is initialized in its excited state and the other is kept in its ground state while both are un-coupled from the waveguide. We assume that $g_1$ and $g_2$, the coupling constant of each qubit to the waveguide, can be tuned independently, and so at $t_0$, $g_1$ is increased adiabatically, so that the excited qubit entangles with the waveguide progressively, to become a bound-state, by means of the Adiabatic Theorem. Then, at $t_1$, $g_2$ is increased diabatically to match $g_1$. This sudden change in the Hamiltonian does not allow the state to evolve quasi-statically into the new eigenstate and instead gives rise to Rabi oscillations between the left and right bound states, whose symmetric and antysimmetric combinations are actually the eigenstates of the new Hamiltonian. Knowing the hopping frequency  ($\tau$), we can interrupt the dynamics, by diabatically zeroing $g_1$ at $t_2$, at the precise moment where the system is fully in the right bound state.  Finally, $g_2$ is lowered adiabatically, so the right bound state transforms into an excited right-most qubit, succesfully completing the state-transfer protocol at $t_f$. This protocol can be applied sequentially to a succession of qubits, to effectively transport a state along the waveguide. 

It is important to note that this method is limited by the fact that the interaction decays exponentially, and this limitation is twofold. Firstly, an exponential decay means that, in order to transport a state between distant qubits, many ancilla quibts are required, placed in close formation, so that there exists an effective interaction amongst every pair of consecutive qubits. For every qubit added to the chain, the system becomes more susceptible to decoherence and losses. In addition, the hopping frequency ($\tau$), which is proportional to the coupling, is what determines the speed at which a single iteration of the protocol can be performed, so it is against our interest that the coupling decays so rapidly. One may even doubt if an exponentially decaying effective interaction would be, at any range, intense enough to not be overpowered by spurious dipole-dipole interactions between the qubits, which decay more slowly, with a power-law. Fortunately, we can assure that the effective interaction is orders of magnitude greater than dipole-dipole interactions, since the latter is of the order of $10^{-4} \ eV$ for nearest neighbours withing a crystal lattice ($d \sim 1 \text{\AA}$) \footnote{To exemplify the weakness of the dipole-dipole interaction, it is insightful to remember that it was not strong enough to explain ferromagnetism in solids. The \textit{exchange interaction} had to be introduced in the study of ferromagnetism for this very reason.}. In our set up, there is a non-negligible interaction up to distances of around $3$ sites, which in experimental realisations of quantum circuits have sizes of millimetres. 
\section{Conclusions}
We have discussed the main  properties of bound states in waveguide QED beyond the RWA paradigm.  
In other words, we have quantified the corrections to the \emph{standard} calculations where the qubits-photons interaction is assumed to be number conserving based on the perturbative character of the latter.
We have shown that the Polaron technique is useful.
It provides a unitary transformation that disentangles  qubits and  waveguide and the interaction, within this picture, is effectively number conserving. 
Therefore,  it allows to export   techniques as the Weisskopf-Wigner theory and intuitions to a broader range of light-matter coupling strengths where the RWA  fails. \\
The main results  discussed  in the paper  are as follows.
We have extended the calculations for the
 spontaneous emission up to moderate light-matter couplings obtaining a renormalization of the rate (due to the qubit-frequency renormalization).
The existence criteria for bound states has been generalised and its role in the thermalization of the qubits has been discussed.
Finally, we have computed the effective spin-spin interactions both  through vacuum fluctuations and bound states.  We  sketched a perfect state transfer  protocol among bound states.
\label{sec:conclusions}
\section{Acknowledgments}
This work has been supported by the EU (COST Ac- tion 15128 MolSpin on Molecular Spintronics, QUAN- TERA SUMO project and the Spanish MICINN  grant  MAT2017- 89993-R. 
Juan Román-Roche  is supported by ICMA through a Pi2 contract. Eduardo Sánchez-Burillo acknowledges ERC Advanced Grant QUENOCOBA under the EU Horizon 2020 program (grant agreement 742102).
\appendix
\begin{widetext}
\section{Single qubit: some details of the calculations}
\label{app:sqc}
\subsection{Derivation of the basic commutation relations}
\label{ap:commutation_relations}

Let $A$, $B$, and $C$ be operators such that $[B, A] = C$. Then, if $[C, A] = 0$ it follows that $[B, A^n] = n C A^{n-1}$.

The proof is by induction:
\begin{align}
    [B, A^n] &= [B, A A^{n-1}] \notag\\
    &= A[B, A^{n-1}] + [B, A] A^{n-1} \notag\\
    \text{by I.H.} \rightarrow &= A (n-1) C A^{n-2} + C A^{n-1} \notag\\
    [C, A] = 0 \rightarrow &= (n-1)CA^{n-1} + CA^{n-1} \notag\\
    &= nCA^{n-1}
\end{align}

From this, one can prove that $[B, e^A] = Ce^A$ also holds.
\begin{align}
    [B, e^A] &= \sum_{n=0} \frac{[B, A^n]^n}{n!} \notag\\
    &=[A, I] + \sum_{n=1} \frac{[B, A^n]^n}{n!} \notag\\
    &=\sum_{n=1} \frac{nCA^{n-1}}{n!} = C \sum_{n=1} \frac{A^{n-1}}{(n-1)!} \notag\\
    &= C \sum_{n=0} \frac{A^n}{n!} = Ce^A
\end{align}

If we express the Polaron transform as $U_P = \exp[A]$, we can apply the properties we just proved to arrive at the basic commutation relations.
\begin{align}
    &[b_k, A] = -\sigma^x \left(f_k [b_k, b^\dagger_k] - f_k^* [b_k, b_k]\right) = -f_k \sigma^x \rightarrow \notag\\
    &[b_k, U_P] =  -f_k \sigma^x U_P \\
    &[b^\dagger_k, A] = -\sigma^x \left(f_k [b^\dagger_k, b^\dagger_k] - f_k^* [b^\dagger_k, b_k]\right) = -f_k \sigma^x \rightarrow \notag\\
    &[b^\dagger_k, U_P] =  -f_k^* \sigma^x U_P 
\end{align}
\subsection{Calculation of $\expval{b_n^\dagger b_n}$ for the GS of a single qubit}
\label{ap:number_photons}
\begin{align}
    \expval{b_n^\dagger b_n} &= \expval{b_n^\dagger b_n}{GS} \notag\\
    &= \expval{\frac{1}{\sqrt{N}} \sum_k e^{i k (n - N/2)} b_k^\dagger \frac{1}{\sqrt{N}} \sum_p e^{-i p (n - N/2)} b_p}{GS} \notag\\ 
    &= \frac{1}{N} \sum_{k, p} e^{i (k - p) (n - N/2)} \expval{U_p^\dagger b_k^\dagger b_p U_p}{0 0}
\end{align}

In order to continue, we must first calculate $U_p^\dagger b_k^\dagger b_p U_p$. Since we have already taken $f_k$ as real in previous calculations, we assume it to be real here as well.
\begin{align}
    U_p^\dagger b_k^\dagger b_p U_p &= U_P^\dagger \left(b^\dagger_k [b_p, U_P] + [b^\dagger_k, U_P] b_p\right) + b_k^\dagger b_p \notag\\
    &=U_P^\dagger \left(-f_p \sigma^x b^\dagger_k U_P -f_k \sigma^x U_P b_p\right) + b_k^\dagger b_p \notag\\
    &=U_P^\dagger \left(-f_p \sigma^x [b^\dagger_k, U_P] -f_p \sigma^x U_P b^\dagger_k -f_k \sigma^x U_P b_p\right) + b_k^\dagger b_p \notag\\
    &=U_P^\dagger \left(U_P f_k f_p - f_p \sigma^x U_P b^\dagger_k -f_k \sigma^x U_P b_p\right) + b_k^\dagger b_p \notag\\
    &= f_k f_p - \sigma^x (f_p b_k^\dagger + f_k b_p) + b_k^\dagger b_p
\end{align}

We are now equipped with the necessary ingredients to compute $\expval{U_p^\dagger b_k^\dagger b_p U_p}{0 0}$. Considering that the state $\ket{0 0}$ does not connect through the second and third terms, the mean value is just $f_k f_p$.

With that, we simply have
\begin{equation}
    \expval{b_n^\dagger b_n}  = \frac{1}{\sqrt{N}} \sum_k e^{i k (n - N/2)} f_k \frac{1}{\sqrt{N}} \sum_p e^{-i p (n - N/2)} f_p = f_n^* f_n = f_n^2.
\end{equation}

\subsection{Exponential localisation of the GS}
\label{ap:1q_exp_localisation}

We recall that
\begin{equation}
    f_k = \frac{g}{\sqrt{N} (\Delta_r + \omega_k)}
\end{equation}
with
\begin{equation}
    \omega_k = \omega_0 - 2 \lambda \cos k
\end{equation}
Noticing that
\begin{equation}
\label{Fekn}
    \mathcal{F} [\; e^{- \kappa_{\rm GS} | n - N/2 |} \; ]
    =
    \frac{1}{\sqrt{N}} 
    \frac{1- e^{- 2 \kappa_{\rm GS} }} {1 + e^{-2 \kappa_{\rm GS}} + 2 e^{-\kappa_{\rm GS}} \cos{k}}
\end{equation}
with \emph{our} convention for the Fourier transform $\mathcal {F} [ g(n) ] = \sum_0^{N-1} e ^{i k (n - N/2)} \; g(n)$.
Therefore,
\begin{equation}
\label{Fekn}
    \mathcal{F} [ \; e^{- \kappa_{\rm GS} | n - N/2 |} \; ]
    \sim
    f_k
\end{equation}
with the identifications
\begin{subequations}
\begin{align}
    \lambda & \sim e^{- \kappa_{\rm GS}}
    \\
    \omega_0 + \Delta_r & \sim 1 + e^{-2 \kappa_{\rm GS}},
\end{align}
\end{subequations}
which yields Eq \eqref{kappa} in the main text.

\subsection{Single qubit effective Hamiltonian}
\label{ap:1q_Heff}
The strict application of the Polaron transform to the original Hamiltonian, $H_P = U_P^\dagger H U_P$, yields the transformed Hamiltonian
\begin{equation}
        H_P = \frac{\Delta}{2} \exp[2 \sigma^x \sum_k f_k b^\dagger_k - f_k^* b_k] \sigma^z + \sum_k \omega_k b_k^\dagger b_k + \Delta_r {\sigma^x} \sum_k f_k \left(b_k^\dagger + b_k\right)
         + E_{ZP}.
\end{equation}

We can further simplify it by expanding the exponential term. Making use of the Baker-Campbell-Hausdorff formula and taking $\{f_k\}$ real we get
\begin{equation}
    \exp[2 \sigma^x \sum_k f_k b^\dagger_k - f_k b_k] = \exp[-2\sum_k f_k^2] \exp[2 \sigma^x \sum_k f_k b^\dagger_k] \exp[-2 \sigma^x\sum_k f_k b_k].
    \label{eq:resultofBCH}
\end{equation}

A power series expansions of the non constant terms gives
\begin{align}
    \exp[2 \sigma^x \sum_k f_k b^\dagger_k] &= 1 + 2 \sigma^x \sum_k f_k b^\dagger_k + \dots, \\
    \exp[-2 \sigma^x\sum_k f_k b_k] &= 1 -2 \sigma^x\sum_k f_k b_k + \dots.
\end{align}

Ignoring higher order terms, the right hand side of  Eq. \eqref{eq:resultofBCH} becomes
\begin{equation}
    \exp[-2\sum_k f_k^2] \left( 1 + 2 \sigma^x \sum_k f_k \left( b^\dagger_k - b_k\right) - 4 \sum_{k, p} f_k f_p b^\dagger_k b_p \right).
\end{equation}

Reintroducing this result in $H_P$ yields
\begin{equation}
\begin{split}
    H_{P} &= \frac{\Delta_r}{2}\sigma^z + \Delta_r \sigma^x \sigma^z \sum_k f_k \left(b^\dagger_k - b_k\right) - 2 \Delta_r \sum_{k, p} f_k f_p b^\dagger_k b_p \\ 
    &+ \sum_k \omega_k b^\dagger_k b_k + \Delta_r {\sigma^x} \sum_k f_k \left(b_k^\dagger + b_k\right) + E_{ZP}.
\end{split}
\end{equation}

Considering that $\sigma^x \sigma^z + \sigma^x = 2 \sigma^-$ and $-\sigma^x \sigma^z + \sigma^x = 2 \sigma^+$ we can combine the second and second-to-last terms to arrive at the final expression for $H_P$
\begin{equation}
    H_{\text{eff}} = \frac{\Delta_r}{2}\sigma^z + \sum_k \omega_k b^\dagger_k b_k + 2 \Delta_r \sum_k f_k  \left(\sigma^- b^\dagger_k - \sigma^+ b_k\right) - 2 \Delta_r \sum_{k, p} f_k f_p b^\dagger_k b_p  + E_{ZP}.
\end{equation}

\subsection{Calculation of $\expval{b_n^\dagger b_n}$ for the SEBS of a single qubit}
\label{ap:n_photons_bound}
The SEBS will be a state of the form
\begin{equation}
    \ket{v} = \lambda_0 \ket{1}\ket{0} + \sum_k \lambda_k \ket{0}\ket{1_k}.
\end{equation}

Thus 
\begin{align}
    \expval{b_n^\dagger b_n} &= \expval{b_n^\dagger b_n}{v} \notag\\
    &= \expval{\frac{1}{\sqrt{N}} \sum_k e^{i  k (n - N/2)} b_k^\dagger \frac{1}{\sqrt{N}} \sum_p e^{-i p (n - N/2)} b_p}{v} \notag\\ 
    &= \frac{1}{N} \sum_{k, p} e^{i (k - p) (n - N/2)} \expval{U_p^\dagger b_k^\dagger b_p U_p}{v}.
    \label{eq:mean_photon_calc}
\end{align}

In App. \ref{ap:number_photons} we saw that $U_p^\dagger b_k^\dagger b_p U_p = f_k f_p - \sigma^x (f_p b_k^\dagger + f_k b_p) + b_k^\dagger b_p$ . The first term is constant so $\ket{v}$ connects entirely yielding $f_k f_p$. The last term only connects the bosonic part of $\ket{v}$ to give $\lambda_k \lambda_p$. Finally, the second term cross-connects the two components of $\ket{v}$ resulting in $\lambda_0 f_p \lambda^*_k + \lambda_0 f_k \lambda_p$. Reintroducing these partial results into Eq. \eqref{eq:mean_photon_calc} one has
\begin{equation}
\label{bndbnapp}
    \expval{b_n^\dagger b_n} = f_n^2 + \lambda_n^2 + \lambda_0 f_n^* \lambda^*_n + \lambda_0 f_n \lambda_n = f_n^2 + \lambda_n^2 + 2 \lambda_0 \Re{f_n \lambda_n}.
\end{equation}

\subsection{Existence of bound states in USC}
\label{ap:existence}
We work in the Polaron picture.  A non-normalized single excitation is,
\begin{equation}
    | \psi _1 \rangle_P = 
   \lambda_0 | 1, 0 \rangle 
    +
    \sum \lambda_k |0, 1_k \rangle
\end{equation}
It is an eigenstate iff 
\begin{subequations}
\begin{align}
\label{E11}
    \Delta_r - \sum \lambda_k  2 \Delta_r f_k
    & = 
    E
    \\ 
\label{E12}
    \lambda_k  \omega_k - 2 \Delta_r f_k + 2 \Delta_r  \lambda_k  \sum_{k^\prime} f_k f_{k^\prime} &= E \lambda_k 
\end{align}
\end{subequations}
The solution for $E$ is found by searching the zeros of the function $F(E)$ [Cf. with the RWA case in Ref. \onlinecite{shi2016}]
\begin{equation}
\label{FE}
    F_1 (E) = E -  \left ( \Delta_r + \sum_k 
    \frac{ ( 2 \Delta_r f_k)^2}{ E - \omega_k - 2 \Delta_r \sum_{k^\prime} f_k f_{k^{\prime}}} \right ) \; .
\end{equation}
\\
If $E< {\rm min} [\omega_k]$, the state is a SEBS.   
Notice that the term in brackets is is a monotonically decreasing  function with $g$.
Therefore if a bound state exists for $g\to 0^+$ then it will exists for any finite value of $g$.  
For our model in the limit $g \to 0^+$ a bound state below the band exists \cite{shi2016}, thus the existence of bound states in the USC is guaranteed.

\subsection{Localization lenght}
\label{ap:localization}
Apart from their existence the key property of bound states is their localization lenght.  From, Eq. \eqref{E12} we obtain that:
\begin{equation}
    \lambda_k = \frac{ 2 \Delta_r f_k }{\omega_k-E_1+ 2 \Delta_r \sum_{k^\prime} f_k f_{k^\prime}}
\end{equation}
In the log $g$-regime $2 \Delta_r f_k \sim g$ and we can neglect the term $2 \Delta_r \sum_{k^\prime} f_k f_{k^\prime} $, therefore by simple inspection we see that 
$\mathcal {F}^{-1} [\lambda_k] = \lambda_n \sim e^{-\kappa n}$ with
\begin{equation}
    \kappa_1 \cong \arccosh \left ( 
    \frac{\omega_0-E_1}{2 \mathcal J}
    \right )
\end{equation}
Looking at Eq. \eqref{bndbnapp}, Section \ref{ap:1q_exp_localisation} and Eq. \eqref{kappa} the localization  is given by 
$
  \kappa_{\rm SEBS}^{-1}= \max   ( \kappa_{\rm GS}^{-1}, \kappa^{-1} )
$ as given by Eq. \eqref{ksebs}.
\\
\section{Calculations for the two-qubit case}
\subsection{Derivation of the basic commutation relations}
\label{ap:2q_commutation_relations}
Recycling much of the work done in Ap. \ref{ap:commutation_relations} we simply see that
\begin{align}
    &[b_k, A_j] = -\sigma_j^x \left( f_k e^{i k x_j} [b_k, b_k^\dagger] - f_k e^{-i k x_j} [b_k, b_k] \right) = - \sigma^x f_k e^{i k x_j} \rightarrow \notag \\
    &[b_k, U_j] = - \sigma^x f_k e^{i k x_j} U_j \label{eq:basic1}\\ 
    &[b_k^\dagger, A_j] = -\sigma_j^x \left( f_k e^{i k x_j} [b_k^\dagger, b_k^\dagger] - f_k e^{-i k x_j} [b_k^\dagger, b_k] \right) = - \sigma^x f_k e^{-i k x_j} \rightarrow \notag \\
    &[b_k^\dagger, U_j] = - \sigma^x f_k e^{-i k x_j} U_j. \label{eq:basic2}
\end{align}

\subsection{Calculation of $H_I$}
\label{ap:H_I}
In an attempt to lighten notation we have omitted the summation signs ($\sum$) in the following calculation. They will be reintroduced when we present the final result. It must be understood that there is summation over all indexes present, for instance

\begin{equation}
    \sigma^x_j c_k \left(b_k^\dagger e^{i k x_j} + b_k e^{-i k x_j} \right) \equiv \sum_j \sigma^x_j \sum_k c_k \left(b_k^\dagger e^{i k x_j} + b_k e^{-i k x_j} \right).
\end{equation}

We thus have
\begin{equation}
    U_P^\dagger H_I U_P = U_2^\dagger U_1^\dagger \left( H_I^1 + H_I^2 \right) U_1 U_2,
\end{equation}
we can focus on $H_I^1$ and the results will be perfectly extensible to $H_I^2$. 

Hence, making use of the basic commutation relations, [Eqs. \eqref{eq:basic1}, \eqref{eq:basic2}],
\begin{align}
U_2^\dagger U_1^\dagger H_I^1 U_1 U_2 &= U_2^\dagger U_1^\dagger \sigma^x_1 c_k \left(b_k^\dagger e^{i k x_1} + b_k e^{-i k x_1} \right) U_1 U_2 \notag \\
&= U_2^\dagger \left( \sigma_1^x c_k \left(-\sigma_1^x e^{ikx_1} f_k e^{-ikx_1} -\sigma_1^x e^{-ikx_1} f_k e^{ikx_1} \right) + H_I^1 \right) U_2 \notag \\
&= U_2^\dagger \left(-2 c_k f_k + H_I^1 \right) U_2 = -2 c_k f_k + U_2^\dagger H_I^1 U_2 \notag \\
&= -2c_k f_k + U_2^\dagger \left( \sigma_1^x c_k \left(b_k^\dagger e^{i k x_1} + b_k e^{-i k x_1} \right) \right) U_2 \notag \\
&= -2 c_k f_k + \sigma_1^x c_k \left( - \sigma_2^x e^{i k x_1} f_k e^{-i k x_2} - \sigma_2^x e^{-i k x_1} f_k e^{i k x_2} \right) + H_I^1 \notag \\
&= - 2c_k f_k - 2 \sigma_1^x \sigma_2^x c_k f_k \cos(kx) + H_I^1.
\end{align}

Likewise,
\begin{equation}
    U_2^\dagger U_1^\dagger H_I^2 U_1 U_2 = - 2c_k f_k - 2 \sigma_1^x \sigma_2^x c_k f_k \cos(kx) + H_I^2.
\end{equation}

And finally,
\begin{equation}
    U_P^\dagger H_I U_P = - 4 \sum_k c_k f_k - 4 \sigma_1^x \sigma_2^x \sum_k c_k f_k \cos(kx) + H_I.
\end{equation}

\subsection{Calculation of $H_B$}
\label{ap:H_B}
Much like in App. \ref{ap:H_I} we have omitted the summation signs for the calculation.
\begin{align}
    U_P^\dagger H_B U_P &= U_2^\dagger U_1^\dagger \omega_k b_k^\dagger b_k U_1 U_2 \notag \\
    &= \omega_k U_2^\dagger \left( U_1^\dagger \left( b_k^\dagger [b_k, U_1] + [b_k^\dagger, U_1] b_k \right)  + H_B / \omega_k \right) \notag \\
    &= \omega_k U_2^\dagger \left( U_1^\dagger \left( - \sigma_1^x b_k^\dagger f_k e^{i k x_1} U_1 - \sigma f_k e^{-i k x_1} U_1 b_k \right) + H_B / \omega_k \right) U_2 \notag \\
    &= \omega_k U_2^\dagger \left(U_1^\dagger \left(- \sigma_1^x f_k e^{i k x_1} \left( U_1 b_k^\dagger - \sigma_1^x f_k e^{-i k x_1} U_1 \right) - \sigma_1^x f_k e^{-i k x_1} U_1 b_k \right) + H_B / \omega_k \right) U_2 \notag \\
    &= \omega_k U_2^\dagger \left( - \sigma_1^x f_k \left( b_k^\dagger e^{i k x_1} + b_k e^{-i k x_1} \right) + f_k^2 + H_B / \omega_k \right)U_2 \notag \\
    &= H_B + \omega_k \left( 2f_k^2 - \sigma_1^x f_k \left( b_k^\dagger e^{i k x_1} + b_k e^{-i k x_1} \right) - \sigma_2^x f_k \left( b_k^\dagger e^{i k x_2} + b_k e^{-i k x_2} \right) \right. \notag \\
    &\left. \quad \quad \quad \quad \quad \quad +\ U_2^\dagger \left[- \sigma_1^x f_k \left( b_k^\dagger e^{i k x_1} + b_k e^{-i k x_1} \right), U_2\right]\right) \notag \\
    &= H_B + \omega_k \left( 2f_k^2 - \sigma_j^x f_k \left( b_k^\dagger e^{i k x_j} + b_k e^{-i k x_j}\right) \right. \notag \\
    &\left. \quad \quad \quad \quad \quad \quad -\ \sigma_1^x f_k \left(-\sigma_2^x e^{i k x_1} f_k e^{-i k x_2} - \sigma_2^x e^{-i k x_1} f_k e^{i k x_2}\right)\right) \notag \\
    &= \omega_k \left( 2 f_k^2 - \sigma_j^x f_k \left( b_k^\dagger e^{i k x_j} + b_k e^{-i k x_j}\right) + 2 \sigma_1^x \sigma_2^x f_k^2 \cos(k(x_2 - x_1)) \right) + H_B
\end{align}

So finally,
\begin{equation}
    \begin{split}
        U_P^\dagger H_B U_P =&\ 2 \sum_k \omega_k f_k^2 + 2 \sigma_1^x \sigma_2^x \sum_k \omega_k f_k^2 \cos(kx) \\
        &- \sum_j \sigma_j^x \sum_k \omega_k f_k \left(b_k^\dagger e^{i k x_j} + b_k e^{-i k x_j}\right) + \sum_k \omega_k b_k^\dagger b_k.
    \end{split}
\end{equation}

\subsection{Calculation of the minimal value of $f_k$}
\label{ap:f_k}
The explicit dependence of $\bar E_{GS}$ with $f_k$ is
\begin{align}
    \bar E_{GS} &= - \mathcal E + 2 \sum_k f_k (\omega_k f_k - 2 c_k)= - \sqrt{\Delta_r^2 + \mathcal J^2} + 2 \sum_k f_k (\omega_k f_k - 2 c_k)  \notag \\
    \begin{split}
        &= -\sqrt{4 \left( \sum_k f_k (2 c_k - \omega_k f_k) \cos(kx) \right)^2 + \Delta^2 \exp[-4\sum_k f_k^2]}\\
    & \quad \quad + 2 \sum_k f_k (\omega_k f_k - 2 c_k).
    \end{split}
\end{align}

Thus
\begin{align}
    &\frac{\partial \bar E_{GS}}{\partial f_k} = - \frac{4 \mathcal J (2c_k - 2 \omega_k f_k) \cos(kx)) - 8 f_k \Delta_r^2}{2 \mathcal E} + 2(2\omega_k f_k - 2c_k) = 0 \rightarrow \notag \\
    &f_k = c_k \frac{\mathcal E + \mathcal J \cos(kx)}{\mathcal E \omega_k + \mathcal J \omega_k \cos(kx) + \Delta_r^2}
\end{align}

\subsection{Calculation of $\expval{b_n^\dagger b_n}$ for the GS of the two-qubit scenario}
\label{ap:GS_nphotons}

The ground state is
\begin{equation}
    \ket{GS} = \left(\alpha \ket{00} + \beta \ket{11} \right) \ket{0}
\end{equation}

Thus 
\begin{align}
    \expval{b_n^\dagger b_n} &= \expval{b_n^\dagger b_n}{GS} \notag\\
    &= \expval{\frac{1}{\sqrt{N}} \sum_k e^{i  k (n - N/2)} b_k^\dagger \frac{1}{\sqrt{N}} \sum_p e^{-i p (n - N/2)} b_p}{GS} \notag\\ 
    &= \frac{1}{N} \sum_{k, p} e^{i (k - p) (n - N/2)} \expval{U_p^\dagger b_k^\dagger b_p U_p}{GS}.
    \label{eq:mean_photon_calc}
\end{align}

We must now calculate $U_p^\dagger b_k^\dagger b_p U_p$ in the two-qubit case.
\begin{align}
    U_p^\dagger b_k^\dagger b_p U_p &= U_2^\dagger U_1^\dagger b^\dagger_k b_p U_1 U_2 \notag\\
    &=U_2^\dagger U_1^\dagger \left( \left(b^\dagger_k [b_p, U_1] + [b^\dagger_k, U_1] b_p\right) + b_k^\dagger b_p \right) U_2 \notag\\
    &=U_2^\dagger U_1^\dagger \left( \left(-f_p e^{i p x_1} \sigma_1^x b^\dagger_k U_1 -f_k e^{-i k x_1} \sigma_1^x U_1 b_p\right) + b_k^\dagger b_p \right) U_2 \notag\\
    &=U_2^\dagger \left( \left(-f_p e^{ip x_1} \sigma_1^x \right) \left(-f_k e^{-ik x_1} \sigma_1^x \right) -f_p  e^{ip x_1} \sigma_1^x b^\dagger_k -f_k  e^{-i k x_1} \sigma_1^x b_p + b_k^\dagger b_p \right) U_2 \notag\\
    &=U_2^\dagger \left( f_k f_p e^{-i k x_1} e^{i p x_1} -f_p  e^{ip x_1} \sigma_1^x b^\dagger_k -f_k  e^{-i k x_1} \sigma_1^x b_p + b_k^\dagger b_p \right) U_2 \notag\\
    &= f_k f_p e^{-i k x_1} e^{i p x_1} + f_k f_p e^{-i k x_2} e^{i p x_2} \notag \\
    & \quad -f_p  e^{ip x_1} \sigma_1^x b^\dagger_k -f_k  e^{-i k x_1} \sigma_1^x b_p \notag \\
    & \quad -f_p  e^{ip x_2} \sigma_2^x b^\dagger_k -f_k  e^{-i k x_2} \sigma_2^x b_p \notag \\
    & \quad + b_k^\dagger b_p + \left[-f_p  e^{ip x_1} \sigma_1^x b^\dagger_k, U_2 \right] + \left[-f_k  e^{-i k x_1} \sigma_1^x b_p, U_2 \right] \notag
\end{align}

The last two terms give,
\begin{align*}
    &= -f_p  e^{ip x_1} \sigma_1^x \left(- \sigma_2^x f_k e^{-i k x_2} \right) -f_k  e^{-i k x_1} \sigma_1^x \left(- \sigma_2^x f_p e^{ip x_2} \right) \\
    &=\sigma_1^x \sigma_2^x f_p f_k e^{i p x_1} e^{- i k x_2} + \sigma_1^x \sigma_2^x f_p f_k e^{i p x_2} e^{- i k x_1}.
\end{align*}

Putting everything together one has
\begin{align}
    U_p^\dagger b_k^\dagger b_p U_p &= f_k f_p e^{-i k x_1} e^{i p x_1} + f_k f_p e^{-i k x_2} e^{i p x_2} \notag \\
    &\quad - \sum_{j} \sigma_j^x \left(f_p e^{i p x_j} b_k^\dagger + f_k e^{-i k x_j} b_p \right) \notag \\
    & \quad + \sigma_1^x \sigma_2^x f_p f_k \left(e^{i p x_1} e^{-i k x_2} + e^{i p x_2} e^{-i k x_1} \right) \notag \\
    & \quad + b_k^\dagger b_p
\end{align}

The ground state has no photons, so it only connects with itself through the first and second-to-last terms of $U_p^\dagger b_k^\dagger b_p U_p$. The first term connects the GS with itself completely, while the other cross-connects the spin terms $\ket{00}$ and $\ket{11}$. This yields
\begin{align}
    \expval{U_p^\dagger b_k^\dagger b_p U_p}{GS} &= f_k f_p e^{-i k x_1} e^{i p x_1} + f_k f_p e^{-i k x_2} e^{i p x_2} \notag \\
    & \quad + 2 \alpha \beta f_p f_k \left(e^{i p x_1} e^{-i k x_2} + e^{i p x_2} e^{-i k x_1} \right).
\end{align}

Completing the Fourier transform one finally arrives at
\begin{equation}
    \expval{b_n^\dagger b_n} = \abs{f_{n, 1}}^2 + \abs{f_{n, 2}}^2 + 4 \alpha \beta \Re{f_{n, 1} f^*_{n, 2}}.
\end{equation}

Where $f_{n, 1}$ is the fourier transform of $f_{k, 1}$, defined as
\begin{equation}
    f_{k, 1} = f_k e^{i k x_1}.
\end{equation}

\subsection{Calculation of $\expval{b_n^\dagger b_n}$ for the bound states of the two-qubit scenario}
\label{ap:bounds_nphotons}

The SEBS will be states of the form
\begin{equation}
    \ket{v} = \lambda_0 \ket{01}\ket{0} + \lambda_1 \ket{10}\ket{0} + \sum_k \lambda_k \left(\alpha \ket{00} + \beta \ket{11} \right) \ket{1_k}.
\end{equation}

Thus 
\begin{align}
    \expval{b_n^\dagger b_n} &= \expval{b_n^\dagger b_n}{v} \notag\\
    &= \expval{\frac{1}{\sqrt{N}} \sum_k e^{i  k (n - N/2)} b_k^\dagger \frac{1}{\sqrt{N}} \sum_p e^{-i p (n - N/2)} b_p}{v} \notag\\ 
    &= \frac{1}{N} \sum_{k, p} e^{i (k - p) (n - N/2)} \expval{U_p^\dagger b_k^\dagger b_p U_p}{v}.
    \label{eq:mean_photon_calc}
\end{align}

We saw in App. \ref{ap:GS_nphotons} that
\begin{align}
    U_p^\dagger b_k^\dagger b_p U_p &= f_k f_p e^{-i k x_1} e^{i p x_1} + f_k f_p e^{-i k x_2} e^{i p x_2} \notag \\
    &\quad - \sum_{j} \sigma_j^x \left(f_p e^{i p x_j} b_k^\dagger + f_k e^{-i k x_j} b_p \right) \notag \\
    & \quad + \sigma_1^x \sigma_2^x f_p f_k \left(e^{i p x_1} e^{-i k x_2} + e^{i p x_2} e^{-i k x_1} \right) \notag \\
    & \quad + b_k^\dagger b_p.
\end{align}

Contrary to what happened with the GS, all terms must now be considered because the SEBS connect through them all in one way or another. Thus, we must study each term individually.

The first two are trivial, as they connect SEBS completely with them selves, so they will not be discussed. 

The second term is more interesting. Through
\begin{equation}
    - \sum_{j} \sigma_j^x \left(f_p e^{i p x_j} b_k^\dagger + f_k e^{-i k x_j} b_p \right),
\end{equation}
the term $\lambda_0 \ket{01}\ket{0}$ in $\ket{v}$ becomes 
\begin{equation}
    -\lambda_0 f_{p,1} \ket{11} \ket{1_k} - \lambda_0 f_{p,2} \ket{00} \ket{1_k},
\end{equation}
which connects with $\lambda_k \left(\alpha \ket{00} + 
\beta \ket{11} \right) \ket{1_k}$ to yield 
\begin{equation}
    -\lambda_0 \lambda_k \left( \beta f_{p, 1} + \alpha f_{p, 2} \right).
\end{equation}
The term $\lambda_1 \ket{10}\ket{0}$ in $\ket{v}$ becomes 
\begin{equation}
    -\lambda_1 f_{p,1} \ket{11} \ket{1_k} - \lambda_1 f_{p,2} \ket{00} \ket{1_k},
\end{equation}
which connects with  $\lambda_k \left(\alpha \ket{00} + \beta \ket{11} \right) \ket{1_k}$ to yield
\begin{equation}
    -\lambda_1 \lambda_k \left( \alpha f_{p, 1} + \beta f_{p, 2} \right).
\end{equation}
Naturally, the term $\left(\alpha \ket{00} + \beta \ket{11} \right) \sum_k \lambda_k \ket{1_k}$ connects with both $\lambda_0 \ket{01}\ket{0}$ and $\lambda_1 \ket{10}\ket{0}$ to yield the complex conjugate of the terms that we just calculated in the opposite direction.

Through the third term,
\begin{equation}
    \quad + \sigma_1^x \sigma_2^x f_p f_k \left(e^{i p x_1} e^{-i k x_2} + e^{i p x_2} e^{-i k x_1} \right),
\end{equation}
the term $\lambda_0 \ket{01}\ket{0}$ in $\ket{v}$ becomes 
\begin{equation}
    \lambda_0 \left(f_{p,1} f^*_{k, 2} + f_{p,2} f^*_{k, 1} \right) \ket{10}\ket{0},
\end{equation}
which connects with $\lambda_1 \ket{10}\ket{0}$ to yield 
\begin{equation}
    \lambda_0 \lambda_1 \left(f_{p,1} f^*_{k, 2} + f_{p,2} f^*_{k, 1} \right).
\end{equation}
Naturally, the term $\lambda_1 \ket{10}\ket{0}$ in $\ket{v}$ becomes
\begin{equation}
    \lambda_1 \left(f_{p,1} f^*_{k, 2} + f_{p,2} f^*_{k, 1} \right) \ket{01}\ket{0},
\end{equation}
which connects with $\lambda_0 \ket{01}\ket{0}$ to yield 
\begin{equation}
    \lambda_0 \lambda_1 \left(f_{p,1} f^*_{k, 2} + f_{p,2} f^*_{k, 1}  \right),
\end{equation}
the complex conjugate of its counterpart.  Lastly, the term $\left(\alpha \ket{00} + \beta \ket{11} \right) \sum_k \lambda_k \ket{1_k}$ becomes
\begin{equation}
    \left(\alpha \ket{11} + \beta \ket{00} \right) \sum_k \lambda_k \ket{1_k} \left(f_{p,1} f^*_{k, 2} + f_{p,2} f^*_{k, 1} \right),
\end{equation}
and connects with $\left(\alpha \ket{00} + \beta \ket{11} \right) \sum_k \lambda_k \ket{1_k}$ to yield
\begin{equation}
    2 \alpha \beta \left( 1 - \lambda_0^2 - \lambda_1^2 \right) \left(\alpha \ket{00} + \beta \ket{11} \right) \left(f_{p,1} f^*_{k, 2} + f_{p,2} f^*_{k, 1} \right).
\end{equation}
Finally, the term $b_k^\dagger b_p$ connects the $p^{\text{th}}$ and $k^{\text{th}}$ photonic terms to yield $\lambda_k^* \lambda_p$.

Summarizing, we have
\begin{align}
    \expval{U_p^\dagger b_k^\dagger b_p U_p}{v} &= \left(f^*_{k,1} f_{p, 2} + f^*_{k,2} f_{p, 1} \right) \notag \\
    & \quad -\lambda_0 \lambda_k \left( \beta f_{p, 1} + \alpha f_{p, 2} \right) -\lambda_1 \lambda_k \left( \alpha f_{p, 1} + \beta f_{p, 2} \right) \notag \\
    & \quad  -\lambda_0 \lambda_p \left( \beta f^*_{k, 1} + \alpha f^*_{k, 2} \right) -\lambda_1 \lambda_p \left( \alpha f^*_{k, 1} + \beta f^*_{k, 2} \right) \notag \\
    & \quad +\lambda_0 \lambda_1 \left(f_{p,1} f^*_{k, 2} + f_{p,2} f^*_{k, 1}  \right) \notag \\
    & \quad +2 \alpha \beta \left( 1 - \lambda_0^2 - \lambda_1^2 \right) \left(\alpha \ket{00} + \beta \ket{11} \right) \left(f_{p,1} f^*_{k, 2} + f_{p,2} f^*_{k, 1} \right) \notag \\
    & \quad + \lambda_k^* \lambda_p.
\end{align}

Completing the Fourier transform, one arrives at
\begin{align}
    \expval{b_n^\dagger b_n} &= \abs{f_{n, 1}}^2 + \abs{f_{n, 2}}^2 \notag \\
    & \quad - 2\lambda_0 \Re{\lambda_n^* \left(\beta f_{n,1} + \alpha f_{n, 2} \right)} \notag \\
    & \quad - 2\lambda_1 \Re{\lambda_n^* \left(\alpha f_{n,1} + \beta f_{n, 2} \right)} \notag \\
    & \quad + 4 \lambda_0 \lambda_1 \Re{f_{n, 1} f^*_{n, 2}}\notag \\
    & \quad+ 4 \alpha \beta \left( 1 - \lambda_0^2 - \lambda_1^2 \right) \Re{f_{n, 1} f^*_{n, 2}} \notag \\
    & \quad + \abs{\lambda_n}^2
\end{align}

Where $f_{n, 1}$ is the fourier transform of $f_{k, 1}$, defined as
\begin{equation}
    f_{k, 1} = f_k e^{i k x_1}.
\end{equation}
\section{A generalized Polaron transform}
\label{ap:PTgeneral}

Let us consider a more general form of Eq. \eqref{Hk} for the two-qubit case, i.e. $N_q=2$.
\begin{equation}
    H = \frac{\epsilon}{2} \sum_{j=1}^2 \sigma_j^x
+ \frac{\Delta}{2} \sum_{j=1}^2 \sigma_j^z + \sum_k \omega_k b^\dagger_k b_k + \sum_{j = 1}^2 \sigma_j^x \sum_k c_k\left( b_k^\dagger e^{ikx_j} + h.c. \right)\end{equation}
In order to acomodate the bias introduced to the qubits, we consider a variation of the Polaron transform presented in Eq. \eqref{UP}
\begin{equation}
\label{eq:UP_gen}
    U_P = \exp[-\sum_{j=1}^{2} \sigma_j^x \hat \alpha_j + \hat \beta_j],
\end{equation}
with, $\hat \alpha = \sum_k (f_k b_k^\dagger e^{i k x_j} - f_k^* b_k e^{-i k x_j})$ and $\hat \beta = \sum_k (l_k b_k^\dagger e^{i k x_j} - l_k^* b_k e^{-i k x_j})$. When the new variational parameters $\{l_k\}$ vanish, Eq. \eqref{eq:UP_gen} reduces to Eq. \eqref{UP}. In fact, provided there is no privileged direction of travel, so that $|f_k| = |f_{-k}|$ and $|l_k| = |l_{-k}|$, and the fact that the sine is odd, it can be seen that the transform factors as
\begin{equation}
    U_P = \bigotimes_{j = 1}^2 U_j \tilde U_j,
\end{equation}
with
\begin{align}
    &U_j = \exp[-\sigma_j^x \sum_k (f_k b_k^\dagger e^{i k x_j} - f_k^* b_k e^{-i k x_j})] \\
    &\tilde U_j = \exp[\sum_k (l_k b_k^\dagger e^{i k x_j} - l_k^* b_k e^{-i k x_j})]
\end{align}
Setting $x_1 - x_2 = x$, the minimization of the ground state energy using this new transform yields the following spin model
\begin{equation}
    \mathcal H_S = \frac{\Delta_r}{2} \left(\sigma_1^z + \sigma_2^z \right) + \frac{\epsilon'}{2} \left(\sigma_1^x + \sigma_2^x \right) - \mathcal J \sigma_1^x \sigma_2^x + 2 \sum_k f_k (w_k f_k - 2 c_k) + 2 \sum_k \omega_k l_k^2 (1 + \cos{kx}),
    \label{eq:HSgen}
\end{equation}
with,
\begin{equation}
    \epsilon' = \epsilon - 2 \sum_k l_k (c_k - \omega_k f_ k) (1 - \cos{kx}).
\end{equation}
This spin model is not exactly solvable, but performing perturbation theory on the term $\frac{\epsilon'}{2} \left(\sigma_1^x + \sigma_2^x \right)$, we find a ground state energy of the form
\begin{equation}
    \bar E_{GS} = \frac{1}{2} \left( - \mathcal J - \mathcal E - \sqrt{(\mathcal E - \mathcal J)^2 + 4 \eta^2}\right) + 2 \sum_k f_k (w_k f_k - 2 c_k) + 2 \sum_k \omega_k l_k^2 (1 + \cos{kx}),
\end{equation}
where
\begin{equation}
    \eta = \frac{\epsilon'}{\sqrt{2}} \frac{\Delta_r + \mathcal E + \mathcal J}{\sqrt{(\Delta_r + \mathcal E)^2 + \mathcal J^2}},
\end{equation}
which is minimum when
\begin{align}
    &l_k = \frac{\eta}{\sqrt{2} \sqrt{(\mathcal E - \mathcal J)^2 + 4 \eta^2}} \frac{\Delta_r + \mathcal E + \mathcal J}{\sqrt{(\Delta_r + \mathcal E)^2 + \mathcal J^2}} \frac{\omega_k f_k - c_k}{\omega_k}, \\
    &f_k = \frac{N_1 + N_2 + N_3 + N_4 + N_5}{ D_1 + D_2 + D_3 + D_4 + D_5 + D_6 + D_7 + D_8}.
\end{align}
We have used the following compact notation to trim the lengthy expression of $f_k$
\begin{align}
    &N_1 = 4 c_k + 2 c_k \cos{kx}(1 + \mathcal J / \mathcal E) \\
    &N_2 = \frac{2 c_k (1 + \cos{kx}) (\mathcal E + \mathcal J + \Delta_r)^2 \eta^2}{((\Delta_r + \mathcal E)^2 + \mathcal J^2)((\mathcal E - \mathcal J)^2 + 4 \eta^2)}, \\
    &N_3 = \frac{2 c_k \cos{kx} (\mathcal E - \mathcal J) (\mathcal J / \mathcal E - 1)}{\sqrt{(\mathcal E - \mathcal J)^2 + 4 \eta^2}}, \\
    &N_4 = \frac{8 c_k \cos{kx} (1 + \mathcal J / \mathcal E) \eta^2}{(\mathcal E + \mathcal J + \Delta_r) \sqrt{(\mathcal E - \mathcal J)^2 + 4 \eta^2}}, \\
    &N_5 = \frac{8 c_k \cos{kx} (2 + \Delta_r / \mathcal E) \eta^2 \mathcal J}{(\mathcal J^2 + (\mathcal E^2 + \Delta_r^2)^2) \sqrt{(\mathcal E - \mathcal J)^2 + 4 \eta^2}}, \\
    &D_1 = 4 \omega_k + 2 \omega_k \cos{kx}(1 + \mathcal J / \mathcal E) + 2 \Delta_r^2 / \mathcal E, \\
    &D_2 = \frac{2 \omega_k (1 + \cos{kx}) (\mathcal E + \mathcal J + \Delta_r)^2 \eta^2}{((\Delta_r + \mathcal E)^2 + \mathcal J^2)((\mathcal E - \mathcal J)^2 + 4 \eta^2)}, \\
    &D_3 = \frac{2 \omega_k \cos{kx} (\mathcal E - \mathcal J) (\mathcal J / \mathcal E - 1)}{\sqrt{(\mathcal E - \mathcal J)^2 + 4 \eta^2}}, \\
    &D_4 = \frac{2 (\mathcal E - \mathcal J) \Delta_r^2}{\mathcal E \sqrt{(\mathcal E - \mathcal J)^2 + 4 \eta^2}}, \\
    &D_5 = \frac{8 c_k \cos{kx} (1 + \mathcal J / \mathcal E) \eta^2}{(\mathcal E + \mathcal J + \Delta_r) \sqrt{(\mathcal E - \mathcal J)^2 + 4 \eta^2}}, \\
    &D_6 = \frac{8 \Delta_r \eta^2 (1 + \Delta_r / \mathcal E)}{(\mathcal E + \mathcal J + \Delta_r) \sqrt{(\mathcal E - \mathcal J)^2 + 4 \eta^2}}, \\
    &D_7 = \frac{8 \omega_k \cos{kx} (2 + \Delta_r / \mathcal E) \eta^2 \mathcal J}{(\mathcal J^2 + (\mathcal E^2 + \Delta_r^2)^2) \sqrt{(\mathcal E - \mathcal J)^2 + 4 \eta^2}}, \\
    &D_8 = \frac{8 \Delta_r \eta^2 (\mathcal E + \Delta_r)(1 + \Delta_r / \mathcal E)}{(\mathcal J^2 + (\mathcal E^2 + \Delta_r^2)^2) \sqrt{(\mathcal E - \mathcal J)^2 + 4 \eta^2}}.
\end{align}
As one can see, the calculations quickly become cumbersome when considering a biased model with the generalized Polaron transform. At the same time, we find (see Fig. \ref{fig:Egs_biased}) that the results in frequency renormalization and ground state energy do not deviate from those obtained with the standard transform in an unbiased model. That is why we have omitted this method in the main body of this paper. Nevertheless, it is important to notice that the introduction of a perturbative bias serves to lift the degeneracy between the ground state and the first excited state of the effective spin model [Eq. \eqref{eq:HSgen}] that arises when one goes beyond the USC regime into a scenario with full frequency renormalization, i.e. $\Delta_r \to 0$.
\begin{figure}[!h]
    \centering
    \includegraphics[width = 0.6 \textwidth]{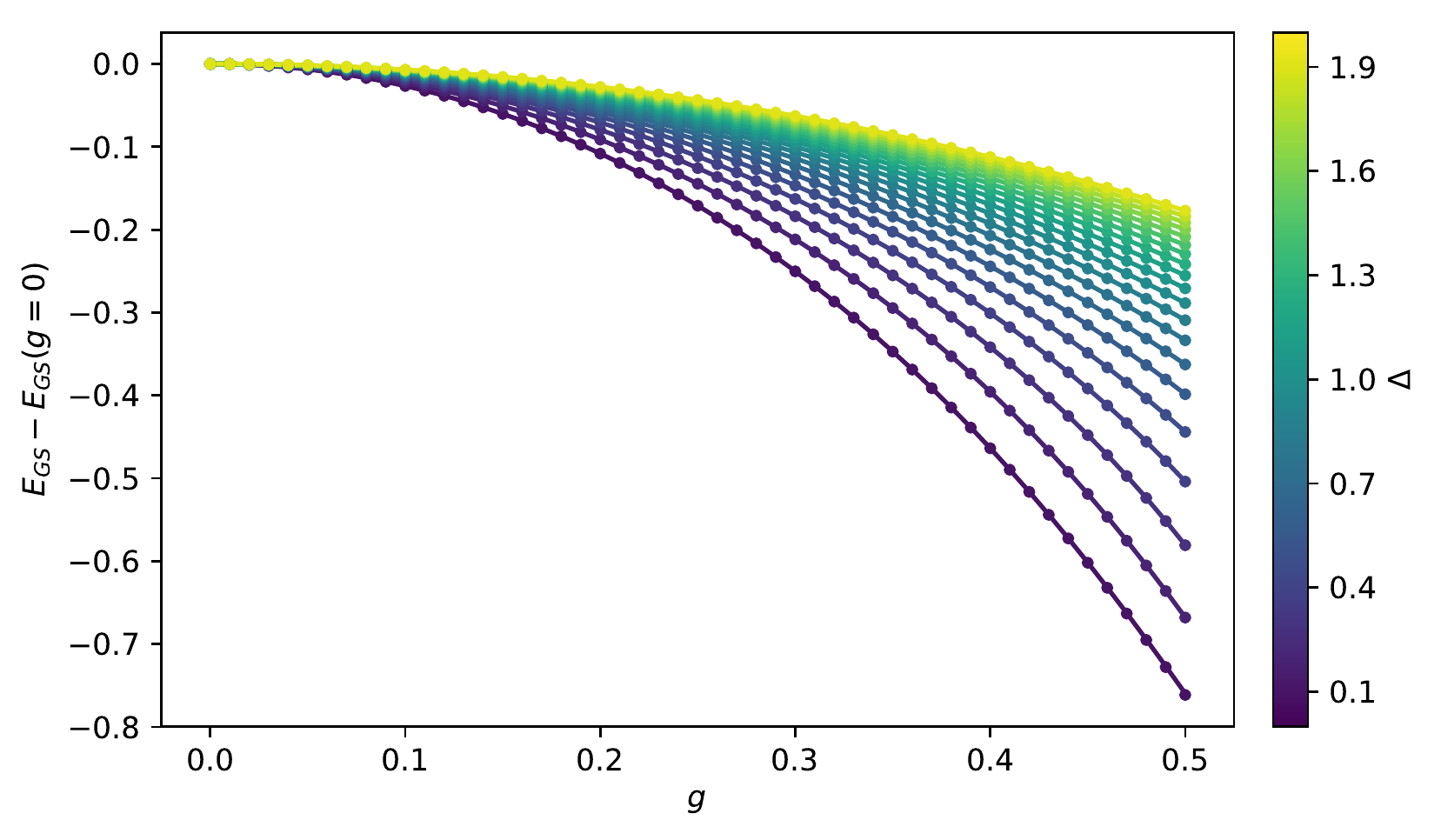}
    \caption{Dependence of the ground state energy with $g$ for several values of $\Delta$, for x = $5$ and $\epsilon = 10^{-4}$. Solid lines represent results for the biased model. For comparison, dotted lines represent results for the unbiased model.}
    \label{fig:Egs_biased}
\end{figure}
\section{Code}
\label{ap:code}

All numerical calculations  can be found \href{https://github.com/chuan97/TFG-Appendix-C}{https://github.com/chuan97/TFG-Appendix-C}.

\end{widetext}

\bibliography{Juan}

\begin{thebibliography}{86}%
\makeatletter
\providecommand \@ifxundefined [1]{%
 \@ifx{#1\undefined}
}%
\providecommand \@ifnum [1]{%
 \ifnum #1\expandafter \@firstoftwo
 \else \expandafter \@secondoftwo
 \fi
}%
\providecommand \@ifx [1]{%
 \ifx #1\expandafter \@firstoftwo
 \else \expandafter \@secondoftwo
 \fi
}%
\providecommand \natexlab [1]{#1}%
\providecommand \enquote  [1]{``#1''}%
\providecommand \bibnamefont  [1]{#1}%
\providecommand \bibfnamefont [1]{#1}%
\providecommand \citenamefont [1]{#1}%
\providecommand \href@noop [0]{\@secondoftwo}%
\providecommand \href [0]{\begingroup \@sanitize@url \@href}%
\providecommand \@href[1]{\@@startlink{#1}\@@href}%
\providecommand \@@href[1]{\endgroup#1\@@endlink}%
\providecommand \@sanitize@url [0]{\catcode `\\12\catcode `\$12\catcode
  `\&12\catcode `\#12\catcode `\^12\catcode `\_12\catcode `\%12\relax}%
\providecommand \@@startlink[1]{}%
\providecommand \@@endlink[0]{}%
\providecommand \url  [0]{\begingroup\@sanitize@url \@url }%
\providecommand \@url [1]{\endgroup\@href {#1}{\urlprefix }}%
\providecommand \urlprefix  [0]{URL }%
\providecommand \Eprint [0]{\href }%
\providecommand \doibase [0]{http://dx.doi.org/}%
\providecommand \selectlanguage [0]{\@gobble}%
\providecommand \bibinfo  [0]{\@secondoftwo}%
\providecommand \bibfield  [0]{\@secondoftwo}%
\providecommand \translation [1]{[#1]}%
\providecommand \BibitemOpen [0]{}%
\providecommand \bibitemStop [0]{}%
\providecommand \bibitemNoStop [0]{.\EOS\space}%
\providecommand \EOS [0]{\spacefactor3000\relax}%
\providecommand \BibitemShut  [1]{\csname bibitem#1\endcsname}%
\let\auto@bib@innerbib\@empty
\bibitem [{\citenamefont {Roy}\ \emph {et~al.}(2017)\citenamefont {Roy},
  \citenamefont {Wilson},\ and\ \citenamefont {Firstenberg}}]{roy2017}%
  \BibitemOpen
  \bibfield  {author} {\bibinfo {author} {\bibfnamefont {D.}~\bibnamefont
  {Roy}}, \bibinfo {author} {\bibfnamefont {C.~M.}\ \bibnamefont {Wilson}}, \
  and\ \bibinfo {author} {\bibfnamefont {O.}~\bibnamefont {Firstenberg}},\
  }\href@noop {} {\bibfield  {journal} {\bibinfo  {journal} {Reviews of Modern
  Physics}\ }\textbf {\bibinfo {volume} {89}},\ \bibinfo {pages} {021001}
  (\bibinfo {year} {2017})}\BibitemShut {NoStop}%
\bibitem [{\citenamefont {Gu}\ \emph {et~al.}(2017)\citenamefont {Gu},
  \citenamefont {Kockum}, \citenamefont {Miranowicz}, \citenamefont {Liu},\
  and\ \citenamefont {Nori}}]{gu2017}%
  \BibitemOpen
  \bibfield  {author} {\bibinfo {author} {\bibfnamefont {X.}~\bibnamefont
  {Gu}}, \bibinfo {author} {\bibfnamefont {A.~F.}\ \bibnamefont {Kockum}},
  \bibinfo {author} {\bibfnamefont {A.}~\bibnamefont {Miranowicz}}, \bibinfo
  {author} {\bibfnamefont {Y.-x.}\ \bibnamefont {Liu}}, \ and\ \bibinfo
  {author} {\bibfnamefont {F.}~\bibnamefont {Nori}},\ }\href@noop {} {\bibfield
   {journal} {\bibinfo  {journal} {Physics Reports}\ }\textbf {\bibinfo
  {volume} {718}},\ \bibinfo {pages} {1} (\bibinfo {year} {2017})}\BibitemShut
  {NoStop}%
\bibitem [{\citenamefont {Astafiev}\ \emph {et~al.}(2010)\citenamefont
  {Astafiev}, \citenamefont {Zagoskin}, \citenamefont {Abdumalikov},
  \citenamefont {Pashkin}, \citenamefont {Yamamoto}, \citenamefont {Inomata},
  \citenamefont {Nakamura},\ and\ \citenamefont {Tsai}}]{astafiev2010}%
  \BibitemOpen
  \bibfield  {author} {\bibinfo {author} {\bibfnamefont {O.}~\bibnamefont
  {Astafiev}}, \bibinfo {author} {\bibfnamefont {A.~M.}\ \bibnamefont
  {Zagoskin}}, \bibinfo {author} {\bibfnamefont {A.}~\bibnamefont
  {Abdumalikov}}, \bibinfo {author} {\bibfnamefont {Y.~A.}\ \bibnamefont
  {Pashkin}}, \bibinfo {author} {\bibfnamefont {T.}~\bibnamefont {Yamamoto}},
  \bibinfo {author} {\bibfnamefont {K.}~\bibnamefont {Inomata}}, \bibinfo
  {author} {\bibfnamefont {Y.}~\bibnamefont {Nakamura}}, \ and\ \bibinfo
  {author} {\bibfnamefont {J.}~\bibnamefont {Tsai}},\ }\href@noop {} {\bibfield
   {journal} {\bibinfo  {journal} {Science}\ }\textbf {\bibinfo {volume}
  {327}},\ \bibinfo {pages} {840} (\bibinfo {year} {2010})}\BibitemShut
  {NoStop}%
\bibitem [{\citenamefont {Van~Loo}\ \emph {et~al.}(2013)\citenamefont
  {Van~Loo}, \citenamefont {Fedorov}, \citenamefont {Lalumi{\`e}re},
  \citenamefont {Sanders}, \citenamefont {Blais},\ and\ \citenamefont
  {Wallraff}}]{van2013}%
  \BibitemOpen
  \bibfield  {author} {\bibinfo {author} {\bibfnamefont {A.~F.}\ \bibnamefont
  {Van~Loo}}, \bibinfo {author} {\bibfnamefont {A.}~\bibnamefont {Fedorov}},
  \bibinfo {author} {\bibfnamefont {K.}~\bibnamefont {Lalumi{\`e}re}}, \bibinfo
  {author} {\bibfnamefont {B.~C.}\ \bibnamefont {Sanders}}, \bibinfo {author}
  {\bibfnamefont {A.}~\bibnamefont {Blais}}, \ and\ \bibinfo {author}
  {\bibfnamefont {A.}~\bibnamefont {Wallraff}},\ }\href@noop {} {\bibfield
  {journal} {\bibinfo  {journal} {Science}\ }\textbf {\bibinfo {volume}
  {342}},\ \bibinfo {pages} {1494} (\bibinfo {year} {2013})}\BibitemShut
  {NoStop}%
\bibitem [{\citenamefont {Liu}\ and\ \citenamefont {Houck}(2017)}]{liu2017}%
  \BibitemOpen
  \bibfield  {author} {\bibinfo {author} {\bibfnamefont {Y.}~\bibnamefont
  {Liu}}\ and\ \bibinfo {author} {\bibfnamefont {A.~A.}\ \bibnamefont
  {Houck}},\ }\href@noop {} {\bibfield  {journal} {\bibinfo  {journal} {Nature
  Physics}\ }\textbf {\bibinfo {volume} {13}},\ \bibinfo {pages} {48} (\bibinfo
  {year} {2017})}\BibitemShut {NoStop}%
\bibitem [{\citenamefont {Faez}\ \emph {et~al.}(2014)\citenamefont {Faez},
  \citenamefont {T{\"u}rschmann}, \citenamefont {Haakh}, \citenamefont
  {G{\"o}tzinger},\ and\ \citenamefont {Sandoghdar}}]{faez2014}%
  \BibitemOpen
  \bibfield  {author} {\bibinfo {author} {\bibfnamefont {S.}~\bibnamefont
  {Faez}}, \bibinfo {author} {\bibfnamefont {P.}~\bibnamefont
  {T{\"u}rschmann}}, \bibinfo {author} {\bibfnamefont {H.~R.}\ \bibnamefont
  {Haakh}}, \bibinfo {author} {\bibfnamefont {S.}~\bibnamefont
  {G{\"o}tzinger}}, \ and\ \bibinfo {author} {\bibfnamefont {V.}~\bibnamefont
  {Sandoghdar}},\ }\href@noop {} {\bibfield  {journal} {\bibinfo  {journal}
  {Physical review letters}\ }\textbf {\bibinfo {volume} {113}},\ \bibinfo
  {pages} {213601} (\bibinfo {year} {2014})}\BibitemShut {NoStop}%
\bibitem [{\citenamefont {Lodahl}\ \emph {et~al.}(2015)\citenamefont {Lodahl},
  \citenamefont {Mahmoodian},\ and\ \citenamefont {Stobbe}}]{lodahl2015}%
  \BibitemOpen
  \bibfield  {author} {\bibinfo {author} {\bibfnamefont {P.}~\bibnamefont
  {Lodahl}}, \bibinfo {author} {\bibfnamefont {S.}~\bibnamefont {Mahmoodian}},
  \ and\ \bibinfo {author} {\bibfnamefont {S.}~\bibnamefont {Stobbe}},\
  }\href@noop {} {\bibfield  {journal} {\bibinfo  {journal} {Reviews of Modern
  Physics}\ }\textbf {\bibinfo {volume} {87}},\ \bibinfo {pages} {347}
  (\bibinfo {year} {2015})}\BibitemShut {NoStop}%
\bibitem [{\citenamefont {Chang}\ \emph {et~al.}(2018)\citenamefont {Chang},
  \citenamefont {Douglas}, \citenamefont {Gonz{\'a}lez-Tudela}, \citenamefont
  {Hung},\ and\ \citenamefont {Kimble}}]{chang2018}%
  \BibitemOpen
  \bibfield  {author} {\bibinfo {author} {\bibfnamefont {D.}~\bibnamefont
  {Chang}}, \bibinfo {author} {\bibfnamefont {J.}~\bibnamefont {Douglas}},
  \bibinfo {author} {\bibfnamefont {A.}~\bibnamefont {Gonz{\'a}lez-Tudela}},
  \bibinfo {author} {\bibfnamefont {C.-L.}\ \bibnamefont {Hung}}, \ and\
  \bibinfo {author} {\bibfnamefont {H.}~\bibnamefont {Kimble}},\ }\href@noop {}
  {\bibfield  {journal} {\bibinfo  {journal} {Reviews of Modern Physics}\
  }\textbf {\bibinfo {volume} {90}},\ \bibinfo {pages} {031002} (\bibinfo
  {year} {2018})}\BibitemShut {NoStop}%
\bibitem [{\citenamefont {Arg{\"u}ello-Luengo}\ \emph
  {et~al.}(2019)\citenamefont {Arg{\"u}ello-Luengo}, \citenamefont
  {Gonz{\'a}lez-Tudela}, \citenamefont {Shi}, \citenamefont {Zoller},\ and\
  \citenamefont {Cirac}}]{arguello2019}%
  \BibitemOpen
  \bibfield  {author} {\bibinfo {author} {\bibfnamefont {J.}~\bibnamefont
  {Arg{\"u}ello-Luengo}}, \bibinfo {author} {\bibfnamefont {A.}~\bibnamefont
  {Gonz{\'a}lez-Tudela}}, \bibinfo {author} {\bibfnamefont {T.}~\bibnamefont
  {Shi}}, \bibinfo {author} {\bibfnamefont {P.}~\bibnamefont {Zoller}}, \ and\
  \bibinfo {author} {\bibfnamefont {J.~I.}\ \bibnamefont {Cirac}},\ }\href@noop
  {} {\bibfield  {journal} {\bibinfo  {journal} {Nature}\ }\textbf {\bibinfo
  {volume} {574}},\ \bibinfo {pages} {215} (\bibinfo {year}
  {2019})}\BibitemShut {NoStop}%
\bibitem [{\citenamefont {Bello}\ \emph {et~al.}(2019)\citenamefont {Bello},
  \citenamefont {Platero}, \citenamefont {Cirac},\ and\ \citenamefont
  {Gonz{\'a}lez-Tudela}}]{bello2019}%
  \BibitemOpen
  \bibfield  {author} {\bibinfo {author} {\bibfnamefont {M.}~\bibnamefont
  {Bello}}, \bibinfo {author} {\bibfnamefont {G.}~\bibnamefont {Platero}},
  \bibinfo {author} {\bibfnamefont {J.~I.}\ \bibnamefont {Cirac}}, \ and\
  \bibinfo {author} {\bibfnamefont {A.}~\bibnamefont {Gonz{\'a}lez-Tudela}},\
  }\href@noop {} {\bibfield  {journal} {\bibinfo  {journal} {Science advances}\
  }\textbf {\bibinfo {volume} {5}},\ \bibinfo {pages} {eaaw0297} (\bibinfo
  {year} {2019})}\BibitemShut {NoStop}%
\bibitem [{\citenamefont {Lodahl}\ \emph {et~al.}(2017)\citenamefont {Lodahl},
  \citenamefont {Mahmoodian}, \citenamefont {Stobbe}, \citenamefont
  {Rauschenbeutel}, \citenamefont {Schneeweiss}, \citenamefont {Volz},
  \citenamefont {Pichler},\ and\ \citenamefont {Zoller}}]{lodahl2017}%
  \BibitemOpen
  \bibfield  {author} {\bibinfo {author} {\bibfnamefont {P.}~\bibnamefont
  {Lodahl}}, \bibinfo {author} {\bibfnamefont {S.}~\bibnamefont {Mahmoodian}},
  \bibinfo {author} {\bibfnamefont {S.}~\bibnamefont {Stobbe}}, \bibinfo
  {author} {\bibfnamefont {A.}~\bibnamefont {Rauschenbeutel}}, \bibinfo
  {author} {\bibfnamefont {P.}~\bibnamefont {Schneeweiss}}, \bibinfo {author}
  {\bibfnamefont {J.}~\bibnamefont {Volz}}, \bibinfo {author} {\bibfnamefont
  {H.}~\bibnamefont {Pichler}}, \ and\ \bibinfo {author} {\bibfnamefont
  {P.}~\bibnamefont {Zoller}},\ }\href@noop {} {\bibfield  {journal} {\bibinfo
  {journal} {Nature}\ }\textbf {\bibinfo {volume} {541}},\ \bibinfo {pages}
  {473} (\bibinfo {year} {2017})}\BibitemShut {NoStop}%
\bibitem [{\citenamefont {S{\'a}nchez-Burillo}\ \emph
  {et~al.}(2019{\natexlab{a}})\citenamefont {S{\'a}nchez-Burillo},
  \citenamefont {Wan}, \citenamefont {Zueco},\ and\ \citenamefont
  {Gonz{\'a}lez-Tudela}}]{sanchez2019b}%
  \BibitemOpen
  \bibfield  {author} {\bibinfo {author} {\bibfnamefont {E.}~\bibnamefont
  {S{\'a}nchez-Burillo}}, \bibinfo {author} {\bibfnamefont {C.}~\bibnamefont
  {Wan}}, \bibinfo {author} {\bibfnamefont {D.}~\bibnamefont {Zueco}}, \ and\
  \bibinfo {author} {\bibfnamefont {A.}~\bibnamefont {Gonz{\'a}lez-Tudela}},\
  }\href@noop {} {\bibfield  {journal} {\bibinfo  {journal} {arXiv preprint
  arXiv:1907.00840}\ } (\bibinfo {year} {2019}{\natexlab{a}})}\BibitemShut
  {NoStop}%
\bibitem [{\citenamefont {Zheng}\ \emph {et~al.}(2013)\citenamefont {Zheng},
  \citenamefont {Gauthier},\ and\ \citenamefont {Baranger}}]{zheng2013}%
  \BibitemOpen
  \bibfield  {author} {\bibinfo {author} {\bibfnamefont {H.}~\bibnamefont
  {Zheng}}, \bibinfo {author} {\bibfnamefont {D.~J.}\ \bibnamefont {Gauthier}},
  \ and\ \bibinfo {author} {\bibfnamefont {H.~U.}\ \bibnamefont {Baranger}},\
  }\href@noop {} {\bibfield  {journal} {\bibinfo  {journal} {Physical review
  letters}\ }\textbf {\bibinfo {volume} {111}},\ \bibinfo {pages} {090502}
  (\bibinfo {year} {2013})}\BibitemShut {NoStop}%
\bibitem [{\citenamefont {{Niemczyk}}\ \emph {et~al.}(2010)\citenamefont
  {{Niemczyk}}, \citenamefont {{Deppe}}, \citenamefont {{Huebl}}, \citenamefont
  {{Menzel}}, \citenamefont {{Hocke}}, \citenamefont {{Schwarz}}, \citenamefont
  {{Garcia-Ripoll}}, \citenamefont {{Zueco}}, \citenamefont {{H{\"u}mmer}},
  \citenamefont {{Solano}}, \citenamefont {{Marx}},\ and\ \citenamefont
  {{Gross}}}]{Niemczyk2010}%
  \BibitemOpen
  \bibfield  {author} {\bibinfo {author} {\bibfnamefont {T.}~\bibnamefont
  {{Niemczyk}}}, \bibinfo {author} {\bibfnamefont {F.}~\bibnamefont {{Deppe}}},
  \bibinfo {author} {\bibfnamefont {H.}~\bibnamefont {{Huebl}}}, \bibinfo
  {author} {\bibfnamefont {E.~P.}\ \bibnamefont {{Menzel}}}, \bibinfo {author}
  {\bibfnamefont {F.}~\bibnamefont {{Hocke}}}, \bibinfo {author} {\bibfnamefont
  {M.~J.}\ \bibnamefont {{Schwarz}}}, \bibinfo {author} {\bibfnamefont {J.~J.}\
  \bibnamefont {{Garcia-Ripoll}}}, \bibinfo {author} {\bibfnamefont
  {D.}~\bibnamefont {{Zueco}}}, \bibinfo {author} {\bibfnamefont
  {T.}~\bibnamefont {{H{\"u}mmer}}}, \bibinfo {author} {\bibfnamefont
  {E.}~\bibnamefont {{Solano}}}, \bibinfo {author} {\bibfnamefont
  {A.}~\bibnamefont {{Marx}}}, \ and\ \bibinfo {author} {\bibfnamefont
  {R.}~\bibnamefont {{Gross}}},\ }\href {\doibase 10.1038/nphys1730} {\bibfield
   {journal} {\bibinfo  {journal} {Nature Physics}\ }\textbf {\bibinfo {volume}
  {6}},\ \bibinfo {pages} {772} (\bibinfo {year} {2010})}\BibitemShut {NoStop}%
\bibitem [{\citenamefont {Forn-D{\'\i}az}\ \emph {et~al.}(2010)\citenamefont
  {Forn-D{\'\i}az}, \citenamefont {Lisenfeld}, \citenamefont {Marcos},
  \citenamefont {Garcia-Ripoll}, \citenamefont {Solano}, \citenamefont
  {Harmans},\ and\ \citenamefont {Mooij}}]{forn2010}%
  \BibitemOpen
  \bibfield  {author} {\bibinfo {author} {\bibfnamefont {P.}~\bibnamefont
  {Forn-D{\'\i}az}}, \bibinfo {author} {\bibfnamefont {J.}~\bibnamefont
  {Lisenfeld}}, \bibinfo {author} {\bibfnamefont {D.}~\bibnamefont {Marcos}},
  \bibinfo {author} {\bibfnamefont {J.~J.}\ \bibnamefont {Garcia-Ripoll}},
  \bibinfo {author} {\bibfnamefont {E.}~\bibnamefont {Solano}}, \bibinfo
  {author} {\bibfnamefont {C.}~\bibnamefont {Harmans}}, \ and\ \bibinfo
  {author} {\bibfnamefont {J.}~\bibnamefont {Mooij}},\ }\href@noop {}
  {\bibfield  {journal} {\bibinfo  {journal} {Physical review letters}\
  }\textbf {\bibinfo {volume} {105}},\ \bibinfo {pages} {237001} (\bibinfo
  {year} {2010})}\BibitemShut {NoStop}%
\bibitem [{\citenamefont {Forn-D{\'\i}az}\ \emph {et~al.}(2017)\citenamefont
  {Forn-D{\'\i}az}, \citenamefont {Garc{\'\i}a-Ripoll}, \citenamefont
  {Peropadre}, \citenamefont {Orgiazzi}, \citenamefont {Yurtalan},
  \citenamefont {Belyansky}, \citenamefont {Wilson},\ and\ \citenamefont
  {Lupascu}}]{forn2017}%
  \BibitemOpen
  \bibfield  {author} {\bibinfo {author} {\bibfnamefont {P.}~\bibnamefont
  {Forn-D{\'\i}az}}, \bibinfo {author} {\bibfnamefont {J.~J.}\ \bibnamefont
  {Garc{\'\i}a-Ripoll}}, \bibinfo {author} {\bibfnamefont {B.}~\bibnamefont
  {Peropadre}}, \bibinfo {author} {\bibfnamefont {J.-L.}\ \bibnamefont
  {Orgiazzi}}, \bibinfo {author} {\bibfnamefont {M.}~\bibnamefont {Yurtalan}},
  \bibinfo {author} {\bibfnamefont {R.}~\bibnamefont {Belyansky}}, \bibinfo
  {author} {\bibfnamefont {C.~M.}\ \bibnamefont {Wilson}}, \ and\ \bibinfo
  {author} {\bibfnamefont {A.}~\bibnamefont {Lupascu}},\ }\href@noop {}
  {\bibfield  {journal} {\bibinfo  {journal} {Nature Physics}\ }\textbf
  {\bibinfo {volume} {13}},\ \bibinfo {pages} {39} (\bibinfo {year}
  {2017})}\BibitemShut {NoStop}%
\bibitem [{\citenamefont {Mart{\'\i}nez}\ \emph {et~al.}(2019)\citenamefont
  {Mart{\'\i}nez}, \citenamefont {L{\'e}ger}, \citenamefont {Gheeraert},
  \citenamefont {Dassonneville}, \citenamefont {Planat}, \citenamefont
  {Foroughi}, \citenamefont {Krupko}, \citenamefont {Buisson}, \citenamefont
  {Naud}, \citenamefont {Hasch-Guichard} \emph {et~al.}}]{martinez2019}%
  \BibitemOpen
  \bibfield  {author} {\bibinfo {author} {\bibfnamefont {J.~P.}\ \bibnamefont
  {Mart{\'\i}nez}}, \bibinfo {author} {\bibfnamefont {S.}~\bibnamefont
  {L{\'e}ger}}, \bibinfo {author} {\bibfnamefont {N.}~\bibnamefont
  {Gheeraert}}, \bibinfo {author} {\bibfnamefont {R.}~\bibnamefont
  {Dassonneville}}, \bibinfo {author} {\bibfnamefont {L.}~\bibnamefont
  {Planat}}, \bibinfo {author} {\bibfnamefont {F.}~\bibnamefont {Foroughi}},
  \bibinfo {author} {\bibfnamefont {Y.}~\bibnamefont {Krupko}}, \bibinfo
  {author} {\bibfnamefont {O.}~\bibnamefont {Buisson}}, \bibinfo {author}
  {\bibfnamefont {C.}~\bibnamefont {Naud}}, \bibinfo {author} {\bibfnamefont
  {W.}~\bibnamefont {Hasch-Guichard}},  \emph {et~al.},\ }\href@noop {}
  {\bibfield  {journal} {\bibinfo  {journal} {npj Quantum Information}\
  }\textbf {\bibinfo {volume} {5}},\ \bibinfo {pages} {19} (\bibinfo {year}
  {2019})}\BibitemShut {NoStop}%
\bibitem [{\citenamefont {Leger}\ \emph {et~al.}(2019)\citenamefont {Leger},
  \citenamefont {Puertas-Martinez}, \citenamefont {Bharadwaj}, \citenamefont
  {Dassonneville}, \citenamefont {Delaforce}, \citenamefont {Foroughi},
  \citenamefont {Milchakov}, \citenamefont {Planat}, \citenamefont {Buisson},
  \citenamefont {Naud}, \citenamefont {Hasch-Guichard}, \citenamefont
  {Florens}, \citenamefont {Snyman},\ and\ \citenamefont {Roch}}]{leger2019}%
  \BibitemOpen
  \bibfield  {author} {\bibinfo {author} {\bibfnamefont {S.}~\bibnamefont
  {Leger}}, \bibinfo {author} {\bibfnamefont {J.}~\bibnamefont
  {Puertas-Martinez}}, \bibinfo {author} {\bibfnamefont {K.}~\bibnamefont
  {Bharadwaj}}, \bibinfo {author} {\bibfnamefont {R.}~\bibnamefont
  {Dassonneville}}, \bibinfo {author} {\bibfnamefont {J.}~\bibnamefont
  {Delaforce}}, \bibinfo {author} {\bibfnamefont {F.}~\bibnamefont {Foroughi}},
  \bibinfo {author} {\bibfnamefont {V.}~\bibnamefont {Milchakov}}, \bibinfo
  {author} {\bibfnamefont {L.}~\bibnamefont {Planat}}, \bibinfo {author}
  {\bibfnamefont {O.}~\bibnamefont {Buisson}}, \bibinfo {author} {\bibfnamefont
  {C.}~\bibnamefont {Naud}}, \bibinfo {author} {\bibfnamefont {W.}~\bibnamefont
  {Hasch-Guichard}}, \bibinfo {author} {\bibfnamefont {S.}~\bibnamefont
  {Florens}}, \bibinfo {author} {\bibfnamefont {I.}~\bibnamefont {Snyman}}, \
  and\ \bibinfo {author} {\bibfnamefont {N.}~\bibnamefont {Roch}},\ }\href@noop
  {} {\bibfield  {journal} {\bibinfo  {journal} {arXiv preprint
  arXiv:1910.08340}\ } (\bibinfo {year} {2019})}\BibitemShut {NoStop}%
\bibitem [{\citenamefont {Shirley}(1965)}]{shirley1965}%
  \BibitemOpen
  \bibfield  {author} {\bibinfo {author} {\bibfnamefont {J.~H.}\ \bibnamefont
  {Shirley}},\ }\href@noop {} {\bibfield  {journal} {\bibinfo  {journal}
  {Physical Review}\ }\textbf {\bibinfo {volume} {138}},\ \bibinfo {pages}
  {B979} (\bibinfo {year} {1965})}\BibitemShut {NoStop}%
\bibitem [{\citenamefont {Leggett}\ \emph {et~al.}(1987)\citenamefont
  {Leggett}, \citenamefont {Chakravarty}, \citenamefont {Dorsey}, \citenamefont
  {Fisher}, \citenamefont {Garg},\ and\ \citenamefont {Zwerger}}]{Leggett1987}%
  \BibitemOpen
  \bibfield  {author} {\bibinfo {author} {\bibfnamefont {A.~J.}\ \bibnamefont
  {Leggett}}, \bibinfo {author} {\bibfnamefont {S.}~\bibnamefont
  {Chakravarty}}, \bibinfo {author} {\bibfnamefont {A.~T.}\ \bibnamefont
  {Dorsey}}, \bibinfo {author} {\bibfnamefont {M.~P.~A.}\ \bibnamefont
  {Fisher}}, \bibinfo {author} {\bibfnamefont {A.}~\bibnamefont {Garg}}, \ and\
  \bibinfo {author} {\bibfnamefont {W.}~\bibnamefont {Zwerger}},\ }\href
  {\doibase 10.1103/revmodphys.59.1} {\bibfield  {journal} {\bibinfo  {journal}
  {Reviews of Modern Physics}\ }\textbf {\bibinfo {volume} {59}},\ \bibinfo
  {pages} {1} (\bibinfo {year} {1987})}\BibitemShut {NoStop}%
\bibitem [{\citenamefont {Ashhab}\ and\ \citenamefont
  {Nori}(2010)}]{Ashhab2010}%
  \BibitemOpen
  \bibfield  {author} {\bibinfo {author} {\bibfnamefont {S.}~\bibnamefont
  {Ashhab}}\ and\ \bibinfo {author} {\bibfnamefont {F.}~\bibnamefont {Nori}},\
  }\href {\doibase 10.1103/physreva.81.042311} {\bibfield  {journal} {\bibinfo
  {journal} {Physical Review A}\ }\textbf {\bibinfo {volume} {81}} (\bibinfo
  {year} {2010}),\ 10.1103/physreva.81.042311}\BibitemShut {NoStop}%
\bibitem [{\citenamefont {Ciuti}\ \emph {et~al.}(2005)\citenamefont {Ciuti},
  \citenamefont {Bastard},\ and\ \citenamefont {Carusotto}}]{Ciuti2005}%
  \BibitemOpen
  \bibfield  {author} {\bibinfo {author} {\bibfnamefont {C.}~\bibnamefont
  {Ciuti}}, \bibinfo {author} {\bibfnamefont {G.}~\bibnamefont {Bastard}}, \
  and\ \bibinfo {author} {\bibfnamefont {I.}~\bibnamefont {Carusotto}},\ }\href
  {\doibase 10.1103/PhysRevB.72.115303} {\bibfield  {journal} {\bibinfo
  {journal} {Phys. Rev. B}\ }\textbf {\bibinfo {volume} {72}},\ \bibinfo
  {pages} {115303} (\bibinfo {year} {2005})}\BibitemShut {NoStop}%
\bibitem [{\citenamefont {Stassi}\ \emph {et~al.}(2013)\citenamefont {Stassi},
  \citenamefont {Ridolfo}, \citenamefont {Stefano}, \citenamefont {Hartmann},\
  and\ \citenamefont {Savasta}}]{Stassi2013}%
  \BibitemOpen
  \bibfield  {author} {\bibinfo {author} {\bibfnamefont {R.}~\bibnamefont
  {Stassi}}, \bibinfo {author} {\bibfnamefont {A.}~\bibnamefont {Ridolfo}},
  \bibinfo {author} {\bibfnamefont {O.~D.}\ \bibnamefont {Stefano}}, \bibinfo
  {author} {\bibfnamefont {M.~J.}\ \bibnamefont {Hartmann}}, \ and\ \bibinfo
  {author} {\bibfnamefont {S.}~\bibnamefont {Savasta}},\ }\href {\doibase
  10.1103/physrevlett.110.243601} {\bibfield  {journal} {\bibinfo  {journal}
  {Physical Review Letters}\ }\textbf {\bibinfo {volume} {110}} (\bibinfo
  {year} {2013}),\ 10.1103/physrevlett.110.243601}\BibitemShut {NoStop}%
\bibitem [{\citenamefont {He}\ \emph {et~al.}(2018)\citenamefont {He},
  \citenamefont {An}, \citenamefont {Song},\ and\ \citenamefont
  {Zhou}}]{QiKai2018}%
  \BibitemOpen
  \bibfield  {author} {\bibinfo {author} {\bibfnamefont {Q.-K.}\ \bibnamefont
  {He}}, \bibinfo {author} {\bibfnamefont {Z.}~\bibnamefont {An}}, \bibinfo
  {author} {\bibfnamefont {H.-J.}\ \bibnamefont {Song}}, \ and\ \bibinfo
  {author} {\bibfnamefont {D.~L.}\ \bibnamefont {Zhou}},\ }\href@noop {}
  {\bibfield  {journal} {\bibinfo  {journal} {ArXiv e-prints}\ } (\bibinfo
  {year} {2018})},\ \Eprint {http://arxiv.org/abs/1810.04523v1}
  {arXiv:1810.04523v1} \BibitemShut {NoStop}%
\bibitem [{\citenamefont {Liberato}(2017)}]{DeLiberato2017}%
  \BibitemOpen
  \bibfield  {author} {\bibinfo {author} {\bibfnamefont {S.~D.}\ \bibnamefont
  {Liberato}},\ }\href {\doibase 10.1038/s41467-017-01504-5} {\bibfield
  {journal} {\bibinfo  {journal} {Nature Communications}\ }\textbf {\bibinfo
  {volume} {8}} (\bibinfo {year} {2017}),\
  10.1038/s41467-017-01504-5}\BibitemShut {NoStop}%
\bibitem [{\citenamefont {Stassi}\ \emph {et~al.}(2017)\citenamefont {Stassi},
  \citenamefont {Macr{\`{\i}}}, \citenamefont {Kockum}, \citenamefont
  {Stefano}, \citenamefont {Miranowicz}, \citenamefont {Savasta},\ and\
  \citenamefont {Nori}}]{Stassi2017}%
  \BibitemOpen
  \bibfield  {author} {\bibinfo {author} {\bibfnamefont {R.}~\bibnamefont
  {Stassi}}, \bibinfo {author} {\bibfnamefont {V.}~\bibnamefont
  {Macr{\`{\i}}}}, \bibinfo {author} {\bibfnamefont {A.~F.}\ \bibnamefont
  {Kockum}}, \bibinfo {author} {\bibfnamefont {O.~D.}\ \bibnamefont {Stefano}},
  \bibinfo {author} {\bibfnamefont {A.}~\bibnamefont {Miranowicz}}, \bibinfo
  {author} {\bibfnamefont {S.}~\bibnamefont {Savasta}}, \ and\ \bibinfo
  {author} {\bibfnamefont {F.}~\bibnamefont {Nori}},\ }\href {\doibase
  10.1103/physreva.96.023818} {\bibfield  {journal} {\bibinfo  {journal}
  {Physical Review A}\ }\textbf {\bibinfo {volume} {96}} (\bibinfo {year}
  {2017}),\ 10.1103/physreva.96.023818}\BibitemShut {NoStop}%
\bibitem [{\citenamefont {Kockum}\ \emph {et~al.}(2019)\citenamefont {Kockum},
  \citenamefont {Miranowicz}, \citenamefont {De~Liberato}, \citenamefont
  {Savasta},\ and\ \citenamefont {Nori}}]{kockum2019}%
  \BibitemOpen
  \bibfield  {author} {\bibinfo {author} {\bibfnamefont {A.~F.}\ \bibnamefont
  {Kockum}}, \bibinfo {author} {\bibfnamefont {A.}~\bibnamefont {Miranowicz}},
  \bibinfo {author} {\bibfnamefont {S.}~\bibnamefont {De~Liberato}}, \bibinfo
  {author} {\bibfnamefont {S.}~\bibnamefont {Savasta}}, \ and\ \bibinfo
  {author} {\bibfnamefont {F.}~\bibnamefont {Nori}},\ }\href@noop {} {\bibfield
   {journal} {\bibinfo  {journal} {Nature Reviews Physics}\ }\textbf {\bibinfo
  {volume} {1}},\ \bibinfo {pages} {19} (\bibinfo {year} {2019})}\BibitemShut
  {NoStop}%
\bibitem [{\citenamefont {Forn-D{\'\i}az}\ \emph {et~al.}(2019)\citenamefont
  {Forn-D{\'\i}az}, \citenamefont {Lamata}, \citenamefont {Rico}, \citenamefont
  {Kono},\ and\ \citenamefont {Solano}}]{forn2019}%
  \BibitemOpen
  \bibfield  {author} {\bibinfo {author} {\bibfnamefont {P.}~\bibnamefont
  {Forn-D{\'\i}az}}, \bibinfo {author} {\bibfnamefont {L.}~\bibnamefont
  {Lamata}}, \bibinfo {author} {\bibfnamefont {E.}~\bibnamefont {Rico}},
  \bibinfo {author} {\bibfnamefont {J.}~\bibnamefont {Kono}}, \ and\ \bibinfo
  {author} {\bibfnamefont {E.}~\bibnamefont {Solano}},\ }\href@noop {}
  {\bibfield  {journal} {\bibinfo  {journal} {Reviews of Modern Physics}\
  }\textbf {\bibinfo {volume} {91}},\ \bibinfo {pages} {025005} (\bibinfo
  {year} {2019})}\BibitemShut {NoStop}%
\bibitem [{\citenamefont {Peropadre}\ \emph {et~al.}(2013)\citenamefont
  {Peropadre}, \citenamefont {Zueco}, \citenamefont {Porras},\ and\
  \citenamefont {Garc{\'\i}a-Ripoll}}]{peropadre2013}%
  \BibitemOpen
  \bibfield  {author} {\bibinfo {author} {\bibfnamefont {B.}~\bibnamefont
  {Peropadre}}, \bibinfo {author} {\bibfnamefont {D.}~\bibnamefont {Zueco}},
  \bibinfo {author} {\bibfnamefont {D.}~\bibnamefont {Porras}}, \ and\ \bibinfo
  {author} {\bibfnamefont {J.~J.}\ \bibnamefont {Garc{\'\i}a-Ripoll}},\
  }\href@noop {} {\bibfield  {journal} {\bibinfo  {journal} {Physical review
  letters}\ }\textbf {\bibinfo {volume} {111}},\ \bibinfo {pages} {243602}
  (\bibinfo {year} {2013})}\BibitemShut {NoStop}%
\bibitem [{\citenamefont {Shi}\ \emph {et~al.}(2018{\natexlab{a}})\citenamefont
  {Shi}, \citenamefont {Chang},\ and\ \citenamefont
  {Garc{\'{\i}}a-Ripoll}}]{Shi2018}%
  \BibitemOpen
  \bibfield  {author} {\bibinfo {author} {\bibfnamefont {T.}~\bibnamefont
  {Shi}}, \bibinfo {author} {\bibfnamefont {Y.}~\bibnamefont {Chang}}, \ and\
  \bibinfo {author} {\bibfnamefont {J.~J.}\ \bibnamefont
  {Garc{\'{\i}}a-Ripoll}},\ }\href {\doibase 10.1103/physrevlett.120.153602}
  {\bibfield  {journal} {\bibinfo  {journal} {Physical Review Letters}\
  }\textbf {\bibinfo {volume} {120}} (\bibinfo {year} {2018}{\natexlab{a}}),\
  10.1103/physrevlett.120.153602}\BibitemShut {NoStop}%
\bibitem [{\citenamefont {Gheeraert}\ \emph {et~al.}(2018)\citenamefont
  {Gheeraert}, \citenamefont {Zhang}, \citenamefont {S{\'e}pulcre},
  \citenamefont {Bera}, \citenamefont {Roch}, \citenamefont {Baranger},\ and\
  \citenamefont {Florens}}]{gheeraert2018}%
  \BibitemOpen
  \bibfield  {author} {\bibinfo {author} {\bibfnamefont {N.}~\bibnamefont
  {Gheeraert}}, \bibinfo {author} {\bibfnamefont {X.~H.}\ \bibnamefont
  {Zhang}}, \bibinfo {author} {\bibfnamefont {T.}~\bibnamefont {S{\'e}pulcre}},
  \bibinfo {author} {\bibfnamefont {S.}~\bibnamefont {Bera}}, \bibinfo {author}
  {\bibfnamefont {N.}~\bibnamefont {Roch}}, \bibinfo {author} {\bibfnamefont
  {H.~U.}\ \bibnamefont {Baranger}}, \ and\ \bibinfo {author} {\bibfnamefont
  {S.}~\bibnamefont {Florens}},\ }\href@noop {} {\bibfield  {journal} {\bibinfo
   {journal} {Physical Review A}\ }\textbf {\bibinfo {volume} {98}},\ \bibinfo
  {pages} {043816} (\bibinfo {year} {2018})}\BibitemShut {NoStop}%
\bibitem [{\citenamefont {S{\'a}nchez-Burillo}\ \emph
  {et~al.}(2014)\citenamefont {S{\'a}nchez-Burillo}, \citenamefont {Zueco},
  \citenamefont {Garcia-Ripoll},\ and\ \citenamefont
  {Martin-Moreno}}]{sanchez2014}%
  \BibitemOpen
  \bibfield  {author} {\bibinfo {author} {\bibfnamefont {E.}~\bibnamefont
  {S{\'a}nchez-Burillo}}, \bibinfo {author} {\bibfnamefont {D.}~\bibnamefont
  {Zueco}}, \bibinfo {author} {\bibfnamefont {J.}~\bibnamefont
  {Garcia-Ripoll}}, \ and\ \bibinfo {author} {\bibfnamefont {L.}~\bibnamefont
  {Martin-Moreno}},\ }\href@noop {} {\bibfield  {journal} {\bibinfo  {journal}
  {Physical review letters}\ }\textbf {\bibinfo {volume} {113}},\ \bibinfo
  {pages} {263604} (\bibinfo {year} {2014})}\BibitemShut {NoStop}%
\bibitem [{\citenamefont {S{\'a}nchez-Burillo}\ \emph
  {et~al.}(2015)\citenamefont {S{\'a}nchez-Burillo}, \citenamefont
  {Garc{\'\i}a-Ripoll}, \citenamefont {Mart{\'\i}n-Moreno},\ and\ \citenamefont
  {Zueco}}]{sanchez2015}%
  \BibitemOpen
  \bibfield  {author} {\bibinfo {author} {\bibfnamefont {E.}~\bibnamefont
  {S{\'a}nchez-Burillo}}, \bibinfo {author} {\bibfnamefont {J.}~\bibnamefont
  {Garc{\'\i}a-Ripoll}}, \bibinfo {author} {\bibfnamefont {L.}~\bibnamefont
  {Mart{\'\i}n-Moreno}}, \ and\ \bibinfo {author} {\bibfnamefont
  {D.}~\bibnamefont {Zueco}},\ }\href@noop {} {\bibfield  {journal} {\bibinfo
  {journal} {Faraday discussions}\ }\textbf {\bibinfo {volume} {178}},\
  \bibinfo {pages} {335} (\bibinfo {year} {2015})}\BibitemShut {NoStop}%
\bibitem [{\citenamefont {S{\'a}nchez-Burillo}\ \emph
  {et~al.}(2019{\natexlab{b}})\citenamefont {S{\'a}nchez-Burillo},
  \citenamefont {Mart{\'\i}n-Moreno}, \citenamefont {Garc{\'\i}a-Ripoll},\ and\
  \citenamefont {Zueco}}]{sanchez2019}%
  \BibitemOpen
  \bibfield  {author} {\bibinfo {author} {\bibfnamefont {E.}~\bibnamefont
  {S{\'a}nchez-Burillo}}, \bibinfo {author} {\bibfnamefont {L.}~\bibnamefont
  {Mart{\'\i}n-Moreno}}, \bibinfo {author} {\bibfnamefont {J.}~\bibnamefont
  {Garc{\'\i}a-Ripoll}}, \ and\ \bibinfo {author} {\bibfnamefont
  {D.}~\bibnamefont {Zueco}},\ }\href@noop {} {\bibfield  {journal} {\bibinfo
  {journal} {Physical review letters}\ }\textbf {\bibinfo {volume} {123}},\
  \bibinfo {pages} {013601} (\bibinfo {year} {2019}{\natexlab{b}})}\BibitemShut
  {NoStop}%
\bibitem [{\citenamefont {Dzsotjan}\ \emph {et~al.}(2011)\citenamefont
  {Dzsotjan}, \citenamefont {K{\"a}stel},\ and\ \citenamefont
  {Fleischhauer}}]{dzsotjan2011}%
  \BibitemOpen
  \bibfield  {author} {\bibinfo {author} {\bibfnamefont {D.}~\bibnamefont
  {Dzsotjan}}, \bibinfo {author} {\bibfnamefont {J.}~\bibnamefont
  {K{\"a}stel}}, \ and\ \bibinfo {author} {\bibfnamefont {M.}~\bibnamefont
  {Fleischhauer}},\ }\href@noop {} {\bibfield  {journal} {\bibinfo  {journal}
  {Physical Review B}\ }\textbf {\bibinfo {volume} {84}},\ \bibinfo {pages}
  {075419} (\bibinfo {year} {2011})}\BibitemShut {NoStop}%
\bibitem [{\citenamefont {Gonzalez-Tudela}\ \emph {et~al.}(2011)\citenamefont
  {Gonzalez-Tudela}, \citenamefont {Martin-Cano}, \citenamefont {Moreno},
  \citenamefont {Martin-Moreno}, \citenamefont {Tejedor},\ and\ \citenamefont
  {Garcia-Vidal}}]{gonzalez2011}%
  \BibitemOpen
  \bibfield  {author} {\bibinfo {author} {\bibfnamefont {A.}~\bibnamefont
  {Gonzalez-Tudela}}, \bibinfo {author} {\bibfnamefont {D.}~\bibnamefont
  {Martin-Cano}}, \bibinfo {author} {\bibfnamefont {E.}~\bibnamefont {Moreno}},
  \bibinfo {author} {\bibfnamefont {L.}~\bibnamefont {Martin-Moreno}}, \bibinfo
  {author} {\bibfnamefont {C.}~\bibnamefont {Tejedor}}, \ and\ \bibinfo
  {author} {\bibfnamefont {F.~J.}\ \bibnamefont {Garcia-Vidal}},\ }\href@noop
  {} {\bibfield  {journal} {\bibinfo  {journal} {Physical review letters}\
  }\textbf {\bibinfo {volume} {106}},\ \bibinfo {pages} {020501} (\bibinfo
  {year} {2011})}\BibitemShut {NoStop}%
\bibitem [{\citenamefont {Zueco}\ \emph {et~al.}(2012)\citenamefont {Zueco},
  \citenamefont {Mazo}, \citenamefont {Solano},\ and\ \citenamefont
  {Garc{\'\i}a-Ripoll}}]{zueco2012}%
  \BibitemOpen
  \bibfield  {author} {\bibinfo {author} {\bibfnamefont {D.}~\bibnamefont
  {Zueco}}, \bibinfo {author} {\bibfnamefont {J.~J.}\ \bibnamefont {Mazo}},
  \bibinfo {author} {\bibfnamefont {E.}~\bibnamefont {Solano}}, \ and\ \bibinfo
  {author} {\bibfnamefont {J.~J.}\ \bibnamefont {Garc{\'\i}a-Ripoll}},\
  }\href@noop {} {\bibfield  {journal} {\bibinfo  {journal} {Physical Review
  B}\ }\textbf {\bibinfo {volume} {86}},\ \bibinfo {pages} {024503} (\bibinfo
  {year} {2012})}\BibitemShut {NoStop}%
\bibitem [{\citenamefont {Zheng}\ and\ \citenamefont
  {Baranger}(2013)}]{zheng2013b}%
  \BibitemOpen
  \bibfield  {author} {\bibinfo {author} {\bibfnamefont {H.}~\bibnamefont
  {Zheng}}\ and\ \bibinfo {author} {\bibfnamefont {H.~U.}\ \bibnamefont
  {Baranger}},\ }\href@noop {} {\bibfield  {journal} {\bibinfo  {journal}
  {Physical review letters}\ }\textbf {\bibinfo {volume} {110}},\ \bibinfo
  {pages} {113601} (\bibinfo {year} {2013})}\BibitemShut {NoStop}%
\bibitem [{\citenamefont {Manzoni}\ \emph {et~al.}(2017)\citenamefont
  {Manzoni}, \citenamefont {Mathey},\ and\ \citenamefont
  {Chang}}]{manzoni2017}%
  \BibitemOpen
  \bibfield  {author} {\bibinfo {author} {\bibfnamefont {M.~T.}\ \bibnamefont
  {Manzoni}}, \bibinfo {author} {\bibfnamefont {L.}~\bibnamefont {Mathey}}, \
  and\ \bibinfo {author} {\bibfnamefont {D.~E.}\ \bibnamefont {Chang}},\
  }\href@noop {} {\bibfield  {journal} {\bibinfo  {journal} {Nature
  communications}\ }\textbf {\bibinfo {volume} {8}},\ \bibinfo {pages} {14696}
  (\bibinfo {year} {2017})}\BibitemShut {NoStop}%
\bibitem [{\citenamefont {John}(1984)}]{John1984}%
  \BibitemOpen
  \bibfield  {author} {\bibinfo {author} {\bibfnamefont {S.}~\bibnamefont
  {John}},\ }\href {\doibase 10.1103/PhysRevLett.53.2169} {\bibfield  {journal}
  {\bibinfo  {journal} {Phys. Rev. Lett.}\ }\textbf {\bibinfo {volume} {53}},\
  \bibinfo {pages} {2169} (\bibinfo {year} {1984})}\BibitemShut {NoStop}%
\bibitem [{\citenamefont {John}(1987)}]{John1987}%
  \BibitemOpen
  \bibfield  {author} {\bibinfo {author} {\bibfnamefont {S.}~\bibnamefont
  {John}},\ }\href {\doibase 10.1103/PhysRevLett.58.2486} {\bibfield  {journal}
  {\bibinfo  {journal} {Phys. Rev. Lett.}\ }\textbf {\bibinfo {volume} {58}},\
  \bibinfo {pages} {2486} (\bibinfo {year} {1987})}\BibitemShut {NoStop}%
\bibitem [{\citenamefont {John}\ and\ \citenamefont {Wang}(1990)}]{john1990}%
  \BibitemOpen
  \bibfield  {author} {\bibinfo {author} {\bibfnamefont {S.}~\bibnamefont
  {John}}\ and\ \bibinfo {author} {\bibfnamefont {J.}~\bibnamefont {Wang}},\
  }\href@noop {} {\bibfield  {journal} {\bibinfo  {journal} {Physical review
  letters}\ }\textbf {\bibinfo {volume} {64}},\ \bibinfo {pages} {2418}
  (\bibinfo {year} {1990})}\BibitemShut {NoStop}%
\bibitem [{\citenamefont {John}\ and\ \citenamefont {Wang}(1991)}]{John1991}%
  \BibitemOpen
  \bibfield  {author} {\bibinfo {author} {\bibfnamefont {S.}~\bibnamefont
  {John}}\ and\ \bibinfo {author} {\bibfnamefont {J.}~\bibnamefont {Wang}},\
  }\href {\doibase 10.1103/PhysRevB.43.12772} {\bibfield  {journal} {\bibinfo
  {journal} {Phys. Rev. B}\ }\textbf {\bibinfo {volume} {43}},\ \bibinfo
  {pages} {12772} (\bibinfo {year} {1991})}\BibitemShut {NoStop}%
\bibitem [{\citenamefont {John}\ and\ \citenamefont {Quang}(1994)}]{john1994}%
  \BibitemOpen
  \bibfield  {author} {\bibinfo {author} {\bibfnamefont {S.}~\bibnamefont
  {John}}\ and\ \bibinfo {author} {\bibfnamefont {T.}~\bibnamefont {Quang}},\
  }\href@noop {} {\bibfield  {journal} {\bibinfo  {journal} {Physical Review
  A}\ }\textbf {\bibinfo {volume} {50}},\ \bibinfo {pages} {1764} (\bibinfo
  {year} {1994})}\BibitemShut {NoStop}%
\bibitem [{\citenamefont {Gonz{\'{a}}lez-Tudela}\ \emph
  {et~al.}(2015)\citenamefont {Gonz{\'{a}}lez-Tudela}, \citenamefont {Hung},
  \citenamefont {Chang}, \citenamefont {Cirac},\ and\ \citenamefont
  {Kimble}}]{GonzalezTudela2015}%
  \BibitemOpen
  \bibfield  {author} {\bibinfo {author} {\bibfnamefont {A.}~\bibnamefont
  {Gonz{\'{a}}lez-Tudela}}, \bibinfo {author} {\bibfnamefont {C.-L.}\
  \bibnamefont {Hung}}, \bibinfo {author} {\bibfnamefont {D.~E.}\ \bibnamefont
  {Chang}}, \bibinfo {author} {\bibfnamefont {J.~I.}\ \bibnamefont {Cirac}}, \
  and\ \bibinfo {author} {\bibfnamefont {H.~J.}\ \bibnamefont {Kimble}},\
  }\href {\doibase 10.1038/nphoton.2015.54} {\bibfield  {journal} {\bibinfo
  {journal} {Nature Photonics}\ }\textbf {\bibinfo {volume} {9}},\ \bibinfo
  {pages} {320} (\bibinfo {year} {2015})}\BibitemShut {NoStop}%
\bibitem [{\citenamefont {Douglas}\ \emph {et~al.}(2015)\citenamefont
  {Douglas}, \citenamefont {Habibian}, \citenamefont {Hung}, \citenamefont
  {Gorshkov}, \citenamefont {Kimble},\ and\ \citenamefont
  {Chang}}]{Douglas2015}%
  \BibitemOpen
  \bibfield  {author} {\bibinfo {author} {\bibfnamefont {J.~S.}\ \bibnamefont
  {Douglas}}, \bibinfo {author} {\bibfnamefont {H.}~\bibnamefont {Habibian}},
  \bibinfo {author} {\bibfnamefont {C.-L.}\ \bibnamefont {Hung}}, \bibinfo
  {author} {\bibfnamefont {A.~V.}\ \bibnamefont {Gorshkov}}, \bibinfo {author}
  {\bibfnamefont {H.~J.}\ \bibnamefont {Kimble}}, \ and\ \bibinfo {author}
  {\bibfnamefont {D.~E.}\ \bibnamefont {Chang}},\ }\href {\doibase
  10.1038/nphoton.2015.57} {\bibfield  {journal} {\bibinfo  {journal} {Nature
  Photonics}\ }\textbf {\bibinfo {volume} {9}},\ \bibinfo {pages} {326}
  (\bibinfo {year} {2015})}\BibitemShut {NoStop}%
\bibitem [{\citenamefont {Shi}\ \emph {et~al.}(2016)\citenamefont {Shi},
  \citenamefont {Wu}, \citenamefont {Gonz{\'a}lez-Tudela},\ and\ \citenamefont
  {Cirac}}]{shi2016}%
  \BibitemOpen
  \bibfield  {author} {\bibinfo {author} {\bibfnamefont {T.}~\bibnamefont
  {Shi}}, \bibinfo {author} {\bibfnamefont {Y.-H.}\ \bibnamefont {Wu}},
  \bibinfo {author} {\bibfnamefont {A.}~\bibnamefont {Gonz{\'a}lez-Tudela}}, \
  and\ \bibinfo {author} {\bibfnamefont {J.~I.}\ \bibnamefont {Cirac}},\
  }\href@noop {} {\bibfield  {journal} {\bibinfo  {journal} {Physical Review
  X}\ }\textbf {\bibinfo {volume} {6}},\ \bibinfo {pages} {021027} (\bibinfo
  {year} {2016})}\BibitemShut {NoStop}%
\bibitem [{\citenamefont {Calaj{\'o}}\ \emph {et~al.}(2016)\citenamefont
  {Calaj{\'o}}, \citenamefont {Ciccarello}, \citenamefont {Chang},\ and\
  \citenamefont {Rabl}}]{calajo2016}%
  \BibitemOpen
  \bibfield  {author} {\bibinfo {author} {\bibfnamefont {G.}~\bibnamefont
  {Calaj{\'o}}}, \bibinfo {author} {\bibfnamefont {F.}~\bibnamefont
  {Ciccarello}}, \bibinfo {author} {\bibfnamefont {D.}~\bibnamefont {Chang}}, \
  and\ \bibinfo {author} {\bibfnamefont {P.}~\bibnamefont {Rabl}},\ }\href@noop
  {} {\bibfield  {journal} {\bibinfo  {journal} {Physical Review A}\ }\textbf
  {\bibinfo {volume} {93}},\ \bibinfo {pages} {033833} (\bibinfo {year}
  {2016})}\BibitemShut {NoStop}%
\bibitem [{\citenamefont {Gonz\'alez-Tudela}\ and\ \citenamefont
  {Cirac}(2017{\natexlab{a}})}]{GonzalezTudela2017}%
  \BibitemOpen
  \bibfield  {author} {\bibinfo {author} {\bibfnamefont {A.}~\bibnamefont
  {Gonz\'alez-Tudela}}\ and\ \bibinfo {author} {\bibfnamefont {J.~I.}\
  \bibnamefont {Cirac}},\ }\href {\doibase 10.1103/PhysRevA.96.043811}
  {\bibfield  {journal} {\bibinfo  {journal} {Phys. Rev. A}\ }\textbf {\bibinfo
  {volume} {96}},\ \bibinfo {pages} {043811} (\bibinfo {year}
  {2017}{\natexlab{a}})}\BibitemShut {NoStop}%
\bibitem [{\citenamefont {Gonz\'alez-Tudela}\ and\ \citenamefont
  {Cirac}(2017{\natexlab{b}})}]{GonzalezTudela2017b}%
  \BibitemOpen
  \bibfield  {author} {\bibinfo {author} {\bibfnamefont {A.}~\bibnamefont
  {Gonz\'alez-Tudela}}\ and\ \bibinfo {author} {\bibfnamefont {J.~I.}\
  \bibnamefont {Cirac}},\ }\href {\doibase 10.1103/PhysRevLett.119.143602}
  {\bibfield  {journal} {\bibinfo  {journal} {Phys. Rev. Lett.}\ }\textbf
  {\bibinfo {volume} {119}},\ \bibinfo {pages} {143602} (\bibinfo {year}
  {2017}{\natexlab{b}})}\BibitemShut {NoStop}%
\bibitem [{\citenamefont {Gonz{\'{a}}lez-Tudela}\ and\ \citenamefont
  {Cirac}(2018)}]{GonzalezTudela2018}%
  \BibitemOpen
  \bibfield  {author} {\bibinfo {author} {\bibfnamefont {A.}~\bibnamefont
  {Gonz{\'{a}}lez-Tudela}}\ and\ \bibinfo {author} {\bibfnamefont {J.~I.}\
  \bibnamefont {Cirac}},\ }\href {\doibase 10.22331/q-2018-10-01-97} {\bibfield
   {journal} {\bibinfo  {journal} {Quantum}\ }\textbf {\bibinfo {volume} {2}},\
  \bibinfo {pages} {97} (\bibinfo {year} {2018})}\BibitemShut {NoStop}%
\bibitem [{\citenamefont {Gonz\'alez-Tudela}\ and\ \citenamefont
  {Cirac}(2018)}]{GonzalezTudela2018b}%
  \BibitemOpen
  \bibfield  {author} {\bibinfo {author} {\bibfnamefont {A.}~\bibnamefont
  {Gonz\'alez-Tudela}}\ and\ \bibinfo {author} {\bibfnamefont {J.~I.}\
  \bibnamefont {Cirac}},\ }\href {\doibase 10.1103/PhysRevA.97.043831}
  {\bibfield  {journal} {\bibinfo  {journal} {Phys. Rev. A}\ }\textbf {\bibinfo
  {volume} {97}},\ \bibinfo {pages} {043831} (\bibinfo {year}
  {2018})}\BibitemShut {NoStop}%
\bibitem [{\citenamefont {Gonz{\'{a}}lez-Tudela}\ and\ \citenamefont
  {Galve}(2018)}]{GonzalezTudela2018d}%
  \BibitemOpen
  \bibfield  {author} {\bibinfo {author} {\bibfnamefont {A.}~\bibnamefont
  {Gonz{\'{a}}lez-Tudela}}\ and\ \bibinfo {author} {\bibfnamefont
  {F.}~\bibnamefont {Galve}},\ }\href {\doibase 10.1021/acsphotonics.8b01455}
  {\bibfield  {journal} {\bibinfo  {journal} {{ACS} Photonics}\ }\textbf
  {\bibinfo {volume} {6}},\ \bibinfo {pages} {221} (\bibinfo {year}
  {2018})}\BibitemShut {NoStop}%
\bibitem [{\citenamefont {Shi}\ \emph {et~al.}(2018{\natexlab{b}})\citenamefont
  {Shi}, \citenamefont {Wu}, \citenamefont {Gonz{\'a}lez-Tudela},\ and\
  \citenamefont {Cirac}}]{shi2018c}%
  \BibitemOpen
  \bibfield  {author} {\bibinfo {author} {\bibfnamefont {T.}~\bibnamefont
  {Shi}}, \bibinfo {author} {\bibfnamefont {Y.~H.}\ \bibnamefont {Wu}},
  \bibinfo {author} {\bibfnamefont {A.}~\bibnamefont {Gonz{\'a}lez-Tudela}}, \
  and\ \bibinfo {author} {\bibfnamefont {J.~I.}\ \bibnamefont {Cirac}},\
  }\href@noop {} {\bibfield  {journal} {\bibinfo  {journal} {New Journal of
  Physics}\ }\textbf {\bibinfo {volume} {20}},\ \bibinfo {pages} {105005}
  (\bibinfo {year} {2018}{\natexlab{b}})}\BibitemShut {NoStop}%
\bibitem [{\citenamefont {Khalfin}(1958)}]{Khalfin1958}%
  \BibitemOpen
  \bibfield  {author} {\bibinfo {author} {\bibfnamefont {L.~A.}\ \bibnamefont
  {Khalfin}},\ }\href {\doibase 10.1103/PhysRevB.35.979} {\bibfield  {journal}
  {\bibinfo  {journal} {Soviet Physics JETP}\ }\textbf {\bibinfo {volume}
  {6}},\ \bibinfo {pages} {1053} (\bibinfo {year} {1958})}\BibitemShut
  {NoStop}%
\bibitem [{\citenamefont {Bykov}(1975)}]{Bykov1975}%
  \BibitemOpen
  \bibfield  {author} {\bibinfo {author} {\bibfnamefont {V.~P.}\ \bibnamefont
  {Bykov}},\ }\href {\doibase 10.1070/qe1975v004n07abeh009654} {\bibfield
  {journal} {\bibinfo  {journal} {Soviet Journal of Quantum Electronics}\
  }\textbf {\bibinfo {volume} {4}},\ \bibinfo {pages} {861} (\bibinfo {year}
  {1975})}\BibitemShut {NoStop}%
\bibitem [{\citenamefont {Fonda}\ \emph {et~al.}(1978)\citenamefont {Fonda},
  \citenamefont {Ghirardi},\ and\ \citenamefont {Rimini}}]{Fonda1978}%
  \BibitemOpen
  \bibfield  {author} {\bibinfo {author} {\bibfnamefont {L.}~\bibnamefont
  {Fonda}}, \bibinfo {author} {\bibfnamefont {G.~C.}\ \bibnamefont {Ghirardi}},
  \ and\ \bibinfo {author} {\bibfnamefont {A.}~\bibnamefont {Rimini}},\ }\href
  {\doibase 10.1088/0034-4885/41/4/003} {\bibfield  {journal} {\bibinfo
  {journal} {Reports on Progress in Physics}\ }\textbf {\bibinfo {volume}
  {41}},\ \bibinfo {pages} {587} (\bibinfo {year} {1978})}\BibitemShut
  {NoStop}%
\bibitem [{\citenamefont {Onley}\ and\ \citenamefont
  {Kumar}(1992)}]{Onley1992}%
  \BibitemOpen
  \bibfield  {author} {\bibinfo {author} {\bibfnamefont {D.~S.}\ \bibnamefont
  {Onley}}\ and\ \bibinfo {author} {\bibfnamefont {A.}~\bibnamefont {Kumar}},\
  }\href {\doibase 10.1119/1.16897} {\bibfield  {journal} {\bibinfo  {journal}
  {American Journal of Physics}\ }\textbf {\bibinfo {volume} {60}},\ \bibinfo
  {pages} {432} (\bibinfo {year} {1992})}\BibitemShut {NoStop}%
\bibitem [{\citenamefont {Gaveau}\ and\ \citenamefont
  {Schulman}(1995)}]{gaveau1995}%
  \BibitemOpen
  \bibfield  {author} {\bibinfo {author} {\bibfnamefont {B.}~\bibnamefont
  {Gaveau}}\ and\ \bibinfo {author} {\bibfnamefont {L.}~\bibnamefont
  {Schulman}},\ }\href {\doibase 10.1088/0305-4470/28/24/029} {\bibfield
  {journal} {\bibinfo  {journal} {Journal of Physics A: Mathematical and
  General}\ }\textbf {\bibinfo {volume} {28}},\ \bibinfo {pages} {7359}
  (\bibinfo {year} {1995})}\BibitemShut {NoStop}%
\bibitem [{\citenamefont {Garmon}\ \emph {et~al.}(2009)\citenamefont {Garmon},
  \citenamefont {Nakamura}, \citenamefont {Hatano},\ and\ \citenamefont
  {Petrosky}}]{Garmon2009}%
  \BibitemOpen
  \bibfield  {author} {\bibinfo {author} {\bibfnamefont {S.}~\bibnamefont
  {Garmon}}, \bibinfo {author} {\bibfnamefont {H.}~\bibnamefont {Nakamura}},
  \bibinfo {author} {\bibfnamefont {N.}~\bibnamefont {Hatano}}, \ and\ \bibinfo
  {author} {\bibfnamefont {T.}~\bibnamefont {Petrosky}},\ }\href {\doibase
  10.1103/PhysRevB.80.115318} {\bibfield  {journal} {\bibinfo  {journal} {Phys.
  Rev. B}\ }\textbf {\bibinfo {volume} {80}},\ \bibinfo {pages} {115318}
  (\bibinfo {year} {2009})}\BibitemShut {NoStop}%
\bibitem [{\citenamefont {Garmon}\ \emph {et~al.}(2013)\citenamefont {Garmon},
  \citenamefont {Petrosky}, \citenamefont {Simine},\ and\ \citenamefont
  {Segal}}]{Garmon2013}%
  \BibitemOpen
  \bibfield  {author} {\bibinfo {author} {\bibfnamefont {S.}~\bibnamefont
  {Garmon}}, \bibinfo {author} {\bibfnamefont {T.}~\bibnamefont {Petrosky}},
  \bibinfo {author} {\bibfnamefont {L.}~\bibnamefont {Simine}}, \ and\ \bibinfo
  {author} {\bibfnamefont {D.}~\bibnamefont {Segal}},\ }\href {\doibase
  10.1002/prop.201200077} {\bibfield  {journal} {\bibinfo  {journal}
  {Fortschritte der Physik}\ }\textbf {\bibinfo {volume} {61}},\ \bibinfo
  {pages} {261} (\bibinfo {year} {2013})}\BibitemShut {NoStop}%
\bibitem [{\citenamefont {Garmon}(2013)}]{Garmon2013b}%
  \BibitemOpen
  \bibfield  {author} {\bibinfo {author} {\bibfnamefont {S.}~\bibnamefont
  {Garmon}},\ }in\ \href {\doibase 10.1364/cqo.2013.m6.06} {\emph {\bibinfo
  {booktitle} {The Rochester Conferences on Coherence and Quantum Optics and
  the Quantum Information and Measurement meeting}}}\ (\bibinfo  {publisher}
  {{OSA}},\ \bibinfo {year} {2013})\BibitemShut {NoStop}%
\bibitem [{\citenamefont {Lombardo}\ \emph {et~al.}(2014)\citenamefont
  {Lombardo}, \citenamefont {Ciccarello},\ and\ \citenamefont
  {Palma}}]{Lombardo2014}%
  \BibitemOpen
  \bibfield  {author} {\bibinfo {author} {\bibfnamefont {F.}~\bibnamefont
  {Lombardo}}, \bibinfo {author} {\bibfnamefont {F.}~\bibnamefont
  {Ciccarello}}, \ and\ \bibinfo {author} {\bibfnamefont {G.~M.}\ \bibnamefont
  {Palma}},\ }\href {\doibase 10.1103/PhysRevA.89.053826} {\bibfield  {journal}
  {\bibinfo  {journal} {Phys. Rev. A}\ }\textbf {\bibinfo {volume} {89}},\
  \bibinfo {pages} {053826} (\bibinfo {year} {2014})}\BibitemShut {NoStop}%
\bibitem [{\citenamefont {S{\'a}nchez-Burillo}\ \emph
  {et~al.}(2017)\citenamefont {S{\'a}nchez-Burillo}, \citenamefont {Zueco},
  \citenamefont {Mart{\'\i}n-Moreno},\ and\ \citenamefont
  {Garc{\'\i}a-Ripoll}}]{sanchez2017b}%
  \BibitemOpen
  \bibfield  {author} {\bibinfo {author} {\bibfnamefont {E.}~\bibnamefont
  {S{\'a}nchez-Burillo}}, \bibinfo {author} {\bibfnamefont {D.}~\bibnamefont
  {Zueco}}, \bibinfo {author} {\bibfnamefont {L.}~\bibnamefont
  {Mart{\'\i}n-Moreno}}, \ and\ \bibinfo {author} {\bibfnamefont {J.~J.}\
  \bibnamefont {Garc{\'\i}a-Ripoll}},\ }\href {\doibase
  10.1103/PhysRevA.96.023831} {\bibfield  {journal} {\bibinfo  {journal}
  {Physical Review A}\ }\textbf {\bibinfo {volume} {96}},\ \bibinfo {pages}
  {023831} (\bibinfo {year} {2017})}\BibitemShut {NoStop}%
\bibitem [{\citenamefont {Ulrich}(1999)}]{Weiss}%
  \BibitemOpen
  \bibfield  {author} {\bibinfo {author} {\bibfnamefont {W.}~\bibnamefont
  {Ulrich}},\ }\href {https://books.google.es/books?id=2WfVCgAAQBAJ} {\emph
  {\bibinfo {title} {Quantum Dissipative Systems (Second Edition)}}},\ Series
  In Modern Condensed Matter Physics\ (\bibinfo  {publisher} {World Scientific
  Publishing Company},\ \bibinfo {year} {1999})\BibitemShut {NoStop}%
\bibitem [{\citenamefont {Prior}\ \emph {et~al.}(2010)\citenamefont {Prior},
  \citenamefont {Chin}, \citenamefont {Huelga},\ and\ \citenamefont
  {Plenio}}]{prior2010}%
  \BibitemOpen
  \bibfield  {author} {\bibinfo {author} {\bibfnamefont {J.}~\bibnamefont
  {Prior}}, \bibinfo {author} {\bibfnamefont {A.~W.}\ \bibnamefont {Chin}},
  \bibinfo {author} {\bibfnamefont {S.~F.}\ \bibnamefont {Huelga}}, \ and\
  \bibinfo {author} {\bibfnamefont {M.~B.}\ \bibnamefont {Plenio}},\
  }\href@noop {} {\bibfield  {journal} {\bibinfo  {journal} {Physical review
  letters}\ }\textbf {\bibinfo {volume} {105}},\ \bibinfo {pages} {050404}
  (\bibinfo {year} {2010})}\BibitemShut {NoStop}%
\bibitem [{\citenamefont {Grifoni}\ and\ \citenamefont
  {H{\"a}nggi}(1998)}]{grifoni1998}%
  \BibitemOpen
  \bibfield  {author} {\bibinfo {author} {\bibfnamefont {M.}~\bibnamefont
  {Grifoni}}\ and\ \bibinfo {author} {\bibfnamefont {P.}~\bibnamefont
  {H{\"a}nggi}},\ }\href@noop {} {\bibfield  {journal} {\bibinfo  {journal}
  {Physics Reports}\ }\textbf {\bibinfo {volume} {304}},\ \bibinfo {pages}
  {229} (\bibinfo {year} {1998})}\BibitemShut {NoStop}%
\bibitem [{\citenamefont {Le~Hur}(2010)}]{lehur2010}%
  \BibitemOpen
  \bibfield  {author} {\bibinfo {author} {\bibfnamefont {K.}~\bibnamefont
  {Le~Hur}},\ }in\ \href@noop {} {\emph {\bibinfo {booktitle} {Understanding
  Quantum Phase Transitions}}}\ (\bibinfo  {publisher} {CRC Press},\ \bibinfo
  {year} {2010})\ pp.\ \bibinfo {pages} {245--268}\BibitemShut {NoStop}%
\bibitem [{\citenamefont {Silbey}\ and\ \citenamefont
  {Harris}(1984)}]{silbey1984}%
  \BibitemOpen
  \bibfield  {author} {\bibinfo {author} {\bibfnamefont {R.}~\bibnamefont
  {Silbey}}\ and\ \bibinfo {author} {\bibfnamefont {R.~A.}\ \bibnamefont
  {Harris}},\ }\href@noop {} {\bibfield  {journal} {\bibinfo  {journal} {The
  Journal of chemical physics}\ }\textbf {\bibinfo {volume} {80}},\ \bibinfo
  {pages} {2615} (\bibinfo {year} {1984})}\BibitemShut {NoStop}%
\bibitem [{\citenamefont {Bera}\ \emph {et~al.}(2014)\citenamefont {Bera},
  \citenamefont {Nazir}, \citenamefont {Chin}, \citenamefont {Baranger},\ and\
  \citenamefont {Florens}}]{Bera2014}%
  \BibitemOpen
  \bibfield  {author} {\bibinfo {author} {\bibfnamefont {S.}~\bibnamefont
  {Bera}}, \bibinfo {author} {\bibfnamefont {A.}~\bibnamefont {Nazir}},
  \bibinfo {author} {\bibfnamefont {A.~W.}\ \bibnamefont {Chin}}, \bibinfo
  {author} {\bibfnamefont {H.~U.}\ \bibnamefont {Baranger}}, \ and\ \bibinfo
  {author} {\bibfnamefont {S.}~\bibnamefont {Florens}},\ }\href {\doibase
  10.1103/physrevb.90.075110} {\bibfield  {journal} {\bibinfo  {journal}
  {Physical Review B}\ }\textbf {\bibinfo {volume} {90}} (\bibinfo {year}
  {2014}),\ 10.1103/physrevb.90.075110}\BibitemShut {NoStop}%
\bibitem [{\citenamefont {D\'{\i}az-Camacho}\ \emph {et~al.}(2016)\citenamefont
  {D\'{\i}az-Camacho}, \citenamefont {Bermudez},\ and\ \citenamefont
  {Garc\'{\i}a-Ripoll}}]{Camacho2016}%
  \BibitemOpen
  \bibfield  {author} {\bibinfo {author} {\bibfnamefont {G.}~\bibnamefont
  {D\'{\i}az-Camacho}}, \bibinfo {author} {\bibfnamefont {A.}~\bibnamefont
  {Bermudez}}, \ and\ \bibinfo {author} {\bibfnamefont {J.~J.}\ \bibnamefont
  {Garc\'{\i}a-Ripoll}},\ }\href {\doibase 10.1103/PhysRevA.93.043843}
  {\bibfield  {journal} {\bibinfo  {journal} {Phys. Rev. A}\ }\textbf {\bibinfo
  {volume} {93}},\ \bibinfo {pages} {043843} (\bibinfo {year}
  {2016})}\BibitemShut {NoStop}%
\bibitem [{\citenamefont {Zueco}\ and\ \citenamefont
  {Garc{\'\i}a-Ripoll}(2019)}]{zueco2019}%
  \BibitemOpen
  \bibfield  {author} {\bibinfo {author} {\bibfnamefont {D.}~\bibnamefont
  {Zueco}}\ and\ \bibinfo {author} {\bibfnamefont {J.}~\bibnamefont
  {Garc{\'\i}a-Ripoll}},\ }\href@noop {} {\bibfield  {journal} {\bibinfo
  {journal} {Physical Review A}\ }\textbf {\bibinfo {volume} {99}},\ \bibinfo
  {pages} {013807} (\bibinfo {year} {2019})}\BibitemShut {NoStop}%
\bibitem [{\citenamefont {Shi}\ \emph {et~al.}(2018{\natexlab{c}})\citenamefont
  {Shi}, \citenamefont {Demler},\ and\ \citenamefont {Cirac}}]{shi2018b}%
  \BibitemOpen
  \bibfield  {author} {\bibinfo {author} {\bibfnamefont {T.}~\bibnamefont
  {Shi}}, \bibinfo {author} {\bibfnamefont {E.}~\bibnamefont {Demler}}, \ and\
  \bibinfo {author} {\bibfnamefont {J.~I.}\ \bibnamefont {Cirac}},\ }\href@noop
  {} {\bibfield  {journal} {\bibinfo  {journal} {Annals of Physics}\ }\textbf
  {\bibinfo {volume} {390}},\ \bibinfo {pages} {245} (\bibinfo {year}
  {2018}{\natexlab{c}})}\BibitemShut {NoStop}%
\bibitem [{\citenamefont {Di~Stefano}\ \emph {et~al.}(2019)\citenamefont
  {Di~Stefano}, \citenamefont {Settineri}, \citenamefont {Macr{\`\i}},
  \citenamefont {Garziano}, \citenamefont {Stassi}, \citenamefont {Savasta},\
  and\ \citenamefont {Nori}}]{di2019}%
  \BibitemOpen
  \bibfield  {author} {\bibinfo {author} {\bibfnamefont {O.}~\bibnamefont
  {Di~Stefano}}, \bibinfo {author} {\bibfnamefont {A.}~\bibnamefont
  {Settineri}}, \bibinfo {author} {\bibfnamefont {V.}~\bibnamefont
  {Macr{\`\i}}}, \bibinfo {author} {\bibfnamefont {L.}~\bibnamefont
  {Garziano}}, \bibinfo {author} {\bibfnamefont {R.}~\bibnamefont {Stassi}},
  \bibinfo {author} {\bibfnamefont {S.}~\bibnamefont {Savasta}}, \ and\
  \bibinfo {author} {\bibfnamefont {F.}~\bibnamefont {Nori}},\ }\href@noop {}
  {\bibfield  {journal} {\bibinfo  {journal} {Nature Physics}\ ,\ \bibinfo
  {pages} {1}} (\bibinfo {year} {2019})}\BibitemShut {NoStop}%
\bibitem [{Note1()}]{Note1}%
  \BibitemOpen
  \bibinfo {note} {Take a function $\protect \mathfrak {f}(\omega )$. Then,
  $\DOTSI \intop \ilimits@ d \omega J(\omega )\protect \mathfrak {f}(\omega )=
  2 \pi g /\protect \sqrt {N} \DOTSB \sum@ \slimits@ \protect \mathfrak
  {f}(\omega _k) \to 2 \pi g \DOTSI \intop \ilimits@ d \omega (d\omega _k/
  dk)^{-1} \protect \mathfrak {f}(\omega )$. This yields Eq. \protect \textup
  {\hbox {\mathsurround \z@ \protect \normalfont (\ignorespaces \ref
  {Jwca}\unskip \@@italiccorr )}} in the main text.}\BibitemShut {Stop}%
\bibitem [{\citenamefont {McCutcheon}\ \emph {et~al.}(2010)\citenamefont
  {McCutcheon}, \citenamefont {Nazir}, \citenamefont {Bose},\ and\
  \citenamefont {Fisher}}]{McCutcheon2010}%
  \BibitemOpen
  \bibfield  {author} {\bibinfo {author} {\bibfnamefont {D.~P.~S.}\
  \bibnamefont {McCutcheon}}, \bibinfo {author} {\bibfnamefont
  {A.}~\bibnamefont {Nazir}}, \bibinfo {author} {\bibfnamefont
  {S.}~\bibnamefont {Bose}}, \ and\ \bibinfo {author} {\bibfnamefont {A.~J.}\
  \bibnamefont {Fisher}},\ }\href {\doibase 10.1103/physrevb.81.235321}
  {\bibfield  {journal} {\bibinfo  {journal} {Phys. Rev. B}\ }\textbf {\bibinfo
  {volume} {81}} (\bibinfo {year} {2010}),\
  10.1103/physrevb.81.235321}\BibitemShut {NoStop}%
\bibitem [{\citenamefont {Zheng}\ \emph {et~al.}(2015)\citenamefont {Zheng},
  \citenamefont {L\"{u}},\ and\ \citenamefont {Zhao}}]{Zheng2015}%
  \BibitemOpen
  \bibfield  {author} {\bibinfo {author} {\bibfnamefont {H.}~\bibnamefont
  {Zheng}}, \bibinfo {author} {\bibfnamefont {Z.}~\bibnamefont {L\"{u}}}, \
  and\ \bibinfo {author} {\bibfnamefont {Y.}~\bibnamefont {Zhao}},\ }\href
  {\doibase 10.1103/physreve.91.062115} {\bibfield  {journal} {\bibinfo
  {journal} {Phys. Rev. E}\ }\textbf {\bibinfo {volume} {91}} (\bibinfo {year}
  {2015}),\ 10.1103/physreve.91.062115}\BibitemShut {NoStop}%
\bibitem [{\citenamefont {Guinea}\ \emph {et~al.}(1998)\citenamefont {Guinea},
  \citenamefont {Bascones},\ and\ \citenamefont {Calderon}}]{guinea1998}%
  \BibitemOpen
  \bibfield  {author} {\bibinfo {author} {\bibfnamefont {F.}~\bibnamefont
  {Guinea}}, \bibinfo {author} {\bibfnamefont {E.}~\bibnamefont {Bascones}}, \
  and\ \bibinfo {author} {\bibfnamefont {M.}~\bibnamefont {Calderon}},\ }in\
  \href@noop {} {\emph {\bibinfo {booktitle} {AIP Conference Proceedings}}},\
  Vol.\ \bibinfo {volume} {438}\ (\bibinfo {organization} {AIP},\ \bibinfo
  {year} {1998})\ pp.\ \bibinfo {pages} {1--82}\BibitemShut {NoStop}%
\bibitem [{Note2()}]{Note2}%
  \BibitemOpen
  \bibinfo {note} {\label {foot:loc} For those who have condensed-matter
  background the spin-boson is paradigmatic in impurity models. In those
  formulations that naturally lead to a double-well interpretation of the TLS,
  the roles of $\sigma ^x$ and $\sigma ^z$ are switched in the Hamiltonian. In
  that case, $c_k = 0$ is viewed as the delocalized regime whereas $\Delta = 0$
  is viewed as the localized regime.}\BibitemShut {Stop}%
\bibitem [{\citenamefont {Spohn1}\ and\ \citenamefont
  {Domcke1}(1985)}]{Spohn1985}%
  \BibitemOpen
  \bibfield  {author} {\bibinfo {author} {\bibfnamefont {H.}~\bibnamefont
  {Spohn1}}\ and\ \bibinfo {author} {\bibfnamefont {R.}~\bibnamefont
  {Domcke1}},\ }\href {\doibase 10.1007/bf01009015} {\bibfield  {journal}
  {\bibinfo  {journal} {Journal of Statistical Physics}\ }\textbf {\bibinfo
  {volume} {41}},\ \bibinfo {pages} {389} (\bibinfo {year} {1985})}\BibitemShut
  {NoStop}%
\bibitem [{\citenamefont {L\"{o}wen}(1988)}]{Lowen1988}%
  \BibitemOpen
  \bibfield  {author} {\bibinfo {author} {\bibfnamefont {H.}~\bibnamefont
  {L\"{o}wen}},\ }\href {\doibase 10.1103/physrevb.37.8661} {\bibfield
  {journal} {\bibinfo  {journal} {Physical Review B}\ }\textbf {\bibinfo
  {volume} {37}},\ \bibinfo {pages} {8661} (\bibinfo {year}
  {1988})}\BibitemShut {NoStop}%
\bibitem [{\citenamefont {Novotny}\ and\ \citenamefont
  {Hecht}(2006)}]{nano_optics}%
  \BibitemOpen
  \bibfield  {author} {\bibinfo {author} {\bibfnamefont {L.}~\bibnamefont
  {Novotny}}\ and\ \bibinfo {author} {\bibfnamefont {B.}~\bibnamefont
  {Hecht}},\ }\href@noop {} {\emph {\bibinfo {title} {Principles of
  Nano-Optics}}}\ (\bibinfo  {publisher} {Cambridge University Press},\
  \bibinfo {year} {2006})\BibitemShut {NoStop}%
\bibitem [{\citenamefont {Longo}\ \emph {et~al.}(2010)\citenamefont {Longo},
  \citenamefont {Schmitteckert},\ and\ \citenamefont {Busch}}]{Longo2010}%
  \BibitemOpen
  \bibfield  {author} {\bibinfo {author} {\bibfnamefont {P.}~\bibnamefont
  {Longo}}, \bibinfo {author} {\bibfnamefont {P.}~\bibnamefont
  {Schmitteckert}}, \ and\ \bibinfo {author} {\bibfnamefont {K.}~\bibnamefont
  {Busch}},\ }\href {\doibase 10.1103/PhysRevLett.104.023602} {\bibfield
  {journal} {\bibinfo  {journal} {Phys. Rev. Lett.}\ }\textbf {\bibinfo
  {volume} {104}},\ \bibinfo {pages} {023602} (\bibinfo {year}
  {2010})}\BibitemShut {NoStop}%
\bibitem [{\citenamefont {Longo}\ \emph {et~al.}(2011)\citenamefont {Longo},
  \citenamefont {Schmitteckert},\ and\ \citenamefont {Busch}}]{Longo2011}%
  \BibitemOpen
  \bibfield  {author} {\bibinfo {author} {\bibfnamefont {P.}~\bibnamefont
  {Longo}}, \bibinfo {author} {\bibfnamefont {P.}~\bibnamefont
  {Schmitteckert}}, \ and\ \bibinfo {author} {\bibfnamefont {K.}~\bibnamefont
  {Busch}},\ }\href {\doibase 10.1103/PhysRevA.83.063828} {\bibfield  {journal}
  {\bibinfo  {journal} {Phys. Rev. A}\ }\textbf {\bibinfo {volume} {83}},\
  \bibinfo {pages} {063828} (\bibinfo {year} {2011})}\BibitemShut {NoStop}%
\bibitem [{\citenamefont {S{\'{a}}nchez-Burillo}\ \emph
  {et~al.}(2018)\citenamefont {S{\'{a}}nchez-Burillo}, \citenamefont {Cadarso},
  \citenamefont {Mart{\'{\i}}n-Moreno}, \citenamefont {Garc{\'{\i}}a-Ripoll},\
  and\ \citenamefont {Zueco}}]{SanchezBurillo2018}%
  \BibitemOpen
  \bibfield  {author} {\bibinfo {author} {\bibfnamefont {E.}~\bibnamefont
  {S{\'{a}}nchez-Burillo}}, \bibinfo {author} {\bibfnamefont {A.}~\bibnamefont
  {Cadarso}}, \bibinfo {author} {\bibfnamefont {L.}~\bibnamefont
  {Mart{\'{\i}}n-Moreno}}, \bibinfo {author} {\bibfnamefont {J.~J.}\
  \bibnamefont {Garc{\'{\i}}a-Ripoll}}, \ and\ \bibinfo {author} {\bibfnamefont
  {D.}~\bibnamefont {Zueco}},\ }\href {\doibase 10.1088/1367-2630/aa9cc2}
  {\bibfield  {journal} {\bibinfo  {journal} {New Journal of Physics}\ }\textbf
  {\bibinfo {volume} {20}},\ \bibinfo {pages} {013017} (\bibinfo {year}
  {2018})}\BibitemShut {NoStop}%
\bibitem [{Note3()}]{Note3}%
  \BibitemOpen
  \bibinfo {note} {To exemplify the weakness of the dipole-dipole interaction,
  it is insightful to remember that it was not strong enough to explain
  ferromagnetism in solids. The \protect \textit {exchange interaction} had to
  be introduced in the study of ferromagnetism for this very
  reason.}\BibitemShut {Stop}%
\end{thebibliography}%

\end{document}